\documentclass[aps,prb,twocolumn,floatfix,showkeys,superscriptaddress,amssymb,nofootinbib,preprintnumbers,10pt]{revtex4-2}
\usepackage[english]{babel}
\usepackage{amsmath}
\usepackage{graphicx}
\usepackage[colorlinks=true, allcolors=blue]{hyperref}
\usepackage{xcolor}
\usepackage{multirow}
\usepackage{textcomp}
\usepackage{comment}
\usepackage{graphicx}
\usepackage{lipsum}
\usepackage{bm}                   
\usepackage{amsmath}
\usepackage{csquotes}        
\usepackage{nicefrac}
\usepackage{multirow}
\usepackage{color, xcolor}
\usepackage{soul}
\usepackage[colorinlistoftodos,textwidth=2cm]{todonotes}

\usepackage{colortbl} 
\usepackage{booktabs} 
\usepackage{bigstrut} 
\usepackage{textcomp}

\usepackage{orcidlink}
\usepackage{balance}

\begin{document}
\title{Impact of interstitial carbon on local lattice distortions in CoCrFeMnNi high-entropy alloys}

\author{Alevtina Smekhova*\orcidlink{0000-0003-0946-2909}}
\affiliation{Helmholtz-Zentrum Berlin für Materialien und Energie (HZB), 12489 Berlin, Germany}
\email{Corresponding authors: alevtina.smekhova@helmholtz-berlin.de (A. Smekhova); a.kuzmin@cfi.lu.lv (A. Kuzmin); yuji.ikeda@imw.uni-stuttgart.de (Y. Ikeda); divin@uni-muenster.de (S. Divinski)}
\author{Alexei Kuzmin*\orcidlink{0000-0003-4641-6354}}
\affiliation{Institute of Solid State Physics, University of Latvia, LV-1063 Riga, Latvia}
\author{G. Mohan Muralikrishna\orcidlink{0000-0002-5841-7496}}
\affiliation{Department of Metallurgical and Materials Engineering, Indian Institute of Technology Patna, 801106 Bihta, India}
\author{Edmund Welter\orcidlink{0000-0003-0698-3151}}
\affiliation{Deutsches Elektronen-Synchrotron DESY, 22603 Hamburg, Germany}
\author{Sergiu Levcenko\orcidlink{0000-0003-2534-5227}}
\affiliation{Deutsches Elektronen-Synchrotron DESY, 22603 Hamburg, Germany}
\author{Fritz K\"ormann\orcidlink{0000-0003-3050-6291}}
\affiliation{Interdisciplinary Centre for Advanced Materials Simulation, Ruhr-Universität Bochum, 44801 Bochum, Germany}
\affiliation{Department of Computational Materials Design, Max-Planck-Institute for Sustainable Materials, 40237 Düsseldorf, Germany}
\author{Yuji Ikeda\orcidlink{0000-0001-9176-3270}}
\affiliation{Institute for Materials Science, University of Stuttgart, 70569 Stuttgart, Germany}
\author{Sergiy V. Divinski*\orcidlink{0000-0003-1935-9542}}
\affiliation{Institute of Materials Physics, University of M\"unster, 48149 M\"unster, Germany}
\date{\today}

\begin{abstract}
Here, we explore component-dependent local lattice distortions in polycrystalline, equiatomic, face-centered cubic CrMnFeCoNi high-entropy alloys and their modifications induced by dilute interstitial carbon. Multi-edge extended X-ray absorption fine structure spectroscopy combined with reverse Monte Carlo analysis reveals that the Cr component experiences the most substantial local distortions, independent of the temperature of prolonged annealing treatments (993~K or 1373~K) and the nominal carbon content (0 to 0.8~at.\%). The static disorder around Cr atoms was found to increase markedly and monotonically upon carbon alloying, whereas Mn, Fe, Co, and Ni demonstrate weaker and non-monotonic tendencies. The carbon-induced lattice distortions extend over several coordination shells, indicating the pronounced effect of the carbon presence on the local environment around Cr absorbers. First-principles density functional theory and finite-temperature molecular dynamics simulations confirm the greater impact of carbon on the local lattice distortions around Cr than around the other $3d$ constituent elements, based on the previous finding that carbon preferentially occupies Cr-rich interstitial sites. These results provide decisive hints towards the atomistic origin of the non-monotonic diffusion behavior previously reported for carbon-doped CrMnFeCoNi alloys, and are noticeable for understanding the carbon-induced phase transitions in compositionally complex systems.
\end{abstract}


\maketitle
\section{Introduction}

Over recent decades, high-entropy alloys (HEAs) have been continuously considered as promising functional materials with high potential in different technological applications, including renewable energy technologies like hydrogen storage, carbon dioxide conversion, oxygen catalysis, rechargeable batteries, and supercapacitors \cite{Cantor2004, Cantor2021, Ma2021, Xie2025}. The chemical complexity of multi-principal-component alloys, achieved through advances in accurate alloying of many different components within a broad range of desired proportions, results in a variety of attractive properties that can surpass the properties of conventional counterparts. A large number of different local atomic configurations available due to several HEA components (five or more) provide not only a great choice of active sites at their surfaces but also huge variations in the local environment of a particular constituent element in the entire volume of compositionally complex systems.

The equiatomic single-phase CrMnFeCoNi (Cantor) alloy with a face-centered-cubic (fcc) crystalline structure is one of the most promising HEAs for applications due to its advantageous combination of mechanical and physical properties \cite{MIRACLE2017448, Gludovatz2014}, which arise from the compositional disorder and the resulting variety of local atomic interactions. Recent experimental studies have clearly indicated that additional alloying with interstitial carbon (and/or nitrogen) further improves the HEA strength and plasticity through the combined effects of twinning-induced plasticity, precipitation, and interstitial solid solution hardening mechanisms \cite{LI2019, Kang2010, Peng2019}. Interstitial atom doping (e.g. with B, C, O, or N) turned out to be an effective strategy for tuning the microstructural stability and mechanical behavior of high-entropy alloys \cite{SHIM2022-int, Semenyuk2022-int, ZHU2023-int}, even in dual-phase
multi-principal element alloys \cite{DONG2026-int}. However, although introducing a carbon content of only 0.5 at.\% into equiatomic CrMnFeCoNi HEA increases the strength compared with the carbon-free alloy \cite{Wu2015}, higher carbon concentrations (around 1.0 at.\%) lead to the noticeable formation of carbides \cite{Guo2019}. In general, this indicates that even small amounts of interstitial carbon significantly influence the thermodynamics and kinetics of the CrMnFeCoNi alloy, triggering phase transformations in the system. Thus, deliberate alloying with carbon performed in a controllable way can be considered as an effective method for tuning the desired properties of multi-principal-component HEAs.

A reliable prediction of the lifetime properties and phase stability of different compositionally complex systems is of key importance in the development of a new generation of HEAs for advanced applications and requires, among other factors, an understanding of their diffusion behavior. In particular, knowing the impact of interstitial carbon incorporation and its correlation with vacancy formation/migration and lattice relaxations is imperative, since interstitial elements have been shown to affect the diffusion jumps of substitutional atoms \cite{LUKIANOVA2020, LUKIANOVA2022}. A strongly non-monotonic dependence of the diffusion coefficients on carbon alloying was observed in (CrMnFeCoNi)$_{1-x}$C$_x$ HEAs, especially for alloys annealed at high temperatures above 1300~K \cite{LUKIANOVA2022}. Moreover, these dependencies vary characteristically among the different atomic species. This effect was assigned to lattice distortions imposed by interstitially dissolved carbon, interactions between interstitial carbon and thermal vacancies, and the predominant short-range order within the vacancy neighborhood in the alloy \cite{LUKIANOVA2022}.

Interstitial elements such as carbon could lead to much higher lattice distortions than are typically provided by substitutional elements, since they can cause a compressive stress field within the alloys \cite{Su2022, Casillas2020}. From an atomistic point of view, the impact of interstitial carbon alloying can be explored theoretically by first-principles density functional theory (DFT) calculations, providing insights into fundamental mechanisms existing at the atomic scale. The first-principles studies have reported the impact of carbon supplement on, e.g., stacking fault energies, phase stability and migration barriers of mono-vacancy defects in CrMnFeCoNi-based HEAs \cite{Ikeda_PRM_2019_Impact, Lu2021}; moreover, an energetically favorable Cr-rich local environment around carbon interstitial sites has been noticed. Experimentally, local atomic environment and component-dependent lattice relaxations in compositionally complex systems can be probed using multi-edge extended X-ray absorption fine structure (EXAFS) spectroscopy combined with reverse Monte Carlo (RMC) simulations \cite{BAKRADZE2023}. It has already been demonstrated that in the case of five-principal-component HEAs the RMC-based EXAFS analysis can unveil the statistically-averaged peculiarities of the component-dependent local environment for each constituent element through radial pair distribution functions \cite{NaRe1, Jalcom, NaRe2, NaRe3, NaRe4}. Recently, it has also been found that for carbon-free nearly equiatomic single-phase single crystalline CrMnFeCoNi alloys prepared at different annealing temperatures, the peculiarities in the local environment of the 3d components are visible in the wavelet transforms (WTs) of the room temperature EXAFS oscillations and can be correlated with the relative diffusivities of the constituents \cite{NaRe3}. The changes obtained in the element-specific local environments were related to vacancy-mediated effects.

In the present work, we use multi-edge EXAFS spectroscopy in combination with RMC simulations to probe the peculiarities of the local environment in the (CrMnFeCoNi)$_{1-x}$C$_x$  HEAs ($x$ = 0, 0.3, and 0.8~at.\%) and to determine the variations in component-dependent lattice relaxations induced by interstitial carbon atoms depending on the fraction of carbon alloying. 
The present work is focused on correlations between the impact of interstitial carbon on local environments and the previous \cite{LUKIANOVA2020, LUKIANOVA2022} tracer diffusion measurements to elucidate the way how dilute interstitial carbon modifies local lattice distortions in equiatomic Cantor alloys. DFT and molecular-dynamics (MD) simulations are employed to support the experimental observations of lattice distortions. Our results indicate that the alloys retain a single-phase fcc crystallographic lattice with a random chemical environment around each constituent element at the atomic scale, while showing clear signs of an incipient transformation phase appearing exclusively in the local environment of Cr atoms depending on the carbon content. Explicit changes in component-dependent lattice distortions and their  correlations with the tracer diffusion data are elucidated.

\section{Materials and methods}
\subsection{Sample preparation}

The polycrystalline (CrMnFeCoNi)$_{1-x}$C$_x$ alloys (with nominal C content $x$ = 0, 0.3 and 0.8~at.\%) were synthesized using pure 3d metal elements (better than 99.9~wt.\% purity) and carbon by melting and casting in a vacuum induction furnace. The thermomechanical processing was performed similarly to that described in \cite{LI2019, LUKIANOVA2020, LUKIANOVA2022}. The cast alloy plates were cold-rolled to a thickness reduction of about 50~\% and subsequently homogenized at 1473~K for 3~h in an atmosphere of argon followed by water quenching. This processing maximized the amount of interstitially dissolved carbon in the fcc matrix. The final bulk chemical composition was determined by a wet-chemical analysis \cite{LUKIANOVA2022}.

Disk- or square-shaped samples of about 5~mm in lateral size and about 0.5~mm in thickness were cut by spark-erosion from the cylindrical ingots. Subsequently, one face was polished to mirror-like quality using standard metallography procedures. The samples were again annealed in a purified Ar atmosphere at 1373~K for 24~hours to remove the induced mechanical stresses. After such annealing treatment, the samples were fine polished using an oxide suspension with silica particle sizes around 50~nm for more than 30~min. Finally, the samples were polished with soap and ethanol for 5~min to remove the nanosilica particles.

Subsequently, two characteristic states of each alloy were deliberately prepared for the present EXAFS spectroscopy studies. A low temperature (LT) annealed state was prepared from the initially homogenized and polished samples by long-term annealing at 993~K for 14~days, while a high temperature (HT) state was prepared analogously by final annealing at 1373~K for 3~days. After annealing at temperatures corresponding to the LT or HT states, the samples were cooled to room temperature by quenching in ice water. The quenching rates are estimated at about 100~K/s. The choice of these particular temperatures is related to the non-monotonic dependence of diffusion coefficients (especially for Mn) on carbon doping, as found earlier in \cite{LUKIANOVA2022} at high annealing temperatures above 1300~K, whereas a monotonic dependence was observed for annealing temperatures below 1200~K. A minor fraction of M$_{23}$C$_6$ carbides, found after annealing treatments and distributed both at grain boundaries (GB) and in grain interiors, as initially reported in \cite{LI2019}, did not significantly influence the results of the tracer diffusion measurements performed by Lukianova et al. \cite{LUKIANOVA2022}. 

Though, by a careful microscopic examination, a few micron-large carbide precipitates were sporadically observed only for the alloy with 0.8~at.\%C in LT conditions at the rod edges, and their local area density was estimated to be less than one such particle per about $100 \times 100$~{\textmu}m$^2$. CALPHAD calculations \cite{Ahmadreza2023} using the Thermocalc software \cite{ANDERSSON2002-Thermocalc} and the available  thermodynamic data \cite{HAASE2017-CALPHAD, KIES2021-CALPHAD, Ahmadreza2023} rate the equilibrium interstitial carbon solubility in CrMnFeCoNi as about 0.01\% and 0.4\% at 993~K and 1373~K, respectively. The present microscopic examinations indicate at least kinetic limitations of the carbide formation in the present samples. Since only the central part of each sample, where no carbides were found, was used for the EXAFS study, carbides are considered to be not critical and there are only the interstitial carbon atoms to be accounted for. Note that the interstitial solubility of carbon in the Cantor alloy is relatively large according to the DFT-based calculations \cite{Baker2023}.

Specific kinks in the Arrhenius dependencies of the tracer diffusion coefficients were reported by Gaertner et al. \cite{GAERTNER2020} at about 1100~K for Co and Ni and at 900~K for Cr and Fe, and were interpreted as a low-temperature modification of the equiatomic solid-solution phase in (C-free) CoCrFeMnNi. Those conclusions correlate with the results of the recent EXAFS study which revealed specific variations in the local environment distortions connected to the different degrees of compositional disorder \cite{NaRe3}.

The chemical composition and grain sizes in the alloys were controlled by energy dispersive X-ray spectroscopy and scanning electron microscopy (SEM), respectively. The microstructure was found to be reasonably coarse-grained and suitable for the EXAFS studies. The grain size was estimated as about $140 \pm 50$~{\textmu}m for all alloys (independently on the C content). In this case, the fraction of GB material is too small, far below $10^{-4}$, to affect the local sampling in the subsequent analysis, since no GB segregation and no measurable fraction of grain boundary carbides were observed in all states.

\subsection{Multi-edge X-ray absorption spectroscopy experiment}

X-ray absorption spectra (XAS) of the (CrMnFeCoNi)$_{1-x}$C$_x$ alloys 
were recorded at the K absorption edges of Cr (5.6~keV), Mn (6.5~keV), Fe (7.1~keV), Co (7.7~keV), and Ni (8.3~keV) at the P65 mini-undulator (11 periods) beamline at the DESY PETRA III synchrotron radiation facility \cite{P65}. 
The water-cooled double crystal monochromator with Si(111) crystals as well as two water-cooled Si-coated plane mirrors mounted prior to the monochromator were employed for all energies.

The intensity of the incoming X-ray photons was continuously monitored using the first ionization chamber filled with pure N$_2$ (to achieve approximately 10\% absorption) for energies around the Cr and Mn K absorption edges or with a mixture of Ar and N$_2$ in a 10:90 proportion for energies around the Fe, Co and Ni K absorption edges, respectively. 
The X-ray flux on the sample was in the range of $\sim$10$^1$$^0$$^-$$^1$$^1$ photons/s.
The samples were fixed inside a closed-cycle helium cryostat and placed in the primary X-ray beam (1.6(horiz.) $\times$ 0.3(vert.) mm$^2$) in an incident geometry of 45$^\circ$. The fluorescent yield (FY) was collected using a four-element silicon drift diode (SDD) fluorescence detector (Hitachi ME4 model) placed at 90$^\circ$ to the X-ray beam.
All measurements were performed first at a near room temperature of 285~K and then at a low temperature of about 18~K.

The raw data were first preprocessed, including necessary incoming intensity normalization, corrections, and integration, and then averaged over the individual channels of the SDD detector. The EXAFS spectra $\chi(k)k^2$ were then extracted from the averaged raw data using a conventional procedure \cite{Kuzmin2014} implemented in the XAESA code \cite{xaesa}. For each sample, the EXAFS spectra recorded at the K absorption edges of all constituents (Cr, Mn, Fe, Co, and Ni) were used in the RMC simulations. The self-absorption effects were found to be negligible in the energy range chosen for the fitting procedure.

\subsection{Reverse Monte Carlo simulations}

In multicomponent systems such as HEAs of five or more principal components, an unbiased and precise analysis of the EXAFS data requires simultaneous fitting of one and the same structural model to all available EXAFS spectra collected independently at the X-ray absorption edges of as many constituents as possible \cite{NaRe1,Jalcom}. 
This capability is currently provided by data analysis based on RMC simulations. In our work, we used the RMC approach implemented in the EvAX code \cite{Timoshenko2014rmc, Timoshenko2012rmc}. The comparison between the experimental and calculated EXAFS spectra was performed simultaneously in both $k$ and $R$ spaces, WTs \cite{Timoshenko2009wt}. To accelerate the optimization procedure, the evolutionary algorithm (EA) 
 \cite{Timoshenko2014rmc} was applied, incorporating three key operations: selection, crossover, and mutation.  These operations were applied to sets of atomic configurations at each step of the RMC simulation, significantly improving the convergence of the structural model. Further details of the method are provided in Ref. \cite{Timoshenko2014rmc}.

To initiate the analysis based on RMC, the simulation box was constructed separately for each system as a $4a \times 4a \times 4a$ supercell, employing periodic boundary conditions with the fcc lattice parameter $a$, determined experimentally by XRD (Table\ \ref{tab:Values}). The Cr, Mn, Fe, Co, and Ni atoms (256 in total) were randomly distributed in appropriate concentrations (20 $\pm$ 2~at.\% for each element, according to the X-ray fluorescence data) at the Wyckoff positions of the fcc lattice. The size of the simulation box remained  constant during the simulations. The number of atomic configurations simultaneously considered in the EA was 32 \cite{Timoshenko2014rmc}. In this study, we followed the procedure described previously in Refs. \cite{NaRe1, Jalcom, NaRe2, NaRe3, NaRe4}, where additional details can be found regarding the general advantages of the RMC method over conventional multi-shell EXAFS analysis, the specific requirements for constructing the initial simulation box, the simulation process, and the main steps at each iteration. Given the relatively large grain sizes of the studied alloys (over 50~{\textmu}m, as estimated from SEM analysis), the simulations were performed without considering the potential influence of grain boundaries, which were assumed to be negligible. The extremely low level of carbon doping in the alloys (0.3 and 0.8~at.\%) corresponds to at most one carbon atom in the constructed simulation box, and thus it was not explicitly included in the simulations. Also, the Warren-Cowley short-range order parameter \cite{Norman1951,Cowley1960} was equal to zero for all structural models due to the random distribution of atoms throughout the simulation box.

\begin{table}[b]
    \centering
    \caption{The unit cell parameter $a$ (in \AA), determined by conventional XRD for the (CrMnFeCoNi)$_{1-x}$C$_x$ ($x$ = 0, 0.3, and 0.8~at.\%) alloys in both the HT and LT states. The uncertainty in the lattice parameters is $\pm 0.0002$~\AA.}\label{tab:Values}
    \vspace{5pt}
    \begin{tabular}{l|ccc}
    \hline
   \multirow{1}{*}{State~~~}  & ~~0~at.\%~C~ & ~~0.3~at.\%~C~ & ~~0.8 at.\%~C~ \\ [0.5ex]
   \cline{2-4}
\hline
 HT & 3.5802 & 3.5750 & 3.5796 \\
 LT & 3.5796 & 3.5678 & 3.5774  \\
    \hline         
    \end{tabular}
\end{table}

At each RMC iteration, the configuration-averaged EXAFS spectra (CA-EXAFS) $\chi(k)k^2$ were calculated for all Cr, Mn, Fe, Co, and Ni atoms in the constructed simulation box using the ab initio real-space multiple-scattering FEFF8.50L code \cite{FEFF8, Rehr2000}. The multiple scattering (MS) effects up to the fourth order were included; these contributions are important when analysis beyond the first coordination shell is performed. The Morlet WTs of the calculated spectra were then compared with those of the experimental spectra, and the best agreement between the WTs of the experimental and calculated CA-EXAFS spectra was used as a criterion for optimizing the model structure. The typical values of residual differences between aforementioned WTs could be found in Appendix~\ref{sec:AppEXAFSRMC} (Table\ \ref{tableKsi}). The total number of RMC iterations was set to 5000 to ensure the convergence of the structural model. The WTs were calculated in the $R$-space range of 1--6~\AA, while in the $k$-space range of 3.0--11.5~\AA$^{-1}$ was used for the Cr K-edge, but 3.0--12.0~\AA$^{-1}$ for the other absorption edges. As a result of each RMC simulation, a final set of atomic coordinates was obtained for each type of atom, which was further used to calculate the total pair distribution functions (PDFs) $g(r)$. To improve statistical accuracy, ten different (independent) starting structural models were considered for the final averaged PDFs.

The statistically averaged mean interatomic distances $\langle R \rangle$ and the mean squared relative displacements (MSRDs) for each atomic pair were evaluated as the first and second moments of the total PDFs $g(r)$, respectively (Table\ \ref{table2} in Section\ \ref{res}). Meanwhile, the mean squared displacements (MSDs) for each atomic type were determined directly from the atomic coordinates. It is important to note that the MSRD provides information on the correlated motion of the atoms, whereas the MSD reflects only uncorrelated displacements, whether determined using diffraction techniques \cite{Dalba1997} or obtained from lattice-dynamics calculations based on DFT. Both MSDs and MSRDs include contributions from structural (static) and thermal (dynamic) disorder.  The differences between the MSD and MSRD values at 285~K and 18~K are given in Table\ \ref{table3} (Section\ \ref{res}), and could be also found in Appendix~\ref{sec:AppEXAFSRMC} (Figs.\ \ref{fig:deltaMSDHTLT}-\ref{fig:consolidatedPlot}).

\subsection{DFT calculations}

First-principles DFT simulations were conducted to analyze the MSDs of the CrMnFeCoNi HEA in the fcc phase.
The alloy was modeled based on the supercell approach in a similar manner to our previous studies \cite{Ikeda_PRM_2019_Impact,Kies_SM_2020_Combined,Ikeda_JPED_2021_Impact}.
Specifically, the supercells contain 54 metal atoms, and the $\langle 111 \rangle$ direction of the fcc phase was set to the third axis, which displays the ``ABCABC'' stacking of six close-packed \{111\} layers.
Ideal mixing of the elements in CrMnFeCoNi was approximated based on special quasirandom structures \cite{Zunger_PRL_1990_Special}, where the first and the second nearest-neighbor pairs are optimized to be close to the ideal mixing state.
In the present study, 20 such configurations were investigated so that the average composition equals the equiatomic CrMnFeCoNi alloy. Alloying with carbon was reproduced using the carbon diluted limit, when the minimum amount of carbon (one C atom) was placed in the supercell; this results in a nominal amount of 1.8 at.\% carbon in the alloy (interstitially dissolved).

The DFT simulations were performed using the VASP  code \cite{Kresse_CMS_1996_Efficiency,Kresse_JNS_1995_Ab,Kresse_PRB_1999_ultrasoft}.
The plane-wave basis set was employed in conjunction with the projector augmented wave method \cite{Bloechl_PRB_1994_Projector}.
The generalized-gradient approximation in the Perdew--Burke--Ernzerhof form \cite{Perdew_PRL_77_3865_1996} was employed as the DFT exchange--correlation functional.
The plane-wave cutoff energy was set to 400\,eV.
The Brillouin zones were sampled by a $\Gamma$-centered $4 \times 4 \times 4$ $k$-point mesh for the 54-atom supercell models in conjunction with the Methfessel--Paxton scheme \cite{Methfessel_PRB_1989_High} with the smearing width of 0.1\,eV.
The 3d4s orbitals of Cr, Mn, Fe, Co, and Ni and the 2s2p orbitals of C were treated as the valence states.
The total energies were minimized until they converged within $10^{-3}$\,eV per simulation cell during the self-consistent-field (SCF) cycle. The convergence tests for these DFT settings are provided in Appendix~\ref{sec:ConvDFT}.
Spin polarization was considered in the DFT calculations.
Following a previous computational study based on the coherent-potential approximation \cite{Ma_AM_2015_Ab}, all the magnetic moments on Cr and Mn were initially set to be antiparallel to those on Fe, Co, and Ni.
As discussed in detail in our previous work \cite{Ikeda_PRM_2019_Impact}, the magnetic orientation on each atom was optimized after the SCF cycle, as well as atomic-position relaxation, optimized based on its local environment. The volumes and the shapes of the supercell models were kept fixed to the fcc lattice parameter of 3.6~\AA, which is close to the experimental value \cite{Cantor_MSEA_2004_Microstructural,Bhattacharjee_JAC_2014_Microstructure,Wu_PhdThesis_2014_Temperature,Laurent-Brocq_AM_2015_Insights,Schuh_AM_2015_Mechanical,Tracy_NC_2017_High,Zhang_NC_2017_Polymorphism}.

For the calculations without interstitial carbon, the metal atoms were initially placed at the exact fcc lattice sites and then relaxed until the residual forces were reduced to less than $5 \times 10^{-2}$ eV/Å.
For each interstitial site considered, a carbon atom was first placed at the geometric center, after which the atomic positions were re-optimized.
A total of 1080 interstitial sites were computed across the 20 configurations.
DFT-MD simulations were further conducted at 300\,K and 600\,K for five selected configurations both without and with the carbon atom.
The MD simulations were performed in the \textit{NVT} ensemble using a Langevin thermostat with a friction parameter of 0.01\,fs$^{-1}$.
The simulations were run with a time step of 5\,fs for 200 steps, and the last 150 steps were used for the MSD calculations, providing sufficiently small standard errors of the means.
Details about magnetic moments of constituent elements during DFT-MD simulations are provided in Appendix~\ref{sec:Magnetism}.

The element-resolved MSDs of the metal atoms were analyzed with separating them into the static and the dynamic contributions.
The static MSD of chemical species X is given by
\begin{align}
\mathrm{MSD}_\mathrm{static}(\mathrm{X})
= \frac{1}{|S_\mathrm{X}|} \sum_{i \in S_\mathrm{X}} (\langle \mathbf{r}_i \rangle_t - \mathbf{r}_i^\mathrm{ideal})^2,
\end{align}
where $S_\mathrm{X}$ is the set of the indices for atoms of chemical species X, $\mathbf{r}_i$ is the position of atom $i$, $\mathbf{r}_i^\mathrm{ideal}$ is the ideal position at the ideal fcc lattice associated with atom $i$, and $\langle \cdots \rangle_t$ denotes the time average.
In DFT at 0~K, $\langle \mathbf{r}_i \rangle_t$ reduce to the position after relaxation.
Note that $\langle \mathbf{r}_i \rangle_t$ were calibrated by a constant shift to satisfy
\begin{align}
\sum_\mathrm{X} \sum_{i \in S_\mathrm{X}} \left(\langle \mathbf{r}_i \rangle_t - \mathbf{r}_i^\mathrm{ideal}\right) = \mathbf{0},
\label{eq:calibration_static}
\end{align}
in order not to have an apparent net drift with respect to the reference ideal fcc lattice.
The dynamic contribution was evaluated in two ways, according to the Einstein model, and within the framework of MD.
In the Einstein model, the dynamic contribution is the vibrational displacements obtained as
\begin{multline}
\mathrm{MSD}_\mathrm{dynamic}^\mathrm{Einstein}(\mathrm{X}) \\
= \frac{1}{|S_\mathrm{X}|} \sum_{i \in S(\mathrm{X})} \sum_{j=1}^{3} \frac{\hbar}{2 m_i \omega^\mathrm{E}_{ij}} \coth\left(\frac{\hbar \omega^\mathrm{E}_{ij}}{2 k_\mathrm{B} T}\right),
\end{multline}
where $m_i$ is the mass of the $i$-th atom, and $\omega_{ij}^\mathrm{E}$ is the $j$-th Einstein frequency of it.
The Einstein frequencies were obtained by the finite-displacement method with a displacement of 0.01\,{\AA}.
Thus obtained MSDs also include zero-point vibrations at 0\,K.
In DFT-MD, the dynamic contribution is evaluated by averaging the atomic fluctuations around the positions averaged over the simulation time as
\begin{align}
\mathrm{MSD}_\mathrm{dynamic}^\mathrm{MD}(\mathrm{X})
= \frac{1}{|S_\mathrm{X}|} \sum_{i \in S_\mathrm{X}}
\left\langle (\mathbf{r}_i - \langle \mathbf{r}_i \rangle_t)^2\right\rangle_t.
\end{align}
Here, $\mathbf{r}_i$ were calibrated by a constant shift to eliminate the net drift of the system center of mass, i.e.,
\begin{align}
\sum_\mathrm{X} \sum_{i \in S_\mathrm{X}} m_i \left(\mathbf{r}_i - \langle \mathbf{r}_i \rangle_t\right) = \mathbf{0}.
\label{eq:calibration_dynamic}
\end{align}
While the MSD\textsubscript{dynamic} obtained from MD simulations cannot represent quantum effects including zero-point vibrations, at finite temperatures it includes anharmonic contributions.
As demonstrated in Appendix~\ref{sec:Einstein_vs_MD}, the zero-point vibrations are found to be substantial, whereas already at room temperature the anharmonic contribution becomes significant.
We therefore took MSD\textsubscript{dynamic} from the Einstein model for 0\,K while the values from DFT-MD for 300\,K and 600\,K.
The total MSD for chemical species X is obtained as
\begin{align}
\mathrm{MSD}(\mathrm{X}) = \mathrm{MSD}_\mathrm{static}(\mathrm{X}) + \mathrm{MSD}_\mathrm{dynamic}(\mathrm{X}).
\end{align}

Note that, in principle, the vibrational contribution should be evaluated from full harmonic phonon calculations, rather than using the Einstein model, as phonon calculations provide a more rigorous treatment of collective atomic vibrations.
However, for the present chemically and magnetically disordered alloy, finite-displacement phonon calculations did not yield numerically robust results; specifically, they often produced spurious imaginary modes, most likely because of the difficulty of keeping the detailed local magnetic states consistent between the reference atomic positions and the corresponding displaced configurations.
By contrast, in the Einstein model, the vibrational frequencies can be obtained for individual atoms independently and are therefore numerically much more robust.
We therefore employed the Einstein model, which should provide a reasonable semi-quantitative description of the zero-point vibrational contribution.

\section{Results}
\label{res}

\subsection{EXAFS and RMC}

\begin{figure*}[t]
\centering
\includegraphics[scale=0.7]{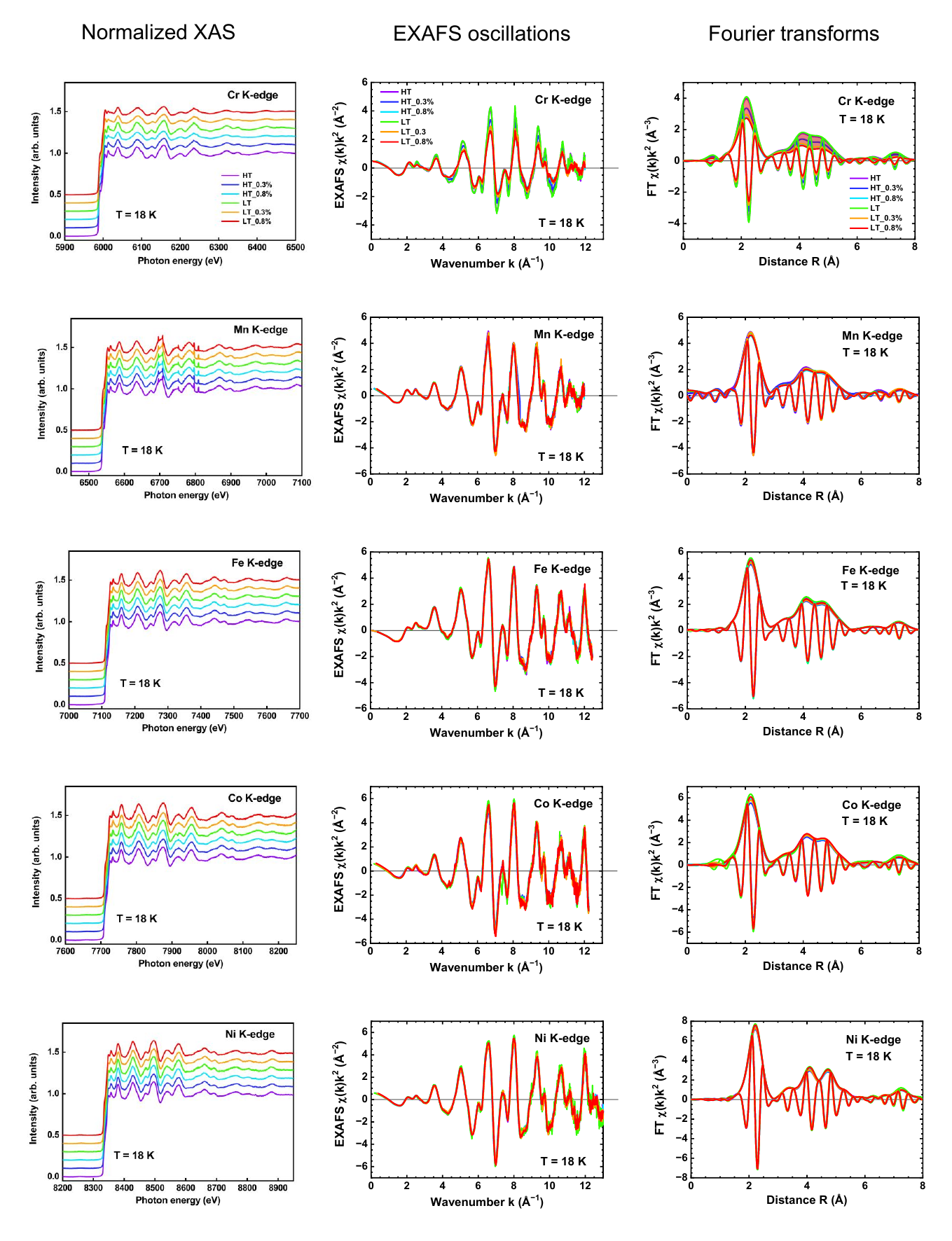}
\caption{
(left panels) The normalized X-ray absorption spectra (XAS) collected at the K absorption edges of Cr ($E_0$ = 5.9~keV), Mn ($E_0$ = 6.5~keV), Fe ($E_0$ = 7.1~keV), Co  ($E_0$ = 7.7~keV), and Ni ($E_0$ = 8.3~keV) at 18~K using fluorescence yield and 45$^{\circ}$  incidence geometry for (CrMnFeCoNi)$_{1-x}$C$_x$ ($x$ = 0, 0.3 and 0.8~at.\%) alloys. The spectra are normalized to unity and shifted vertically for clarity. The color scheme is the same for all plots. (middle panels) The experimental EXAFS spectra $\chi(k)k^2$ and (right panels) their Fourier transforms (FTs) for (CrMnFeCoNi)$_{1-x}$C$_x$ alloys collected at the K-absorption edges of Cr, Mn, Fe, Co, and Ni at 18~K. XAS, experimental EXAFS and their FTs at 285~K could be found in Appendix~\ref{sec:AppEXAFSRMC} (Fig.\ \protect\ref{fig:SM_EXAFS285K}).}
\label{fig:EXAFS}
\end{figure*}

The short-range local environment of each principal component, averaged over the probing volume, was examined using element-specific X-ray absorption spectroscopy at the K absorption edges of all constituents. The EXAFS oscillations observed above each absorption edge reflect the local coordination of each type of atom in the crystallographic lattice and allow conclusions to be drawn about the structural homogeneity of the systems under study. The normalized X-ray absorption spectra recorded at 18~K for the (CrMnFeCoNi)$_{1-x}$C$_x$ ($x$ = 0, 0.3 and 0.8~at.\%) alloys in the HT and LT states are shown in Fig.\ \ref{fig:EXAFS} (left panels) and demonstrate very similar shapes, positions, and amplitudes of the EXAFS oscillations for the same principal components in all systems. The same behaviour was observed for the EXAFS spectra collected at 285~K (see Appendix~\ref{sec:AppEXAFSRMC}, Fig.\ \ref{fig:SM_EXAFS285K}). Possible contributions from the sub-surface oxide layer to the recorded XAS can be neglected due to the relatively large probing depth of FY detection mode in the  energy ranges considered (approximately 10--15~{\textmu}m).

The experimental EXAFS spectra $\chi(k)k^2$ and their Fourier transforms (FTs) used in the RMC fit at 18~K are shown in Fig.\ \ref{fig:EXAFS} (middle and right panels, respectively). The corresponding data for 285~K are provided in the Appendix~\ref{sec:AppEXAFSRMC}, Fig.\ \ref{fig:SM_EXAFS285K}. The most pronounced changes in the amplitudes of the EXAFS oscillations and their FTs are observed for the Cr absorbers at both temperatures, indicating that the chemical environment of Cr atoms is the most strongly affected by carbon doping compared with the other principal components. 

As demonstrated previously \cite{NaRe1, Jalcom, NaRe2, NaRe3, NaRe4}, the collected experimental EXAFS data can be used to extract a reliable 3D structural model of the material through RMC-based analysis. For multi-principal-component systems, the RMC approach enables a self-consistent analysis of the experimental EXAFS spectra in the most unbiased manner, without introducing additional interdependencies among the fitted parameters. Reconstruction of the short-range local environment is generally possible within the first several coordination shells around each type of absorber (distances up to 6.0~\AA). In the specific case of the fcc structure, the first peak in the experimental FTs is well isolated and appears between 1.0 and 3.5~\AA; therefore, any quantitative estimates related to the nearest neighbors around each type of absorber can be obtained with reliable accuracy.

The RMC fitting procedure yields a set of PDFs $g(r)$ that account for both static (structural, radial and angular) and dynamic (thermal) lattice relaxations. The resulting PDFs capture the most significant statistically averaged features of the atom distributions up to the fifth coordination shell for absorbers of each type, even though only a limited number of initial configurations can be considered in the RMC simulations.

\begin{figure}[t]
\centering
\includegraphics[width=\linewidth]{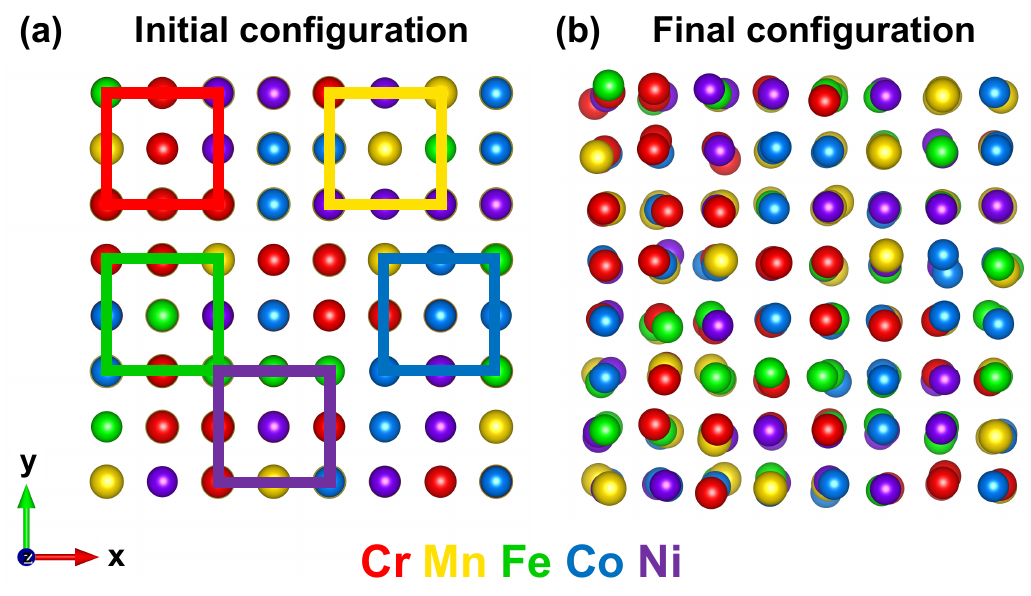}
\caption{Examples of (a) initial and (b) final (relaxed) atom configurations (simulation boxes) used in the RMC simulations for simultaneous fit to the EXAFS spectra of the (CrMnFeCoNi)$_{1-x}$C$_x$ alloys at the K absorption edges of Cr, Mn, Fe, Co, and Ni constituents at 285~K. The initial simulation box (a) was randomly filled with 3d atoms according to the alloy composition; periodic boundary conditions were applied to minimize the influence of surface effects. Colour scheme: Cr (red), Mn (gold/orange), Fe (light green), Co (blue), and Ni (violet). The top views of the simulation boxes are shown along the $z$-direction in chosen $xyz$ coordinates as depicted in (a). The schematic border of the fcc unit cells around each principal component is displayed in the figure related to the initial configuration.}
\label{fig:Configs}
\end{figure}

\begin{figure*}[b]
\centering
\includegraphics[width=0.32\textwidth]{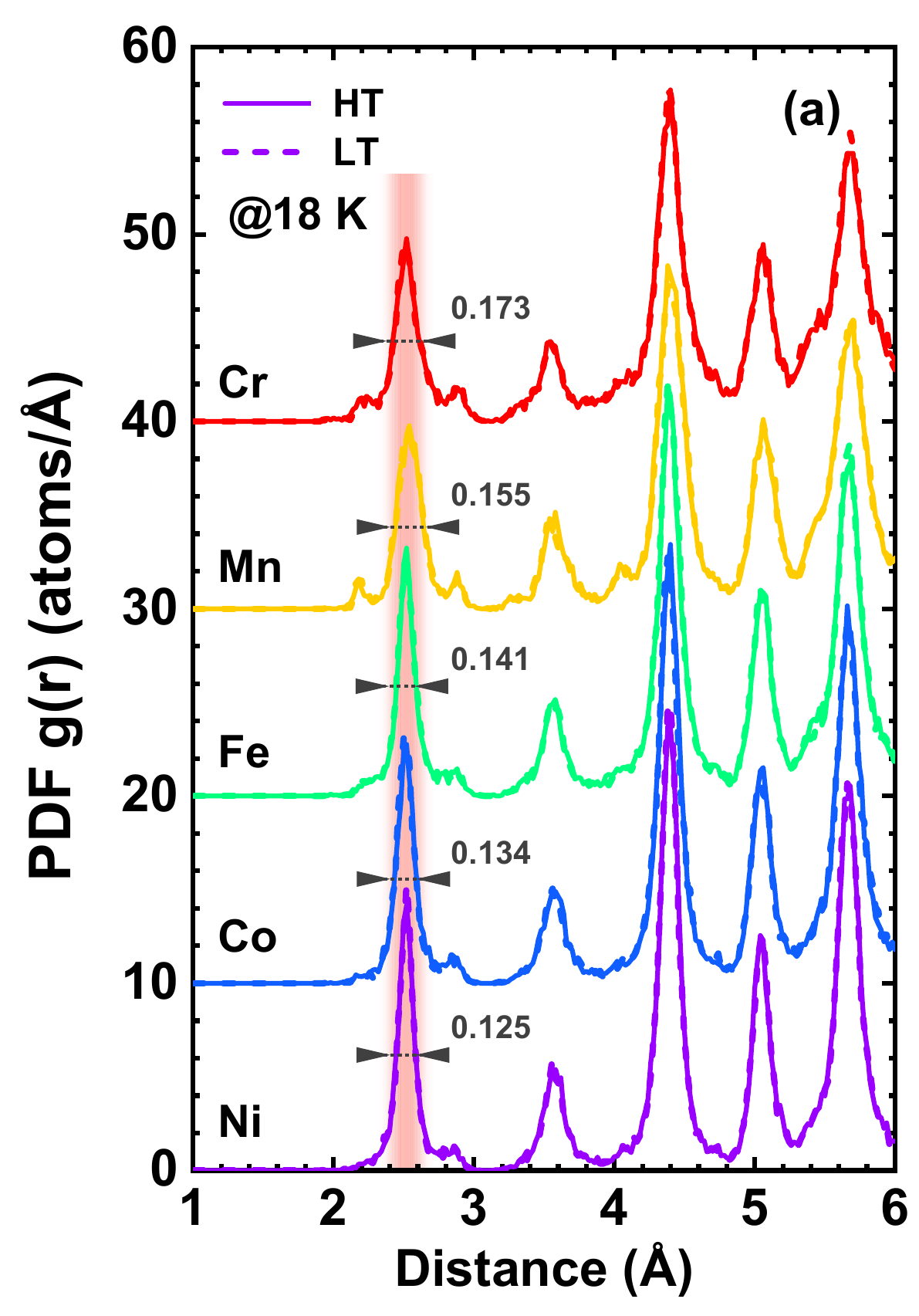}
\includegraphics[width=0.32\textwidth]{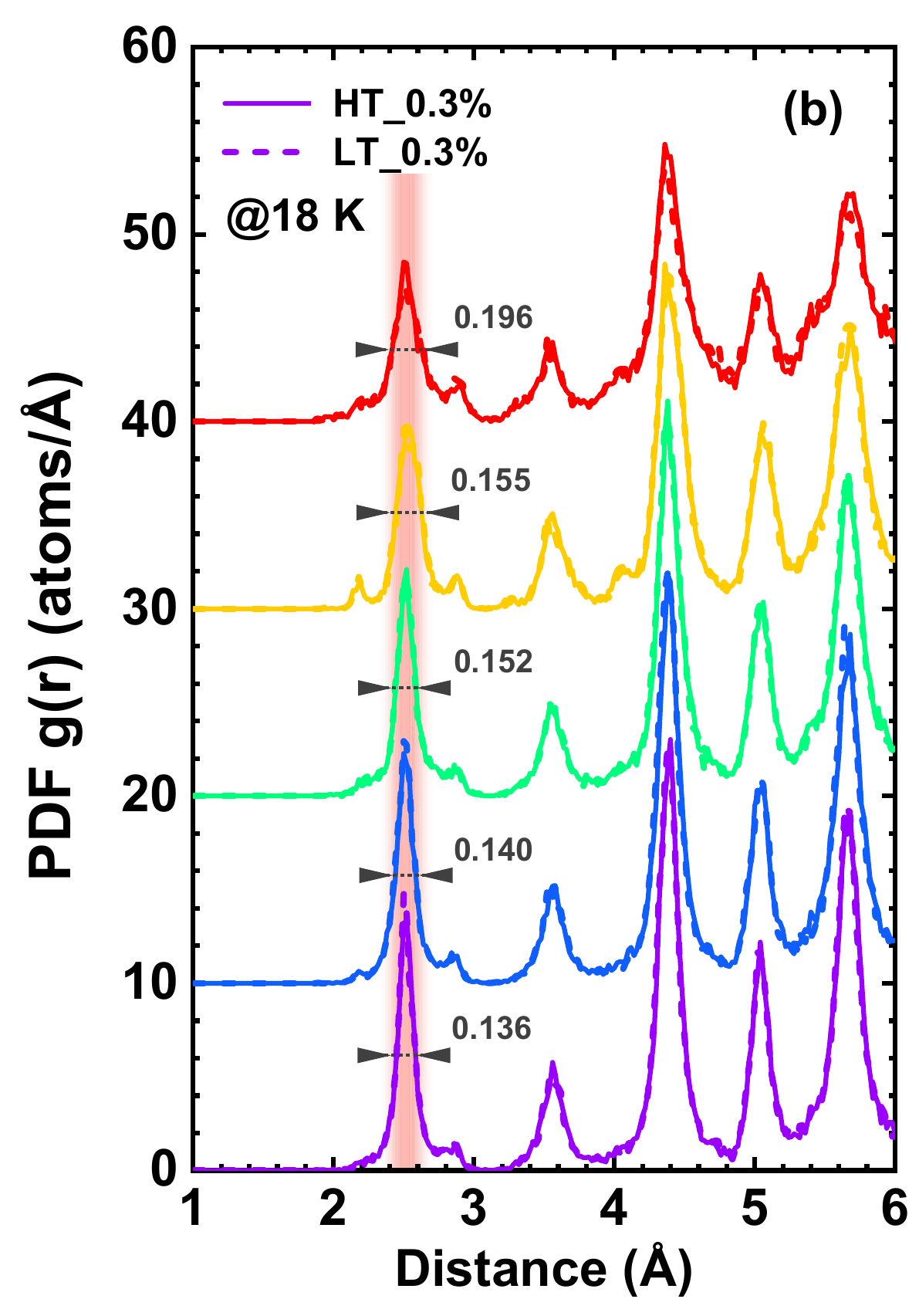}
\includegraphics[width=0.32\textwidth]{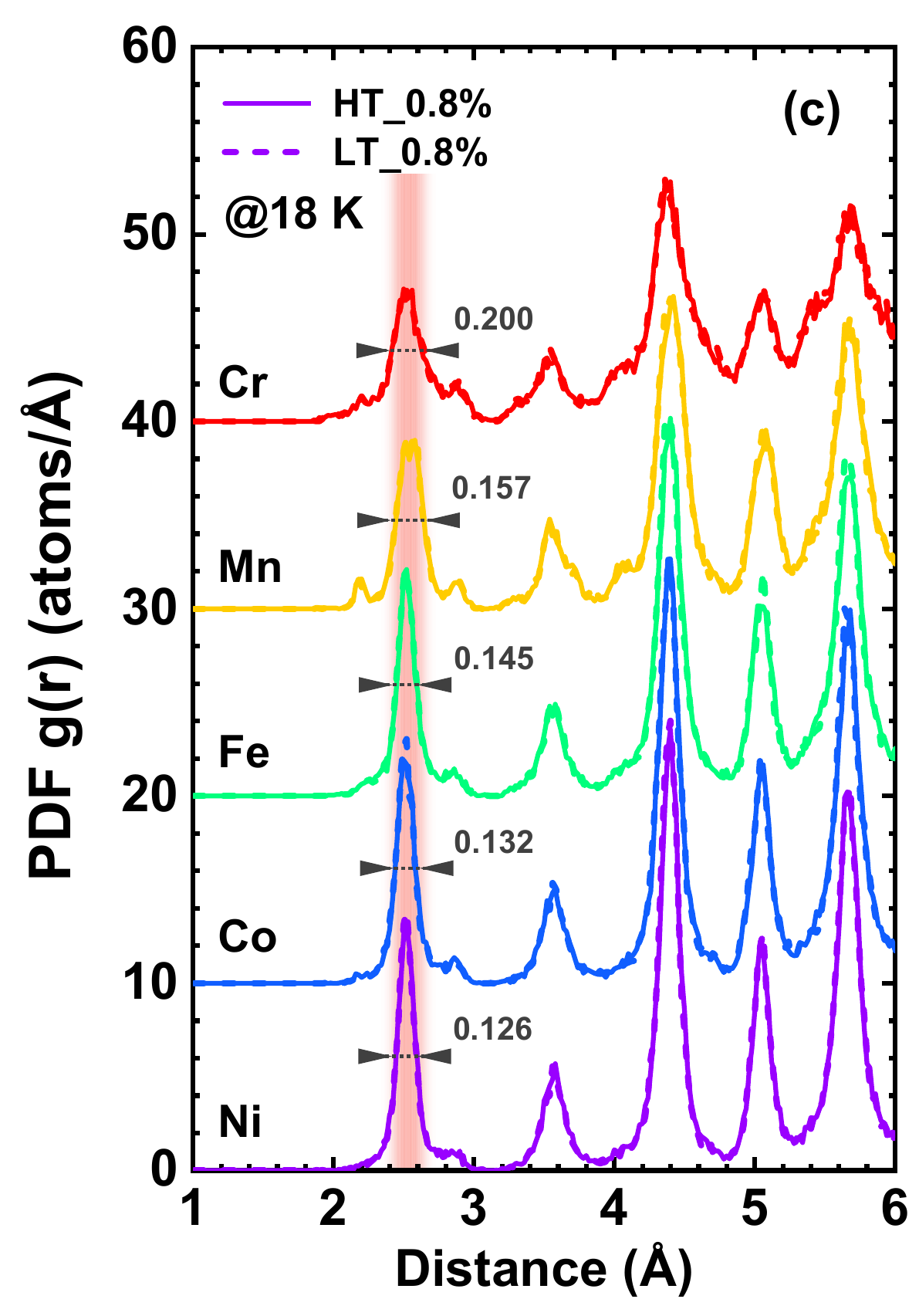}

\caption{Comparison of the total pair distribution functions (PDFs) for the (CrMnFeCoNi)$_{1-x}$C$_x$ alloys with different carbon doping ($x$ = 0 (a), 0.3 (b) and 0.8 (c)~at.\%) annealed at high (HT states) and low (LT states) temperatures. PDFs were obtained using the RMC simulations of EXAFS spectra collected at 18~K. The peaks related to the first coordination shell around a particular type of absorbers are highlighted; the values for the averaged MSRDs taken from Table\ \protect\ref{table2} (in \AA) are indicated for a quick cross reference.}
\label{fig:fig_PDF-18K}
\end{figure*}

An example of the initial (starting) and final (relaxed) configurations of the 3d atoms used in the simulation boxes for the (CrMnFeCoNi)$_{1-x}$C$_x$ alloys during the RMC-based fitting is shown in Fig.\ \ref{fig:Configs}. Each structural model, represented by a simulation box, was simultaneously fitted to the available EXAFS spectra $\chi(k)k^2$  collected independently at the K absorption edges of all principal components, along with their corresponding FTs for each alloy. The atomic coordinates of all relaxed atoms in the constructed supercells were subsequently used to compute the \textit{total} and \textit{partial} component-specific PDFs. However, in this work we focus only on the \textit{total} PDFs, which capture the most essential features of the nearest-neighbor arrangements in the first coordination shell of the corresponding absorbers.

\begin{figure*}[ht]
\centering

\includegraphics[width=0.32\textwidth]{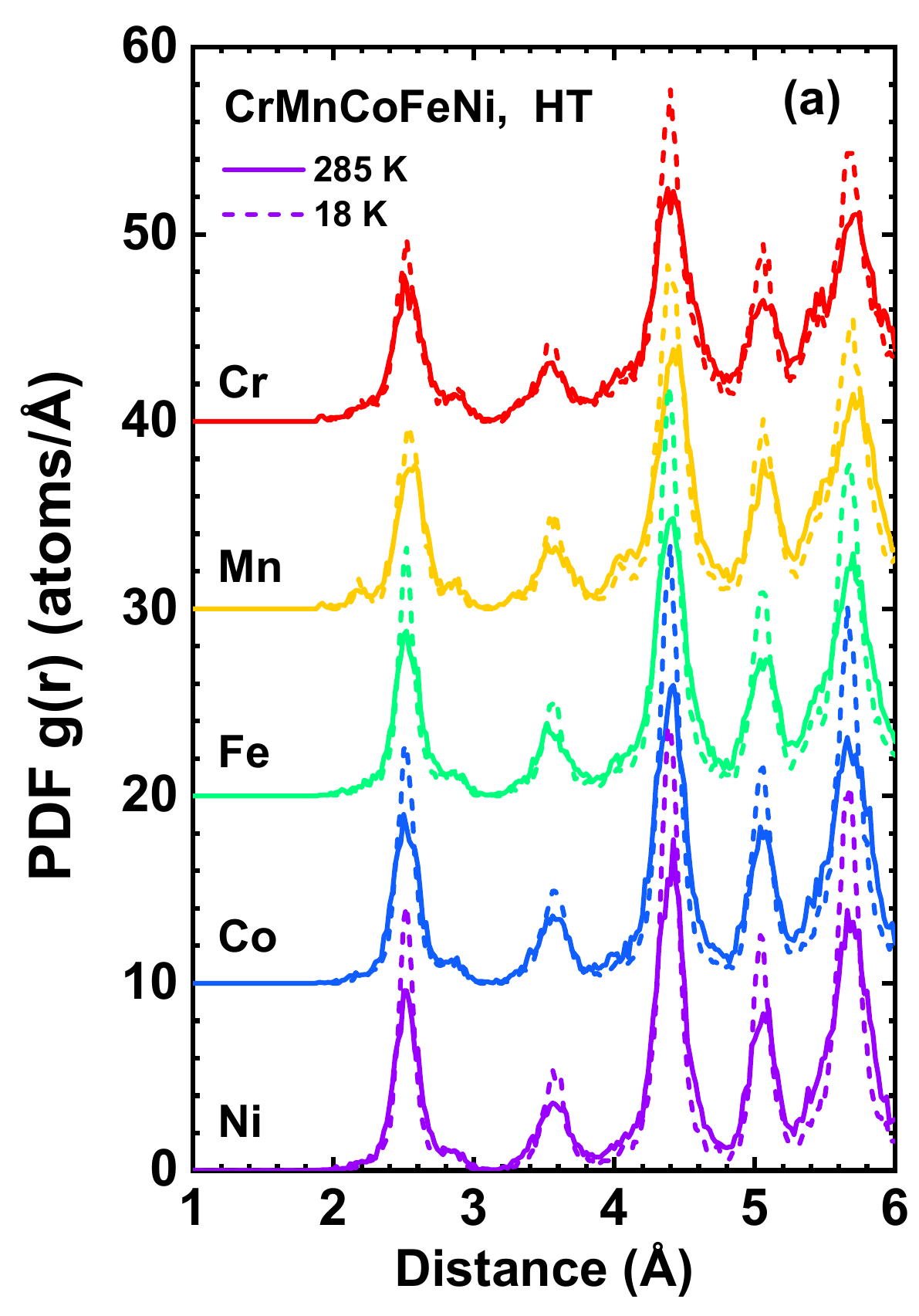}
\includegraphics[width=0.32\textwidth]{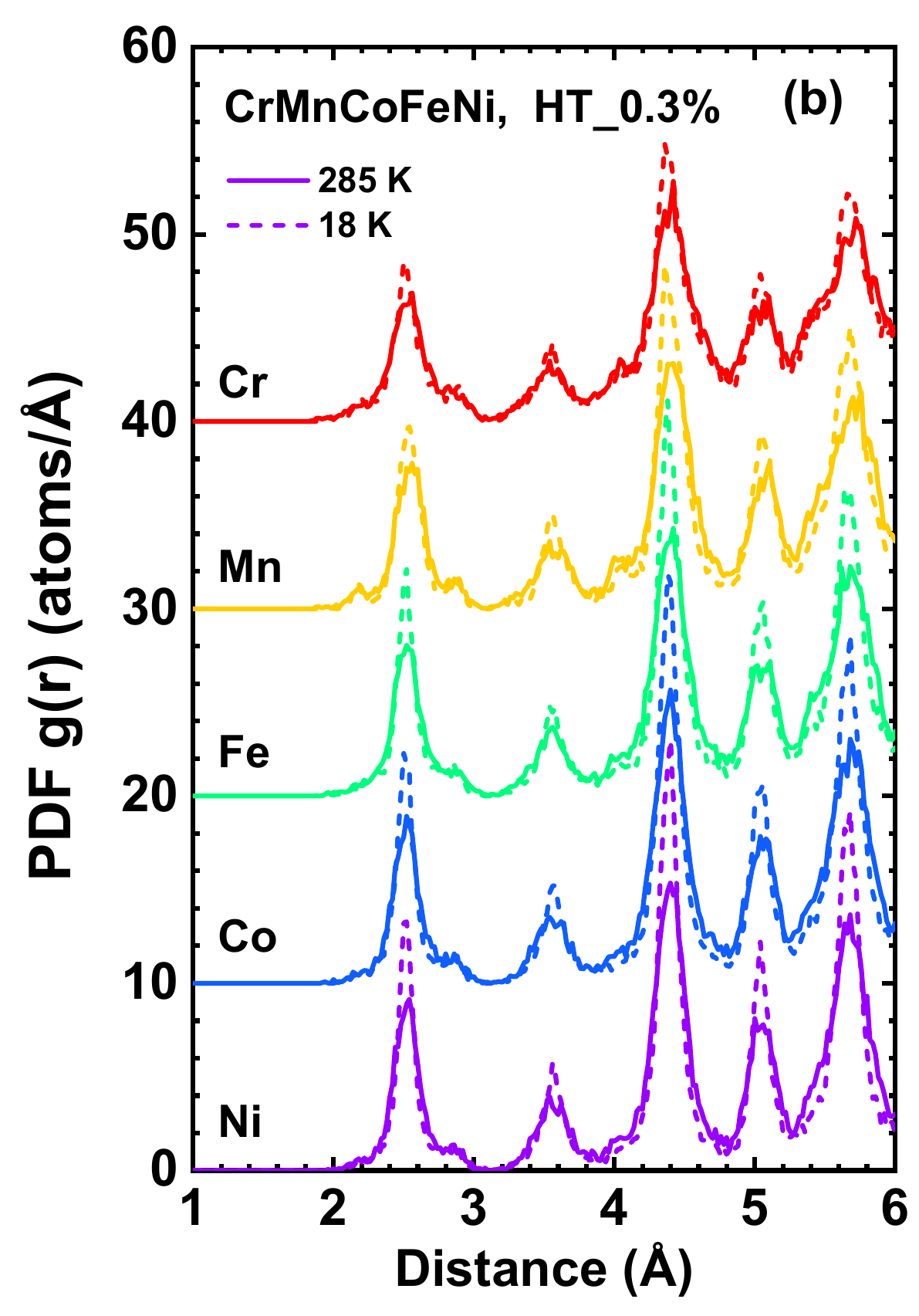}
\includegraphics[width=0.32\textwidth]{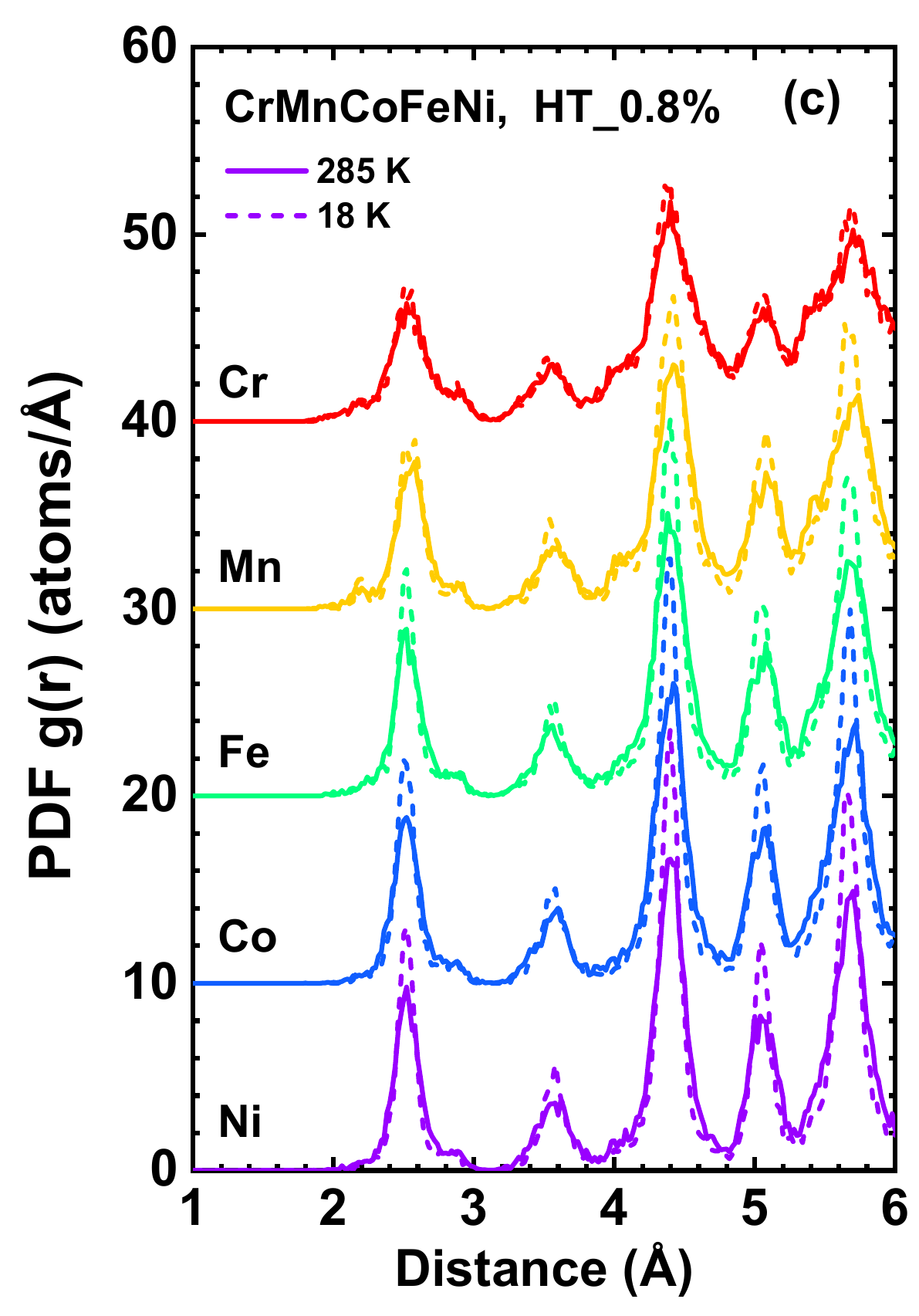}
\hfill
\includegraphics[width=0.32\textwidth]{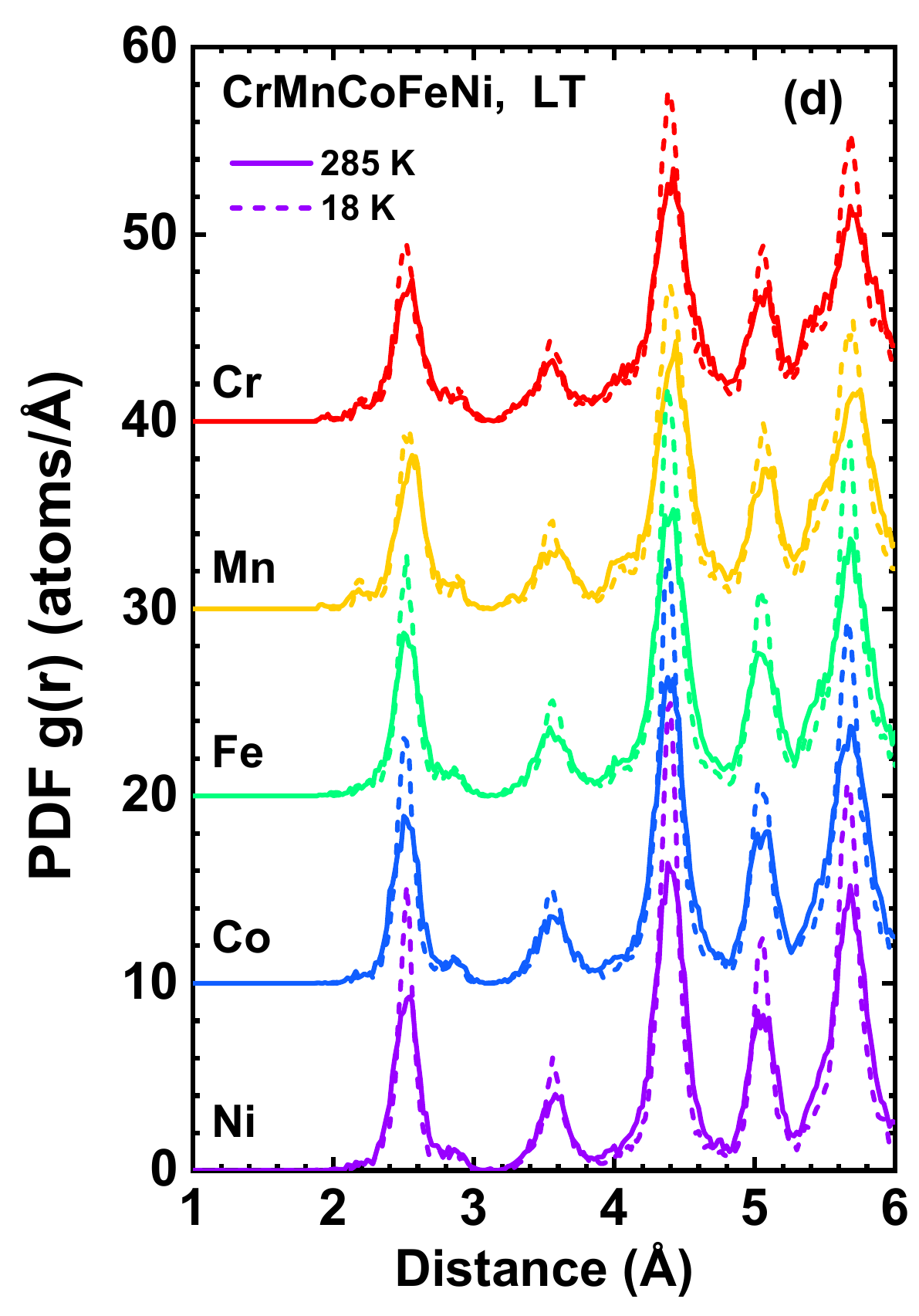}
\includegraphics[width=0.32\textwidth]{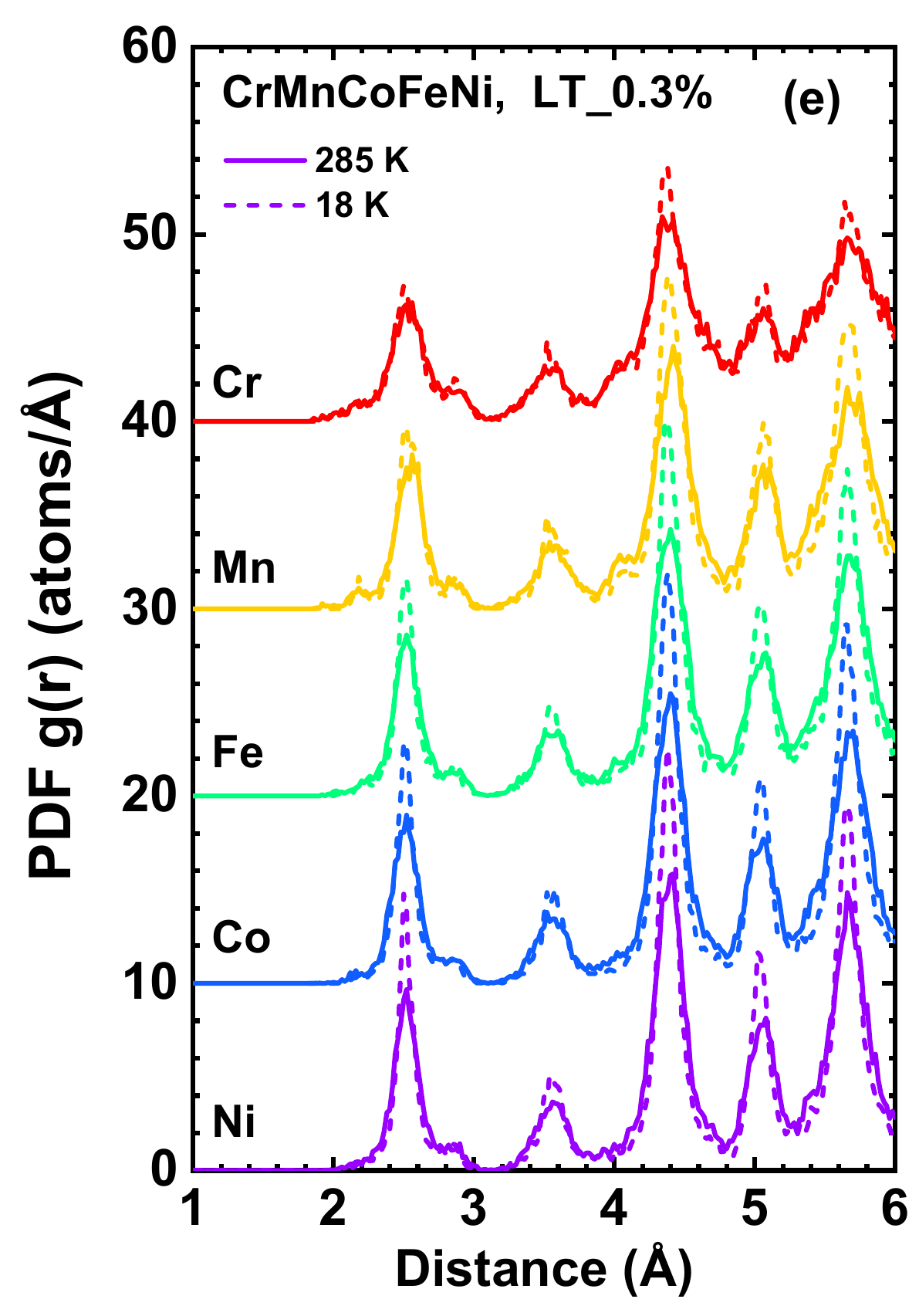}
\includegraphics[width=0.32\textwidth]{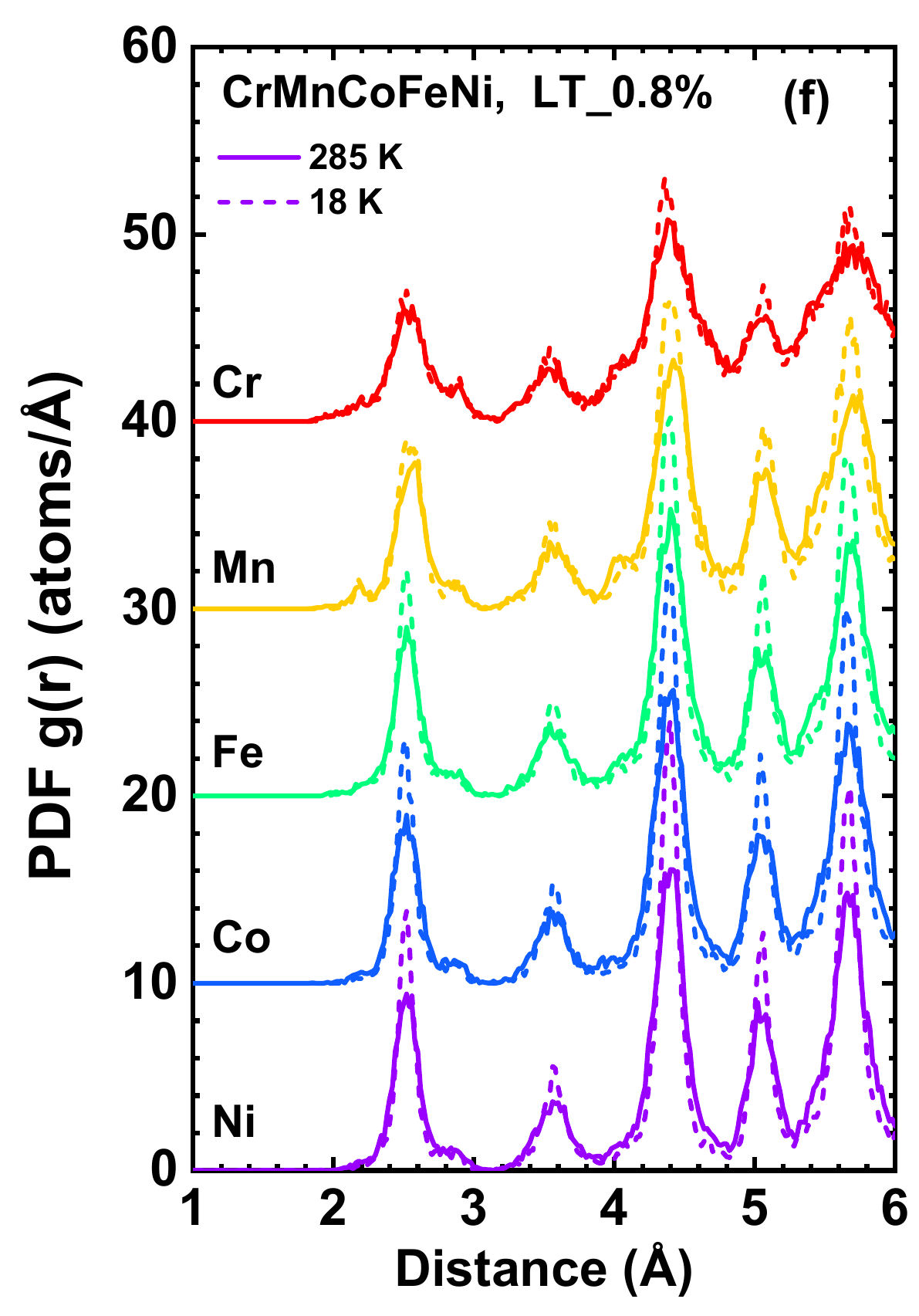}

\caption{Total pair distribution functions (PDFs) for the (CrMnFeCoNi)$_{1-x}$C$_x$ alloys ($x$ = 0, 0.3 and 0.8~at.\%) obtained using the RMC simulations of EXAFS spectra collected at 18 and 285~K for the HT (a, b, and c) and LT (d, e, and f) states.}
\label{fig:PDF-RT}
\end{figure*}

\begin{figure*}[tb]
\centering
\includegraphics[width=0.35\textwidth]{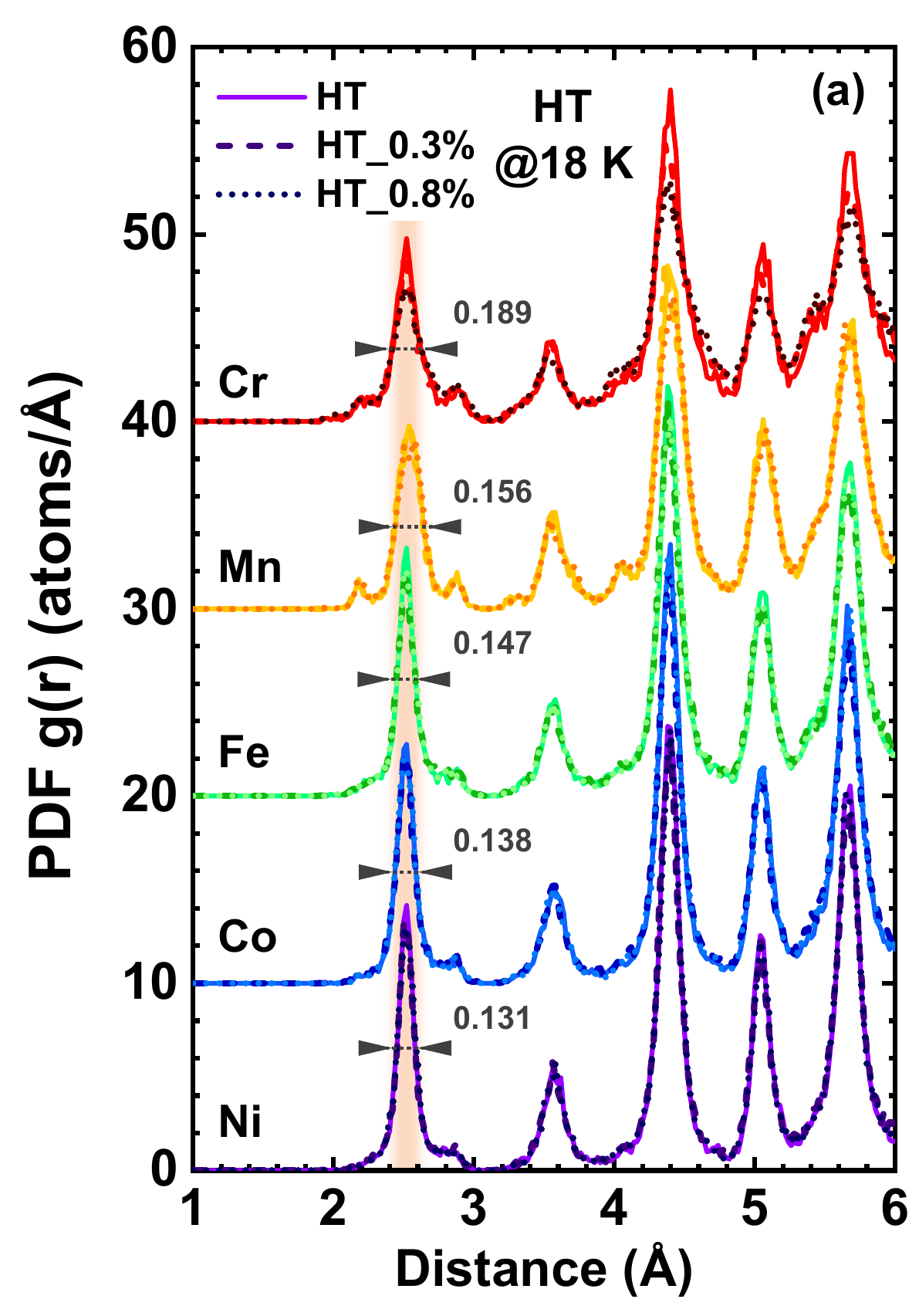}
\includegraphics[width=0.35\textwidth]{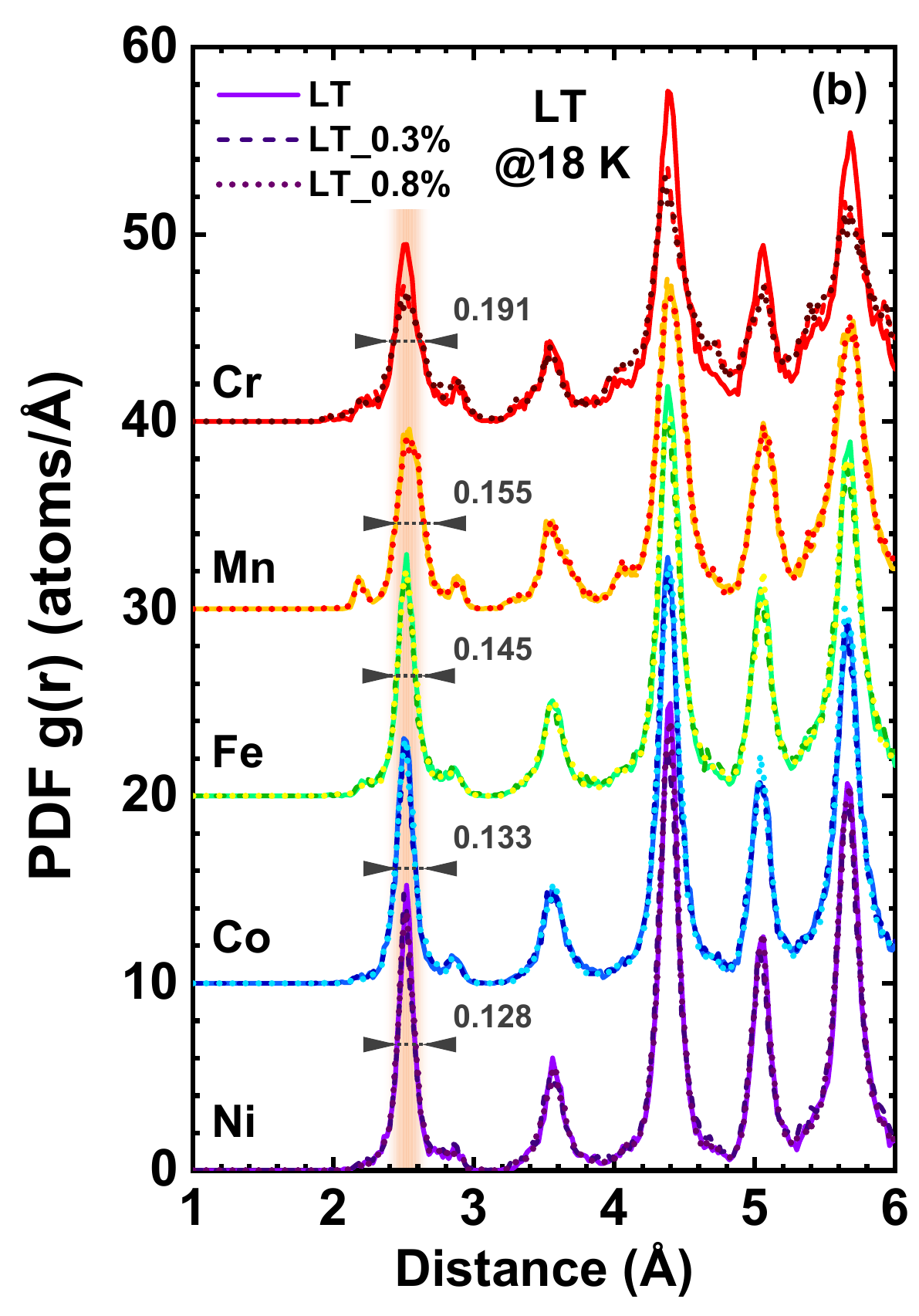}

\caption{Effect of the carbon doping on the total pair distribution functions (PDFs) in the (CrMnFeCoNi)$_{1-x}$C$_x$ alloys ($x$ = 0, 0.3 and 0.8~at.\%) at 18~K in the HT (a) and LT (b) states: 
 HT and LT (solid lines), HT\textunderscore0.3\% and LT\textunderscore0.3\% (dashed lines), HT\textunderscore0.8\% and LT\textunderscore0.8\% (dotted lines). The peaks related to the first coordination shell around a particular type of absorbers are highlighted; the values for the averaged MSRDs taken from Table\ \protect\ref{table2} (in \AA) are indicated for a quick cross reference.}
\label{fig:PDF-carbon}
\end{figure*}

The total PDFs calculated from the EXAFS data for the carbon-free and carbon-doped alloys prepared in their HT and LT states are shown in Figs.\ \ref{fig:fig_PDF-18K}--\ref{fig:PDF-carbon}. They demonstrate a chemically random (within the context of our model) fcc local environment around absorbers of each type in all considered coordination shells (up to 6~\AA) at both temperatures. Qualitatively, the alloys prepared in the HT and LT states behave similarly at the low temperature of 18~K (see Fig.\ \ref{fig:fig_PDF-18K}): the PDF peaks of the Cr and Mn components are visibly broader and exhibit lower intensity compared with those of Fe, Co and Ni, regardless of the doping level. This observation confirms the pattern previously reported for Cantor alloys in Refs.\ \cite{NaRe1, Jalcom, NaRe2, NaRe3, NaRe4}. However, the influence of carbon doping is  most pronounced  in the PDF peaks of the Cr component (see also Fig.\ \ref{fig:PDF-carbon}), which show additional broadening and a further reduction in intensity with increasing carbon concentration, whereas the peaks associated with the other components do not exhibit such pronounced changes.

Performing the EXAFS experiments at two distinct temperatures (18 and 285~K) allowed us to additionally observe the effect of temperature (see Fig.\ \ref{fig:PDF-RT}), which manifests as higher and narrower peaks for all PDFs  at 18~K compared with those at 285~K. Temperature significantly affects the PDFs of all constituent elements in both the  carbon-free and carbon-doped alloys, although this effect is reduced in the alloys containing interstitial carbon. The smallest temperature influence is observed for the highest level of carbon doping (0.8~at.\%). Moreover, the temperature-dependent variations in peak intensities in all PDFs for the carbon-doped alloys in both the HT and LT states exhibit much smaller changes for Cr atoms compared with the other components.

A more detailed comparison of the total PDFs at 18~K for the carbon-doped alloys (see Fig.\ \ref{fig:PDF-carbon}) reveals that all coordination shells of Cr atoms (up to the fifth) are visibly affected by the presence of interstitial carbon, with extremely large variations in peak intensities throughout the more distant shells. These variations are as strong as those observed for the first coordination shell, but only for the Cr component. In contrast, only small variations are observed in the PDFs of Mn, which remain pronounced up to the third coordination shell, whereas negligible differences are seen for Fe, Co and Ni constituents. This tendency in the PDFs of the Cr and Mn components is present in alloys prepared in both the HT and LT states; nevertheless, the changes observed in the LT state are slightly larger than  those in the HT state. The same behavior of the total PDFs for the Cr component is also observed at 285~K (see Appendix~\ref{sec:AppEXAFSRMC}, Fig.\ \ref{fig:PDF-carbonRT}), although the doping effect is less pronounced.

\subsection{DFT simulations}

\begin{figure*}[tb]
\centering
\includegraphics{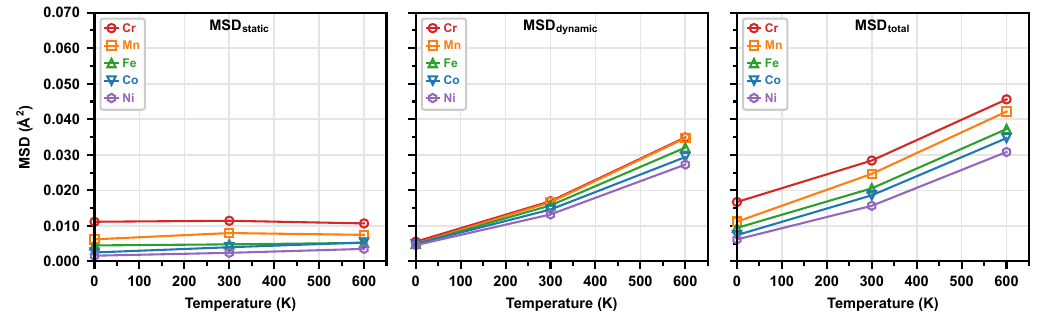}
\caption{Element-resolved MSDs of CrMnFeCoNi without carbon as functions of temperature obtained from DFT.}
\label{fig:DFT_MSDs_vs_T}
\end{figure*}

Figure~\ref{fig:DFT_MSDs_vs_T} presents the element-resolved MSDs of equiatomic CrMnFeCoNi without carbon as functions of temperature obtained from DFT calculations (with \textit{a} = 3.6\,{\AA}).
MSD\textsubscript{static} shows a clear dependence on the chemical element, with the largest MSDs for Cr ({0.011}\,{\AA\textsuperscript{2}} at {0}\,{K}), followed by Mn ({0.006}\,{\AA\textsuperscript{2}}), Fe ({0.005}\,{\AA\textsuperscript{2}}), Co ({0.003}\,{\AA\textsuperscript{2}}), and Ni ({0.002}\,{\AA\textsuperscript{2}}).
MSD\textsubscript{static} is also found to be nearly independent of temperature.
At 0\,K, MSD\textsubscript{dynamic}, which arises solely from zero-point vibrations, is around {0.005}\,{\AA\textsuperscript{2}}, comparable to MSD\textsubscript{static} and therefore substantial.
All metal elements show similar MSD\textsubscript{dynamic} at 0\,K, while at higher temperatures, Cr exhibits substantially larger values than the other elements.
Consequently, Cr shows substantially larger MSD\textsubscript{total} than the other elements both at 0\,K and at higher temperatures.

\begin{figure*}[tb]
\centering
\includegraphics{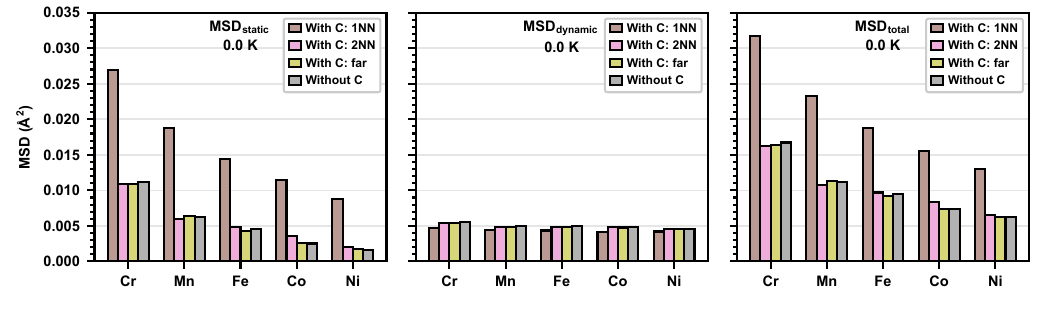}
\caption{Impact of carbon alloying on the element-resolved MSDs of (CrMnFeCoNi)$_{1-x}$C$_x$ ($x \approx 1.8$~at.\%) at 0\,K obtained from DFT.}
\label{fig:DFT_MSDs_C_impact_0.0}
\end{figure*}

The impact of carbon on the element-resolved MSDs in CrMnFeCoNi obtained from DFT calculations at 0\,K is presented in Fig.~\ref{fig:DFT_MSDs_C_impact_0.0}.
The carbon impacts essentially only on MSD\textsubscript{static} and is mostly limited to the six atoms within the first nearest neighbor (1NN) shell around each carbon atom, and it does not essentially impact on MSD\textsubscript{dynamic}.
For the 1NN atoms around carbon, Cr shows the largest MSD\textsubscript{static} increase compared with the case without carbon, followed by Mn, Fe, Co, and Ni in descending order. The impact of carbon on element-resolved MSDs at other temperatures is presented in Appendix~\ref{sec:CarbonImpact}.

\section{Discussion}

\subsection{The 1st shell PDF analysis}

Access to the atomic coordinates of all elements in the final (relaxed) configurations of the simulation boxes allows one to calculate the component-dependent MSD values directly from the coordinates of the 3d atoms for each alloy at both temperatures (see Table\ \ref{table2}) and to estimate their temperature variations (Table\ \ref{table3}). The largest MSD values were found for Cr atoms in both the carbon-free and carbon-bearing  alloys in both states (HT and LT), indicating that the local environment of Cr is the most disordered in terms of lattice relaxations and exhibits the highest degree of static disorder at low temperature in the alloy containing 0.8~at.\% carbon. The local environment of Mn reveals the second largest MSD values for this type of disorder, followed sequentially by the less disordered environments of Fe, Co and Ni. These results are consistent with previous findings that the Cr component exhibits the largest MSD/MSRD values in Cantor alloys \cite{NaRe1, Jalcom, NaRe2, NaRe3, NaRe4}. The \textit{relative} atomic displacements, reflected by the MSRDs calculated from the total PDFs, show behavior similar to that of the  MSDs at both temperatures (Tables\ \ref{table2} and \ref{table3}).

\begin{table*}[t]
    \centering
    \caption{Component-dependent structural parameters (MSD and MSRD) for the (CrMnFeCoNi)$_{1-x}$C$_x$ HEA alloys ($x$ = 0, 0.3 and 0.8~at.\%) at 18 and 285~K. 
    The MSD values were calculated from the coordinates of atoms, while the MSRD values were extracted from the total PDFs.
    The average values of interatomic distances {\textlangle\textit{R}\textrangle} = $2.541 \pm 0.007$~\AA\  at 18~K and $2.545 \pm 0.007$~\AA\ at 285~K. } 
    \begin{tabular}{lcccccc}
    \\
    \hline
         & \multicolumn{6}{c}{ (CrMnFeCoNi)$_{1-x}$C$_x$ alloys } \\    
         & \multicolumn{3}{c}{HT state} & \multicolumn{3}{c}{LT state} \\
         & x = 0 at.\% & x = 0.3 at.\% & x = 0.8 at.\% & x = 0 at.\% & x = 0.3 at.\% & x = 0.8 at.\% \\
         & (HT) & (HT\textunderscore0.3\%) & (HT\textunderscore0.8\%) & (LT) & (LT\textunderscore0.3\%) & (LT\textunderscore0.8\%) \\
    \hline         
         & \multicolumn{6}{c}{MSD (18~K)  (\AA$^2$) } \\ [0.4ex]
Cr&0.037 $\pm$ 0.005&0.048 $\pm$ 0.007&0.059 $\pm$ 0.007&0.036 $\pm$ 0.006&0.059 $\pm$ 0.008&0.061 $\pm$ 0.007  \\
Mn&0.025 $\pm$ 0.004&0.023 $\pm$ 0.003&0.026 $\pm$ 0.003&0.025 $\pm$ 0.003&0.021 $\pm$ 0.003&0.024 $\pm$ 0.003  \\
Fe&0.018 $\pm$ 0.003&0.021 $\pm$ 0.003&0.017 $\pm$ 0.002&0.016 $\pm$ 0.002&0.020 $\pm$ 0.003&0.016 $\pm$ 0.002  \\
Co&0.012 $\pm$ 0.002&0.016 $\pm$ 0.002&0.011 $\pm$ 0.001&0.014 $\pm$ 0.002&0.013 $\pm$ 0.001&0.011 $\pm$ 0.001  \\
Ni&0.011 $\pm$ 0.001&0.012 $\pm$ 0.002&0.010 $\pm$ 0.001&0.010 $\pm$ 0.001&0.012 $\pm$ 0.002&0.009 $\pm$ 0.001  \\
\textlangle{MSD}\textrangle & 0.020  & 0.024 &  0.025 &  0.020 &  0.025 &  0.024 \\
         &  &  &  &  &  & \\
         & \multicolumn{6}{c}{MSRD (18~K) (\AA$^2$), $\pm$0.002~\AA$^2$ } \\  [0.4ex]       
  Cr& 0.030  & 0.037  & 0.040  & 0.030  & 0.040  & 0.040  \\ 
  Mn& 0.024  & 0.024  & 0.025  & 0.024  & 0.024  & 0.024  \\
  Fe& 0.021  & 0.023  & 0.021  & 0.019  & 0.023  & 0.021  \\
  Co& 0.018  & 0.021  & 0.018  & 0.018  & 0.018  & 0.017  \\
  Ni& 0.016  & 0.019  & 0.016  & 0.015  & 0.018  & 0.016  \\   
\textlangle{MSRD}\textrangle & 0.022 & 0.025 & 0.024 & 0.021 & 0.025 &  0.023 \\
         &  &  &  &  &  & \\
         & \multicolumn{6}{c}{MSD (285~K) (\AA$^2$)} \\  [0.4ex]       
  Cr & 0.048 $\pm$ 0.005 & 0.054 $\pm$ 0.005 & 0.062 $\pm$ 0.005 & 0.047 $\pm$ 0.005 & 0.064 $\pm$ 0.006 & 0.066 $\pm$ 0.005 \\
  Mn & 0.039 $\pm$ 0.004 & 0.038 $\pm$ 0.004 & 0.040 $\pm$ 0.004 & 0.040 $\pm$ 0.004 & 0.037 $\pm$ 0.004 & 0.038 $\pm$ 0.004 \\
  Fe & 0.030 $\pm$ 0.004 & 0.033 $\pm$ 0.003 & 0.028 $\pm$ 0.003 & 0.028 $\pm$ 0.003 & 0.028 $\pm$ 0.003 & 0.027 $\pm$ 0.003 \\
  Co & 0.024 $\pm$ 0.003 & 0.025 $\pm$ 0.002 & 0.020 $\pm$ 0.002 & 0.021 $\pm$ 0.002 & 0.023 $\pm$ 0.002 & 0.022 $\pm$ 0.002 \\
  Ni & 0.021 $\pm$ 0.002 & 0.023 $\pm$ 0.002 & 0.018 $\pm$ 0.002 & 0.019 $\pm$ 0.002 & 0.021 $\pm$ 0.002 & 0.018 $\pm$ 0.002 \\
\textlangle{MSD}\textrangle & 0.033  & 0.035  & 0.034  & 0.031 &  0.034 &  0.034    \\    
         &  &  &  &  &  & \\
         & \multicolumn{6}{c}{MSRD (285~K) (\AA$^2$), $\pm$0.002~\AA$^2$} \\ [0.4ex] 
  Cr & 0.034 &  0.038 &   0.041 &   0.034 &   0.043 &   0.043 \\
  Mn & 0.030 &  0.030 &   0.031 &   0.029 &   0.030 &   0.031 \\
  Fe & 0.027 &  0.030 &   0.027 &   0.026 &   0.029 &   0.027 \\
  Co & 0.025 &  0.027 &   0.023 &   0.023 &   0.026 &   0.025 \\
  Ni & 0.023 &  0.025 &   0.021 &   0.022 &   0.025 &   0.022 \\
\textlangle{MSRD}\textrangle & 0.028  & 0.030 &  0.029  & 0.027 & 0.030 & 0.029 \\
    \hline         
    \end{tabular}
    \label{table2}
\end{table*}

\begin{table*}[t]
    \centering
    \caption{The effect of thermal disorder for the (CrMnFeCoNi)$_{1-x}$C$_x$ HEA alloys ($x$ = 0, 0.3 and 0.8~at.\%) in the first coordination shell. 
    $\Delta$MSD and $\Delta$MSRD were calculated using the data from Table\ \protect\ref{table2}. }    
    \begin{tabular}{lcccccc}
    \\
    \hline
         & \multicolumn{6}{c}{ (CrMnFeCoNi)$_{1-x}$C$_x$ alloys } 
         \\
         & \multicolumn{3}{c}{HT state} & \multicolumn{3}{c}{LT state} \\
         & x = 0 at.\% & x = 0.3 at.\% & x = 0.8 at.\% & x = 0 at.\% & x = 0.3 at.\% & x = 0.8 at.\% \\
         & (HT) & (HT\textunderscore0.3\%) & (HT\textunderscore0.8\%) & (LT) & (LT\textunderscore0.3\%) & (LT\textunderscore0.8\%) \\
    \hline         
         & \multicolumn{6}{c}{$\Delta$MSD = MSD (285~K) -- MSD (18~K) (\AA$^2$)} \\ [0.4ex]
  Cr & 0.012  & 0.006  & 0.003  & 0.011  & 0.005  & 0.004 \\
  Mn & 0.014  & 0.015  & 0.014  & 0.015  & 0.016  & 0.014 \\
  Fe & 0.012  & 0.012  & 0.011  & 0.012  & 0.009  & 0.011 \\
  Co & 0.012  & 0.010  & 0.009  & 0.007  & 0.011  & 0.011 \\
  Ni & 0.010  & 0.011  & 0.008  & 0.009  & 0.008  & 0.008 \\ 
         &  &  &  &  &  & \\
         & \multicolumn{6}{c}{$\Delta$MSRD = MSRD (285~K) -- MSRD (18~K) (\AA$^2$)} \\ [0.2ex]
  Cr & 0.004  & 0.001  & 0.001  & 0.004  & 0.003  & 0.002 \\
  Mn & 0.006  & 0.006  & 0.006  & 0.006  & 0.007  & 0.007 \\
  Fe & 0.006  & 0.007  & 0.006  & 0.006  & 0.006  & 0.006 \\
  Co & 0.007  & 0.006  & 0.005  & 0.006  & 0.007  & 0.008 \\
  Ni & 0.007  & 0.006  & 0.005  & 0.006  & 0.007  & 0.007 \\   
    \hline         
    \end{tabular}
    \label{table3}
\end{table*}


There is a correlation between the extracted MSD values for the other 3d elements and the expected lattice stress induced by the differences between the initial crystallographic packing of pure elements and the fcc crystallographic lattice of the Cantor alloy. The crystallographic structures of pure Mn and Fe are $\alpha$-Mn and body-centered-cubic (bcc) Fe, respectively; therefore, Mn and Fe atoms are expected to exhibit larger MSD/MSRD values compared with Co and Ni, which are initially arranged in fcc (or hexagonal-close-packed (hcp)) lattices. This correlation, linked to the initial crystallographic lattice packing of the pure elements comprising the high-entropy alloy, was previously noticed in fcc Al$_{0.3}$-CrFeCoNi \cite{NaRe1} and in single-crystalline fcc CrMnFeCoNi prepared in the HT state \cite{NaRe3}.

\begin{figure}[ht!]
\centering
\includegraphics[width=0.48\textwidth]{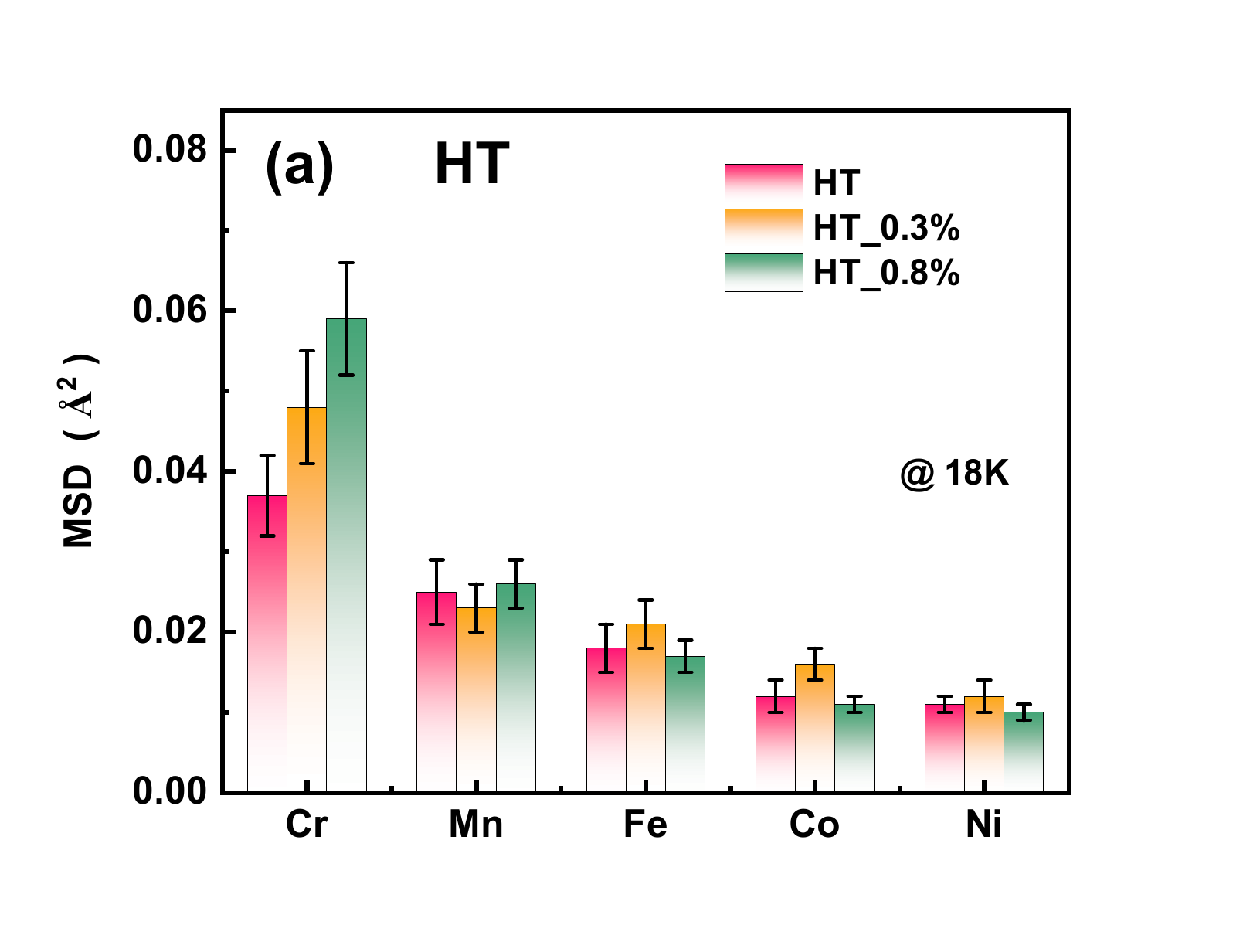}
\hfill
\includegraphics[width=0.48\textwidth]{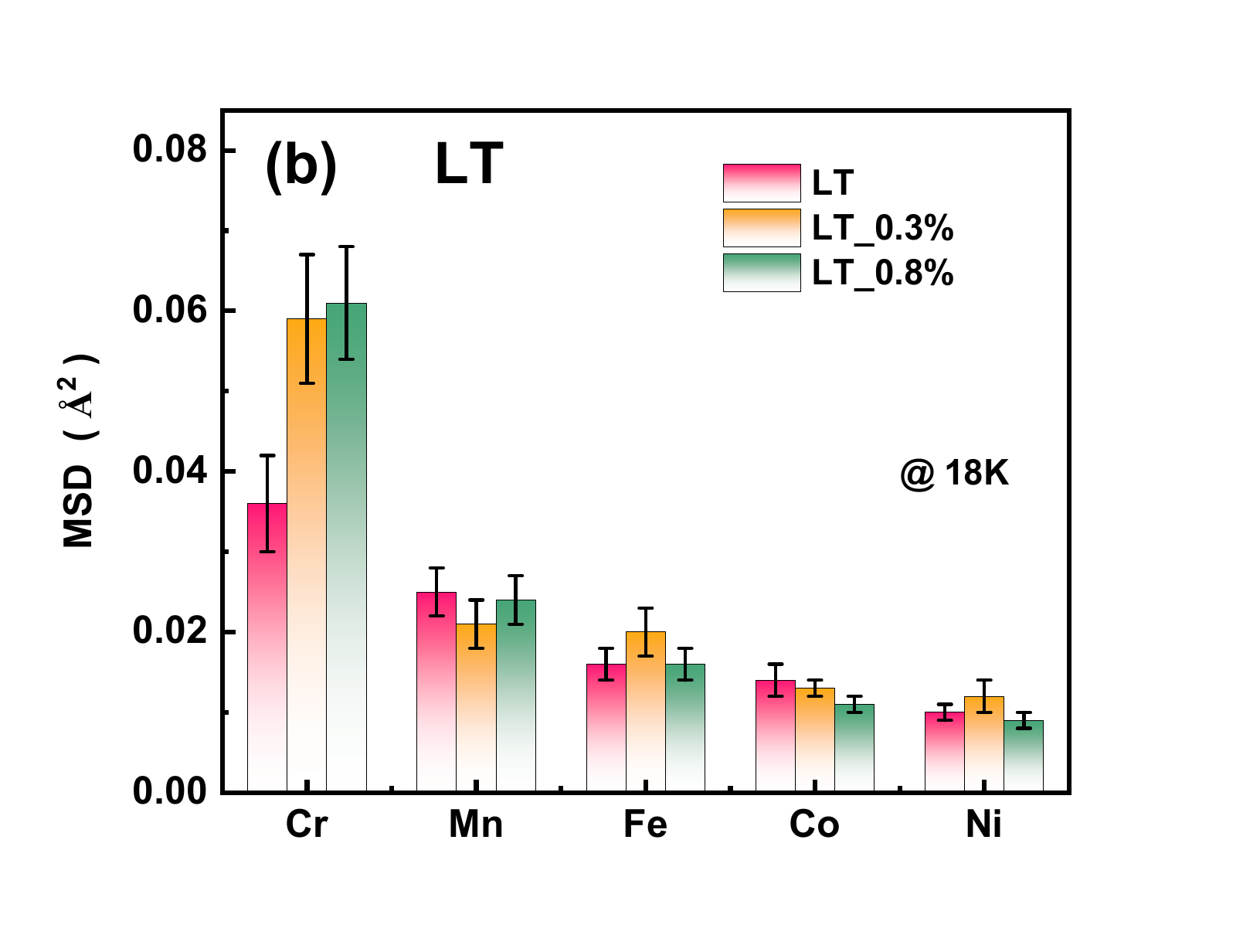}
\caption{Impact of carbon alloying on the component-dependent MSDs extracted from the experimental $\chi(k)k^2$ EXAFS spectra recorded at 18~K from the (CrMnFeCoNi)$_{1-x}$C$_x$ HEAs in the HT (a) and LT (b) states. The monotonic increase of MSDs was found solely for the Cr component.}
\label{fig:CarbonContent18K}
\end{figure}

Carbon alloying produces two general trends in the component-dependent MSDs and MSRDs. First, Cr exhibits a clear and monotonic increase in both quantities with increasing carbon content in both the HT and LT states (Table\ \ref{table2}). The corresponding MSD evolution is illustrated in Fig.\ \ref{fig:CarbonContent18K}. The other elements show only weak and non-monotonic changes. For Fe, Co, and Ni, the MSDs and MSRDs tend to increase at 0.3~at.\% of carbon and decrease again at 0.8~at.\%, whereas Mn shows the opposite tendency.  Although the absolute magnitudes of these variations are small and comparable to the associated uncertainties, their systematic non-monotonic character mirrors the carbon dependence of the self-diffusion coefficients reported in radiotracer experiments \cite{LUKIANOVA2022}. The monotonic increase observed only for Cr supports the vision that interstitial carbon preferentially occupies Cr-rich local environments, leading to the most substantial lattice distortions around Cr and indicating the onset of carbide-related local rearrangements. It is important to underline that the averaged parameters $\langle\mathrm{MSD}\rangle$ and $\langle\mathrm{MSRD}\rangle$ (Table\ \ref{table2}) mask these element-specific relaxations, seriously questioning the applicability of the single displacement parameter to a correct description of the disorder induced, e.g., by interstitial carbon atoms. Similar concerns regarding reliance on a single averaged displacement parameter were raised in an earlier work on AuCuNiPdPt alloys \cite{NaRe4}.

Correlations between element-specific lattice distortions (in terms of MSDs) and diffusion characteristics were recently highlighted for hcp AlScHfTiZr or ScHfTiZr HEAs \cite{SEN2023-Ti, SEN2024-Sc, mohan2024-HCP} and bcc HfTiZrNbTa HEAs \cite{ZHANG2022-Zr, ZHANG2025-imp}. Specifically, the large values of element-specific MSDs in bcc and, especially, in hcp HEAs were shown to influence diffusion behavior, enhancing diffusion rates relative to those expected from a simple average of unary diffusivities.
As a result, the concept of ``anti-sluggish'' diffusion in HEAs with a high level of lattice distortions was formulated \cite{SEN2023-Ti}.

The effect of thermal (dynamic) disorder visible in the PDFs for each constituent element in the considered HEAs (Fig.\ \ref{fig:PDF-RT}) can be expressed through $\Delta$MSDs and $\Delta$MSRDs (see Table\ \ref{table3} and Figs.\ \ref{fig:deltaMSDHTLT}-\ref{fig:consolidatedPlot}), and it is significantly reduced upon carbon alloying,  particularly for the Cr component. This is consistent with the highest degree of static disorder found in the local environment of Cr atoms: if the local atomic surrounding of Cr is highly distorted by interstitial carbon, thermal effects are unlikely to induce substantial additional lattice relaxations for this particular species.

Regarding the MSDs obtained from DFT, based on the local character of the carbon-induced changes, we construct two simple models to estimate the element-resolved MSDs at low carbon concentrations.
Specifically, in the Cantor alloy with \textit{x} at.\% carbon, a fraction of \textit{x}/(100 {\textminus} \textit{x}) of octahedral sites is occupied by carbon. Then, a fraction of 6\textit{x}/(100 {\textminus} \textit{x}) of metal atoms are at the 1NNs of carbon, while the remaining metal atoms are beyond the 1NN shell of any carbon atoms, and their MSDs are approximately equal to those without carbon.
Then, in the ``random'' model, carbon atoms are assumed to occupy octahedral interstitial sites randomly. Consequently, a fraction of 6\textit{x}/(100 {\textminus} \textit{x}) of each metal element is assigned the MSD of the 1NN atoms, while the remaining atoms retain the MSDs without carbon.
In the ``Cr-rich'' model, in contrast, carbon atoms are assumed to occupy only Cr-rich octahedral interstitial sites, consistent with our previous study~\cite{Ikeda_PRM_2019_Impact}, which showed that the carbon solution energy decreases with increasing Cr content in the local environment.
In this model, a fraction of 30\textit{x}/(100 {\textminus} \textit{x}) of Cr shows the MSD of the 1NNs, while the other metal atoms show the values without carbon.

\begin{figure*}[tb]
\centering
\includegraphics{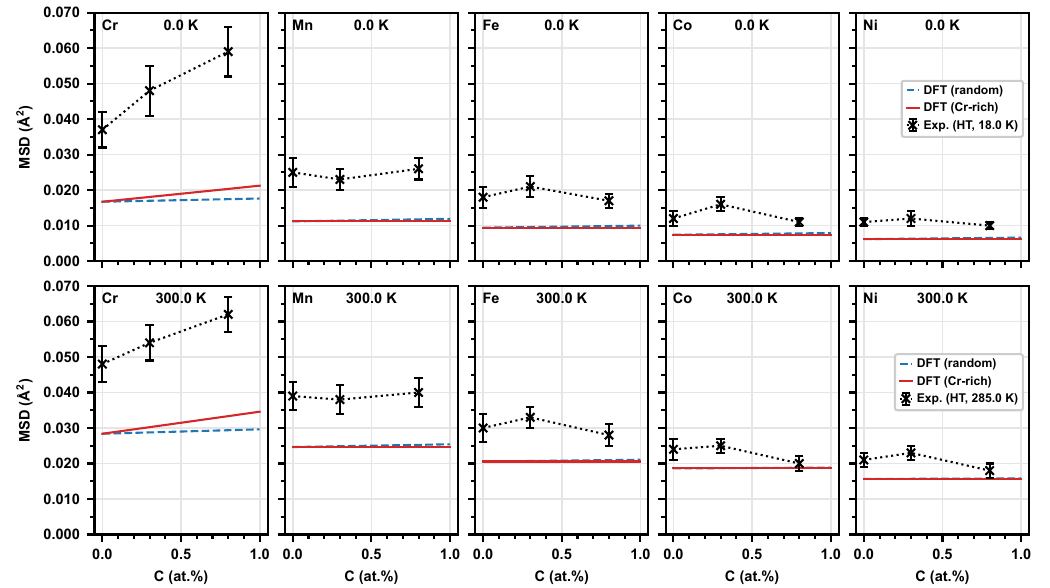}
\caption{MSDs of (CrMnFeCoNi)$_{1-x}$C$_x$ as a function of the carbon concentration (at.\%) at 0\,K and 300\,K estimated from DFT results. Experimental values in the HT state at 18\,K and 285\,K (Table~\ref{table2}) are shown for comparison.}
\label{fig:DFT_MSDs_vs_C_2}
\end{figure*}

Figure~\ref{fig:DFT_MSDs_vs_C_2} presents the resulting element-resolved MSDs as functions of the carbon fraction.
For the random-occupation model, the carbon-concentration dependence of the element-resolved MSDs is barely visible up to 1 at.\% carbon. In contrast, when carbon preferentially occupies Cr-rich octahedral interstitial sites, the Cr MSD increases substantially with increasing carbon concentration, qualitatively reproducing the experimental trend.
Thus, the MSDs extracted from the EXAFS measurements provide independent support for the preferential occupation of Cr-rich octahedral interstitial sites by carbon atoms.
Note that the MSDs estimated from DFT are substantially smaller than the EXAFS-RMC MSDs. Although there are several differences between the DFT calculations and the experiments, for example the lattice parameter (3.57--3.58\ {\AA} in the experiments versus 3.6\ {\AA} in DFT), these differences are too small to account for the observed discrepancy. The origin of the quantitative difference therefore remains to be clarified. Nevertheless, the present discussion is based on the relative changes in the element-resolved MSDs with increasing carbon concentration, which are reproduced consistently by the DFT calculations. A more quantitative comparison between theory and experiment will be the subject of future work.

\subsection{Correlation with diffusion}

\begin{figure}[ht]
\centering
\includegraphics[width=\linewidth]{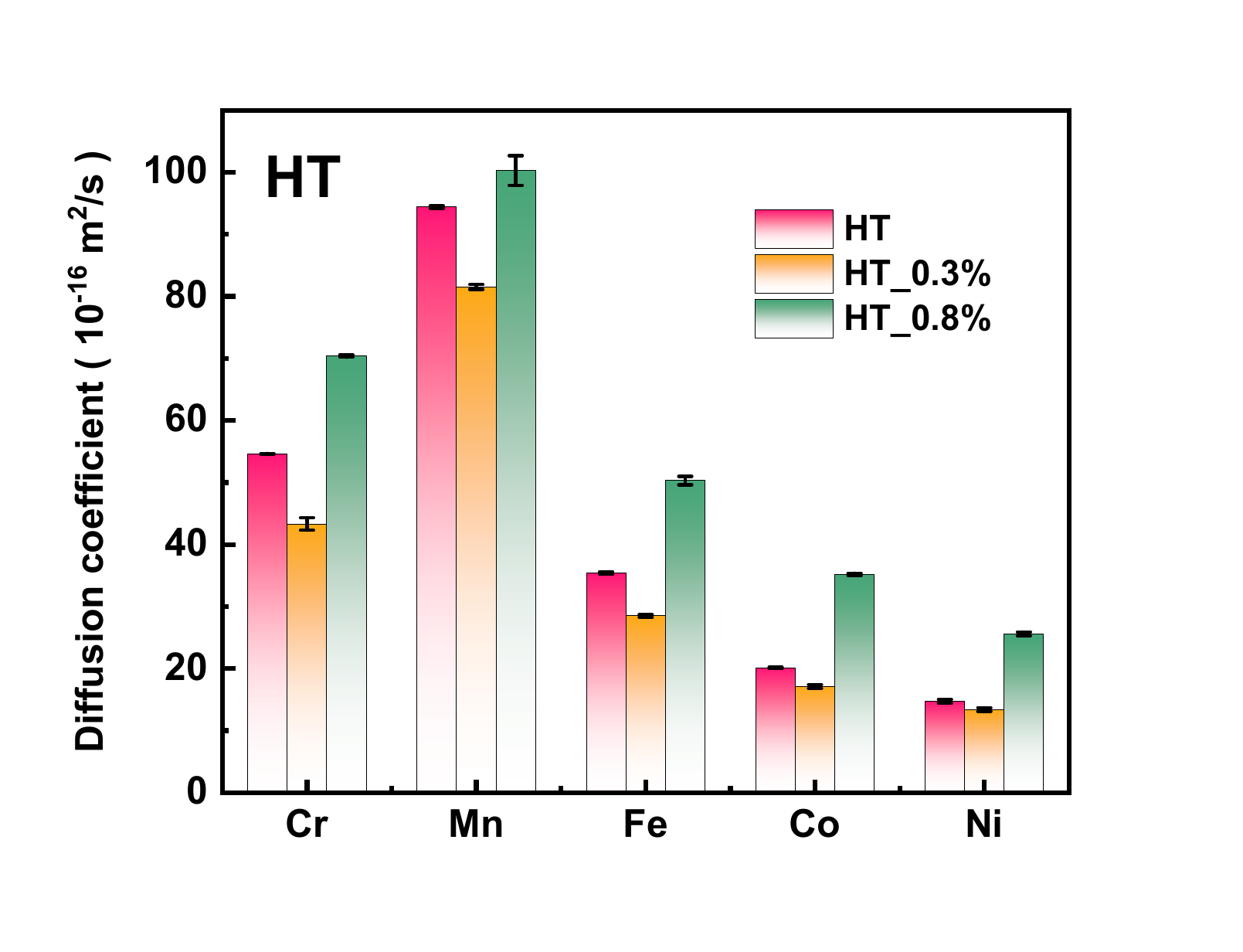}
\caption{Tracer diffusion coefficients \textit{D} (in 10$^{-16}$ m$^{2}$/s) recorded for in the polycrystalline (CrMnFeCoNi)$_{1-x}$C$_x$ HEAs at 1373~K \cite{LUKIANOVA2020}.}
\label{fig:DiffusionHT}
\end{figure}

\begin{figure}[hb]
\centering
\includegraphics[width=\linewidth]{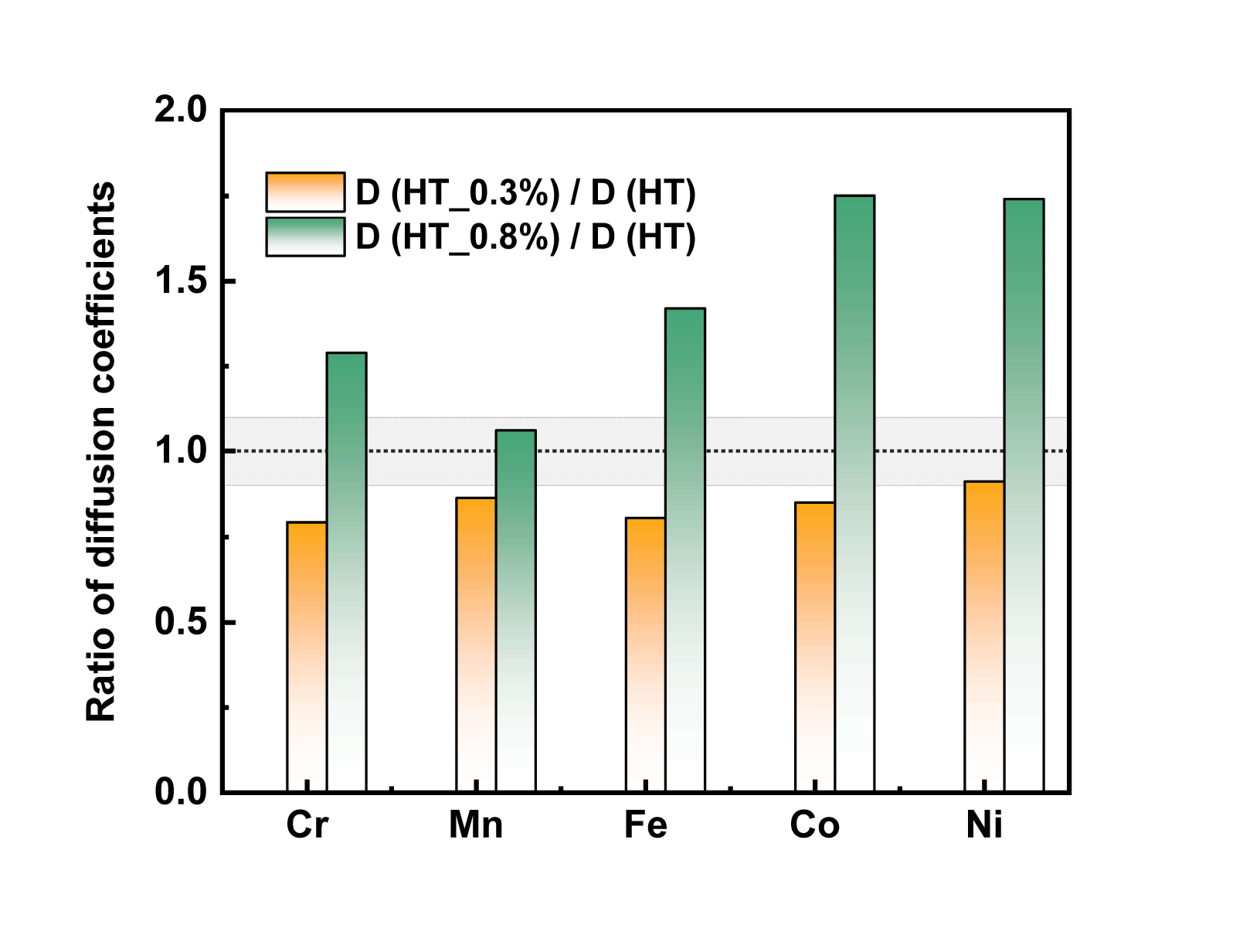}
\caption{Tracer diffusion coefficients in the carbon-alloyed HEAs (CrMnFeCoNi)$_{1-x}$C$_x$ HEAs (HT state) normalized on the corresponding values in the carbon-free alloys, $D(x)/D(x=0)$. The diffusion coefficients are taken from the work of Lukianova et al. \cite{LUKIANOVA2020}. The patterned area represents \textcolor{black}{$1\pm 0.15$} interval that corresponds to typical uncertainties of the tracer diffusion measurements.}
\label{fig:DiffusionHTRatio}
\end{figure}

In Fig.~\ref{fig:DiffusionHT}, the tracer diffusion coefficients, \textit{D}, for all constituent (substitutional) elements reported by Lukianova et al.~\cite{LUKIANOVA2020} for the polycrystalline (CrMnFeCoNi)$_{1-x}$C$_x$ HEAs at $T = 1373$~K are shown. These diffusion data correspond to the state equivalent to the HT state considered in the present work. A non-monotonic dependence of the diffusion coefficients on carbon alloying was found for all constituent elements, with Mn and partially Cr being the fastest elements featuring the largest \textit{D} values. Ni and Co components appear to be the slowest elements in these HEAs at $T = 1373$~K, and Fe stands in between. The relative changes in the diffusion coefficients due to the carbon alloying, i.e. $D(x)/D(x=0)$, where $D(x=0)$ are the corresponding volume diffusion coefficients in the carbon-free alloy (Fig.~\ref{fig:DiffusionHTRatio}), demonstrate that after a small (but distinct) retardation found for all constituent elements at the 0.3~at.\% of carbon addition, a further increase in the carbon content to 0.8~at.\% results in more than a $\sim$50\% enhancement of the diffusion rates of Fe, Co, and Ni, while self-diffusion of Mn remains almost unchanged and Cr becomes faster only by about $\sim$25\%.

\begin{figure}[ht]
\centering
\includegraphics[scale=0.45]{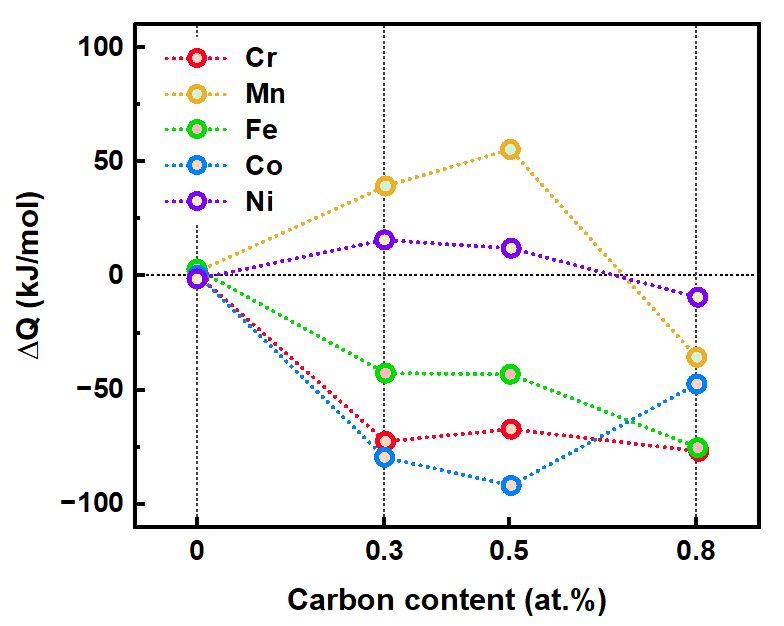}
\caption{Impact of carbon content on the activation enthalpy of element diffusion in the (CrMnFeCoNi)$_{1-x}$C$_x$ HEAs pre-annealed at 1373~K.}
\label{fig:diff}
\end{figure}

Variations of the activation enthalpies of volume diffusion, $Q_Y$, with carbon addition, $\Delta Q_Y=Q_Y(x)-Q_Y(x=0)$ ($Y=$ Cr, Mn, Fe, Co, or Ni; $x$ = 0, 0.3, 0.5 and 0.8 at.\%), are shown for each constituent element in Fig.~\ref{fig:diff}. Strongly non-monotonic $\Delta Q_Y$ dependencies upon increasing carbon content qualitatively agree with the non-monotonic tendencies seen by EXAFS, Fig.~\ref{fig:CarbonContent18K}, with respect to the carbon-induced lattice distortions in terms of MSDs. The largest change in diffusion activation enthalpy is found for the Cr component at $x$ = 0.8 at.\%, while the activation enthalpy for Ni diffusion is less affected at all concentrations.

Lukianova et al.~\cite{LUKIANOVA2022} adopted the model of Overhauser~\cite{Overhauser1953} and correlated the changes of the tracer diffusion coefficients of substitutional atoms with their displacements induced by an interstitial carbon atom considered as a spherical elastic singularity at the octahedral position in the fcc lattice. In compositionally complex systems such as HEAs, we cannot expect a simple direct dependence of the tracer diffusion rates on the lattice distortions and, thus, on the carbon alloying, since the energy barriers for vacancy formation/migration, vibrational frequencies, vacancy–solute binding, correlation factors, and atomically-resolved lattice distortions of each component are influenced by the interstitial carbon atoms in complex and different ways. Local atomic distortions around a particular vacancy (not probed by the EXAFS experiment) are important for a vacancy-mediated atomic transport as well. Though, some qualitative correlations between the element-resolved atomic MSDs and the related changes of the tracer diffusion coefficients induced by the presence of interstitial carbon atoms can be outlined.

\begin{figure}[hb]
\centering
\includegraphics[width=\linewidth]{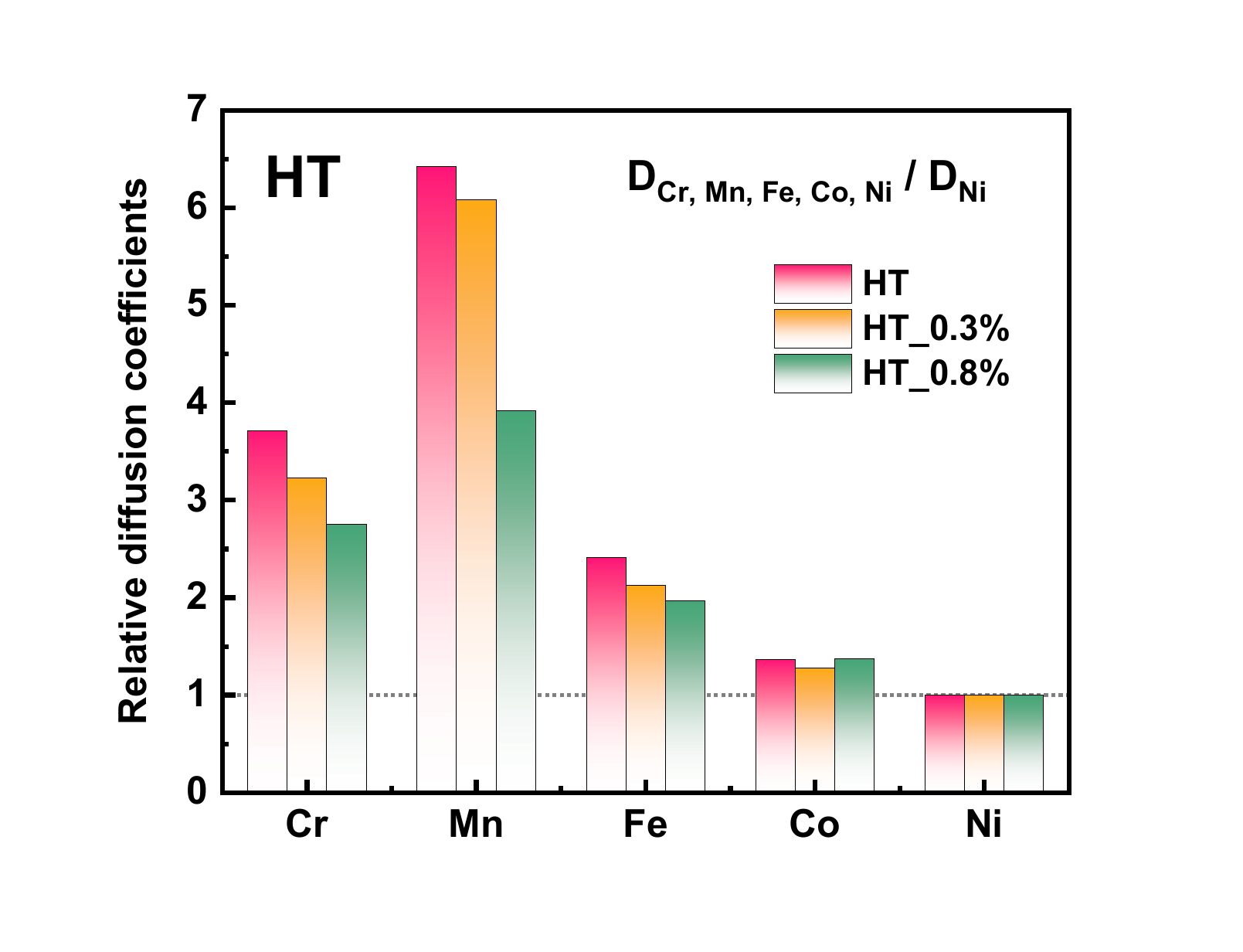}
\caption{Relative tracer diffusion coefficients $D_{\rm Y}/D_{\rm Ni}$ ($Y=$ Cr, Mn, Fe, Co, or Ni) in the carbon-alloyed HEAs (CrMnFeCoNi)$_{1-x}$C$_x$ HEAs (HT state). The diffusion coefficients are taken from the work of Lukianova et al. \cite{LUKIANOVA2020}.}
\label{fig:RelativeDiffHTstate}
\end{figure}

\begin{figure*}[]
\centering
\includegraphics[width=0.49\linewidth]{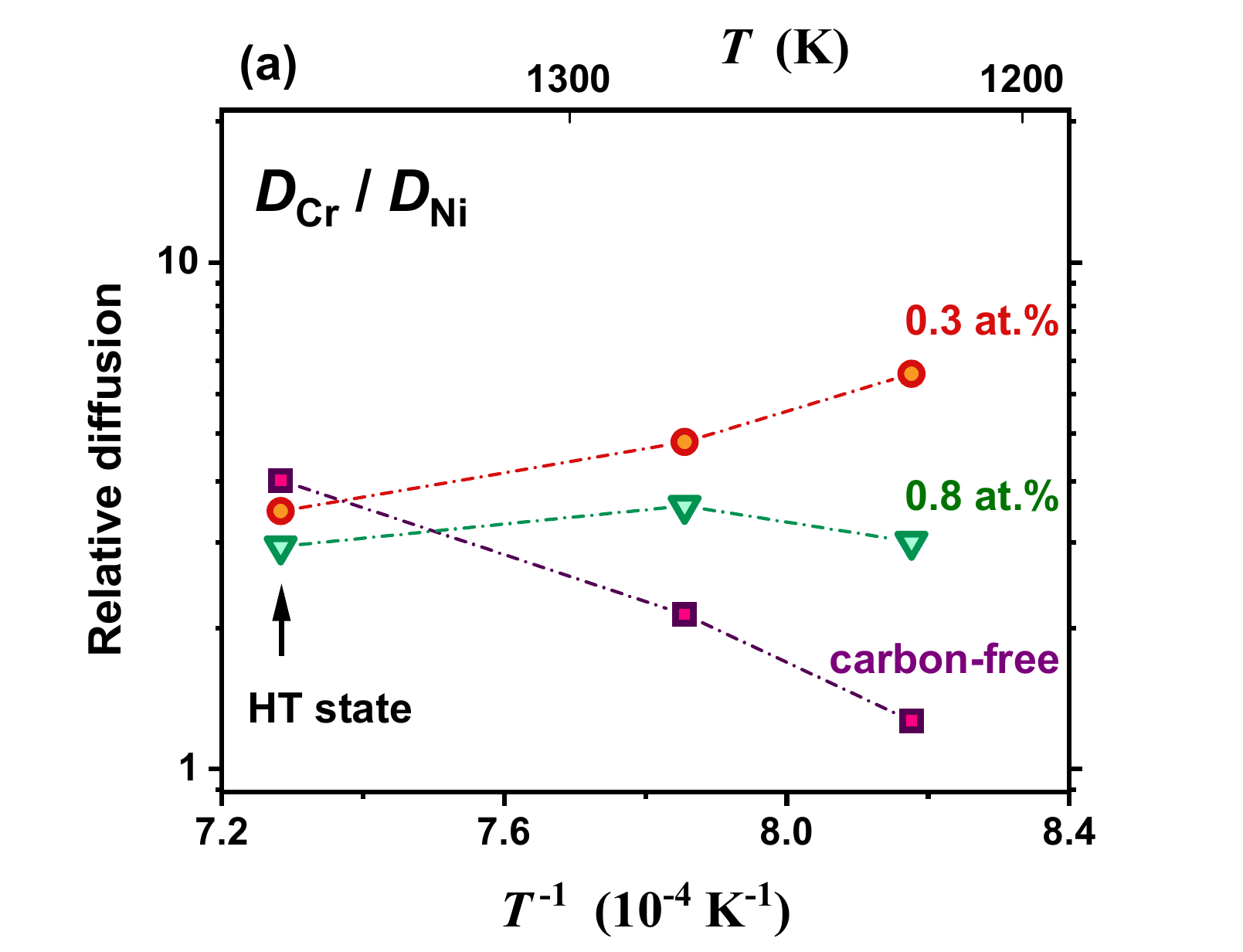}
\hfill
\includegraphics[width=0.49\linewidth]{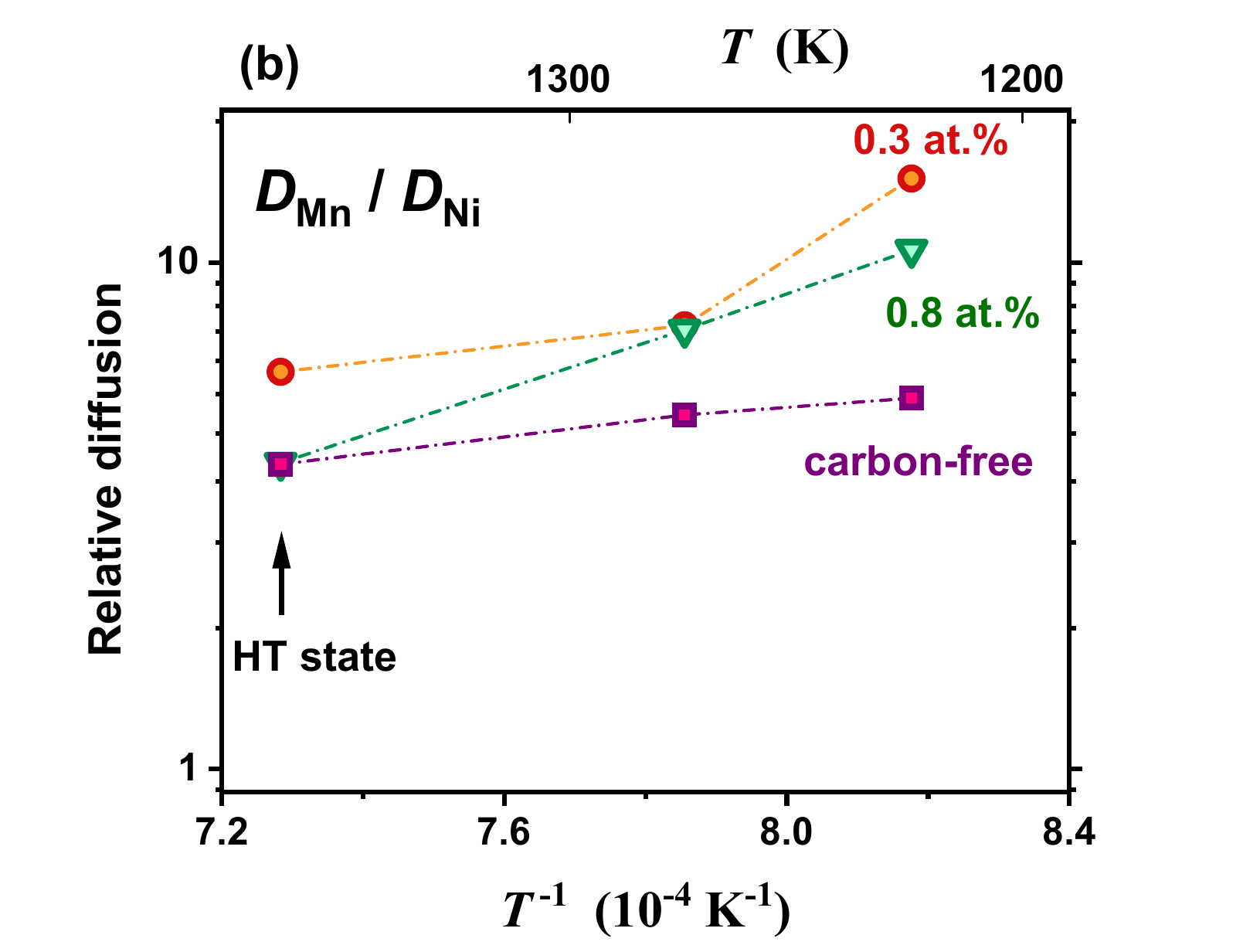}
\caption{Impact of carbon content on the relative Cr (a) and Mn (b) diffusion coefficients, normalized on the corresponding values for Ni diffusion, $D_{\rm Cr}/D_{\rm Ni}$ (left) and $D_{\rm Mn}/D_{\rm Ni}$ (right), respectively. The relative diffusion coefficients are shown as functions of the inverse absolute temperature $1/T$ and $x$ is the carbon content (in at. fractions) in the (CrMnFeCoNi)$_{1-x}$C$_x$ HEAs. The diffusion coefficients are taken from the work of Lukianova et al. \cite{LUKIANOVA2020}.}
\label{fig:diff-norm}
\end{figure*}

Since the activation enthalpy of Ni diffusion is barely affected by the carbon alloying, Fig.~\ref{fig:diff}, Ni can be used as a suitable reference element to address even tiny differences in the behavior of matrix elements induced by interstitial carbon atoms.  In Fig.~\ref{fig:RelativeDiffHTstate}, the impact of carbon content on the relative tracer diffusion coefficients $D_{Y}/D_{\rm Ni}$ ($Y=$ Cr, Mn, Fe, Co, or Ni) for the HT state is shown. One sees that the normalized diffusion coefficients, $D_{Y}/D_{\rm Ni}$, are the largest ones for Mn and Cr and the carbon alloying induces the most prominent changes of the $D_{Y}/D_{\rm Ni}$ values for Mn and Cr. Furthermore, whereas $D_{\rm Cr}/D_{\rm Ni}$ demonstrates almost uniform changes with the carbon alloying, $D_{\rm Mn}/D_{\rm Ni}$ varies more irregularly with carbon concentration, in a striking similarity to the corresponding MSD values (Fig.~\ref{fig:CarbonContent18K}).

A further difference between the Cr and Mn responses to carbon alloying is related to the corresponding temperature dependencies. In Fig.~\ref{fig:diff-norm}, the Cr and Mn tracer diffusion coefficients normalized on that of Ni are shown for different amounts of interstitial carbon atoms limiting the data to the temperature interval, which is representative for the HT state. Here, the diffusion data reported by Lukianova et al. \cite{LUKIANOVA2022} are used. In carbon-free HEAs, the temperature dependence of the relative diffusion coefficients of Cr and Mn, i.e. normalized on those of Ni, demonstrates opposite behaviors: while Mn becomes faster with respect to Ni with decreasing temperature $T$ (at which diffusion experiments were performed), Cr atoms become progressively less mobile with respect to Ni. The carbon alloying drastically modifies the temperature dependence of the reduced (i.e., Ni-normalized) Cr diffusion, while that of Mn remains qualitatively unchanged with only small variations in the absolute values for the diffusion coefficients. This situation can be related to stronger modifications of the Cr local environments and, thus, of the Cr MSDs, while MSDs for other elements remain largely the same. 

An extended analysis of the diffusion behavior in the fcc \cite{VAIDYA2018}, bcc \cite{ZHANG2022-Zr, ZHANG2025-imp}, and hcp \cite{SEN2023-Ti, SEN2024-Sc} HEAs revealed a strong correlation between the lattice distortions and the sluggishness of diffusion in these systems, as it was discussed by Dash et al. \cite{Dash2022}. In fact, diffusion might be considered  as sluggish in alloys with relatively low values of the lattice distortions (like in the fcc CrMnFeCoNi alloy), but no sluggishness is seen in bcc HfTiZrNbTa and HfTiZrNbV alloys with elevated distortions, and even ”anti-sluggish” diffusion was observed for hcp AlScHfTiZr HEAs with large lattice distortions. It was found that the element-specific lattice distortions correlate with the diffusion behavior of Ti and Sc atoms in the hcp HEAs \cite{SEN2023-Ti, SEN2024-Sc}. So, the diffusion enhancement between different alloys follows, in general, the MSD values.

There are at least two known mechanisms that influence self-diffusion coefficients in the (CrMnFeCoNi)$_{1-x}$C$_x$ HEAs upon carbon alloying \cite{LUKIANOVA2020}: vacancy–carbon interactions, which immobilize vacancies at low doping levels, and carbon-induced lattice expansion, which increases the vacancy concentration at higher carbon contents. The formation of carbides could additionally contribute to lattice expansion by providing extra vacancy sources at carbide/matrix interfaces. According to the present experimental data, Table~\ref{tab:Values}, the carbon-induced lattice expansion is at most marginal. Therefore, the role of interstitial carbon-induced lattice distortions is likely the most significant. However, a direct, one-to-one comparison between the radiotracer self-diffusion data and the carbon-induced changes in component-dependent atomic-scale lattice distortions cannot be expected, since thermally activated atomic jumps required for diffusion primarily depend on variations in the predominant vacancy environment of the components. Nevertheless, the experimental EXAFS data, together with the DFT results, suggest that carbon atoms preferentially occupy interstitial positions with a local surplus of particular Cr atoms, influencing the diffusion coefficients of all constituent elements in a complex and strongly non-monotonic manner due to the high atomic heterogeneity of these compositionally complex alloys.

\section{Summary and Conclusions}

In this study, the impact of carbon alloying on the local environment in the polycrystalline (CrMnFeCoNi)$_{1-x}$C$_x$ HEAs with nominal carbon contents of $x$ = 0, 0.3 and 0.8 at.\% was explored at the atomic scale using temperature-dependent, multi-edge X-ray absorption spectroscopy combined with RMC simulations. The results are supported by  first-principles DFT (0\,K) and DFT-MD (300\,K) calculations.

The component-dependent, statistically- and configuration-averaged PDFs for each alloy were obtained from a simultaneous fit provided through  RMC-based analysis applied to the experimental EXAFS spectra collected at the K absorption edges of all five constituent 3d elements. The results indicate that the local atomic environment of Cr atoms is the most disordered, regardless of the degree of carbon alloying, the final annealing temperature maintained during alloy preparation (1373~K for the HT or 993~K for the LT states), and the temperature held during the EXAFS experiment (18~K or 285~K). The second most disordered local environment was found for Mn, followed by Fe, and finally by Co and Ni.

The static disorder around Cr atoms, represented by Cr MSDs, was found to increase monotonically with carbon content and is well visible in the PDFs over all five coordination shells around Cr absorbers, up to 6~\AA. This observation is interpreted as a signature of the emerging initial transformation phase towards the formation of Cr carbides expected at  higher carbon doping levels. This effect is slightly larger for alloys in the LT state compared with those in the HT state. The MSDs of other components do not show a clear monotonic increase upon carbon alloying and preferentially remain within the error bars of the RMC method, although they exhibit non-monotonic variations that might be indirectly correlated with prior findings from radiotracer diffusion experiments. The observed increase in thermal disorder upon heating during the EXAFS experiment is visible for all components in the studied alloys; however, it is significantly reduced in the carbon-doped ones specifically for the Cr constituent, as reflected by its $\Delta$MSD/$\Delta$MSRD parameters.

The DFT and DFT-MD calculations qualitatively reproduce the element-specific MSDs obtained from the EXAFS–RMC data and likewise identify Cr as the element whose local environment is most strongly affected by interstitial carbon. The calculations also confirm that carbon preferentially occupies Cr-rich octahedral interstitial sites. The remaining quantitative differences between the theoretical and experimentally obtained MSDs may arise from several factors. First, the MSDs derived from EXAFS measurements may vary depending on the analysis procedure used to extract them~(see Ref.~\cite{Oh_AM_2021_Element}). Indeed, the range of reported experimental values encompasses the present DFT results. Second, methodological aspects of the theoretical treatment—including, but not limited to, magnetic interactions beyond the collinear approximation—may also contribute to the discrepancies. Further experimental and theoretical investigations will be needed to clarify the relative importance of these factors.

Our results emphasize how dilute interstitial carbon influences component-dependent lattice displacements in (CrMnFeCoNi)$_{1-x}$C$_x$ alloys at the atomic scale, which are important for understanding the carbon-initiated phase transitions in different compositionally complex systems, and link the observed tendencies in lattice distortions to the previously reported non-monotonic self-diffusion behavior of components. The results also reinforce that combining multi-edge EXAFS spectroscopy with RMC-based analysis and first-principles simulations provides a valuable approach for understanding and, ultimately, tailoring carbon-strengthened high-entropy alloys.

\section*{Acknowledgment}

The authors acknowledge DESY (Hamburg, Germany), a member of the Helmholtz Association HGF, for the provision of experimental facilities. A part of this research was carried out on the P65 (EXAFS) beamline of PETRA III (the beamtime was allocated within the proposal I-20240074). Partial funding from the German research Foundation (DFG) is acknowledged through the DI 24-1 research grant (project number 509804947).
A. K. is grateful for the financial support from the Latvian Council of Science project no. lzp-2023/1-0476. 
F.K. acknowledges funding through the DFG Heisenberg Programme, project number 541649719.
Y.I. acknowledges funding through the DFG under IK125/1-1 (Project No. 519607530).

\section*{Data availability}

The raw EXAFS data and RMC final configurations that support the findings of this study are available in Zenodo at https://doi.org/10.5281/zenodo.20784672 \cite{datasetEXAFSRMC}.

\section*{Declaration of competing interest}

The authors declare that they have no known competing financial interests or personal relationships that could have appeared to influence the work reported in this paper.

\section*{Declaration of generative AI use}

The authors declare that they have not used any generative AI in the manuscript preparation process.

\section*{Author contribution statement}

Alevtina Smekhova: Investigation (EXAFS), writing – original draft, writing – review \& editing, visualization.
Alexei Kuzmin: Conceptualization, formal analysis (EXAFS and RMC), writing – original draft, writing – review \& editing, visualization, resources.
G. Mohan Muralikrishna: Investigation (Processing of the samples, Characterization), writing – review \& editing.
Edmund Welter: Resources, Software, writing – review \& editing.
Sergiu Levcenko: Software, writing – review \& editing.
Fritz K\"ormann: DFT data analysis, writing – original draft, writing – review \& editing.
Yuji Ikeda: DFT and DFT-MD calculations, DFT data analysis, writing – original draft, writing – review \& editing.
Sergiy V. Divinski: Conceptualization, writing – original draft, writing – review \& editing, resources.

\appendix
\section{EXAFS and RMC-based fit at the near room temperature}
\label{sec:AppEXAFSRMC}

For the sake of comparison with the low-temperature EXAFS data, the normalized X-ray absorption spectra recorded at $T$ = 285~K for the
(CrMnFeCoNi)$_{1-x}$C$_x$ ($x$ = 0, 0.3 and 0.8~at.\%) alloys in the HT and LT states are
shown in Fig.\ \ref{fig:SM_EXAFS285K} (left panels) along with the $k^2$-weighted experimental EXAFS spectra $\chi(k)k^2$ (middle panels) and their Fourier transforms (right panels) that were used in the RMC-based fit. Qualitatively, the results of the RMC-based analysis are similar to those obtained at $T$ = 18~K.

\setcounter{figure}{0}
\renewcommand\thefigure{A\arabic{figure}} 
\setcounter{table}{0}
\renewcommand\thetable{A\arabic{table}} 
\begin{figure*}[t]
\centering
\includegraphics[scale=0.65]{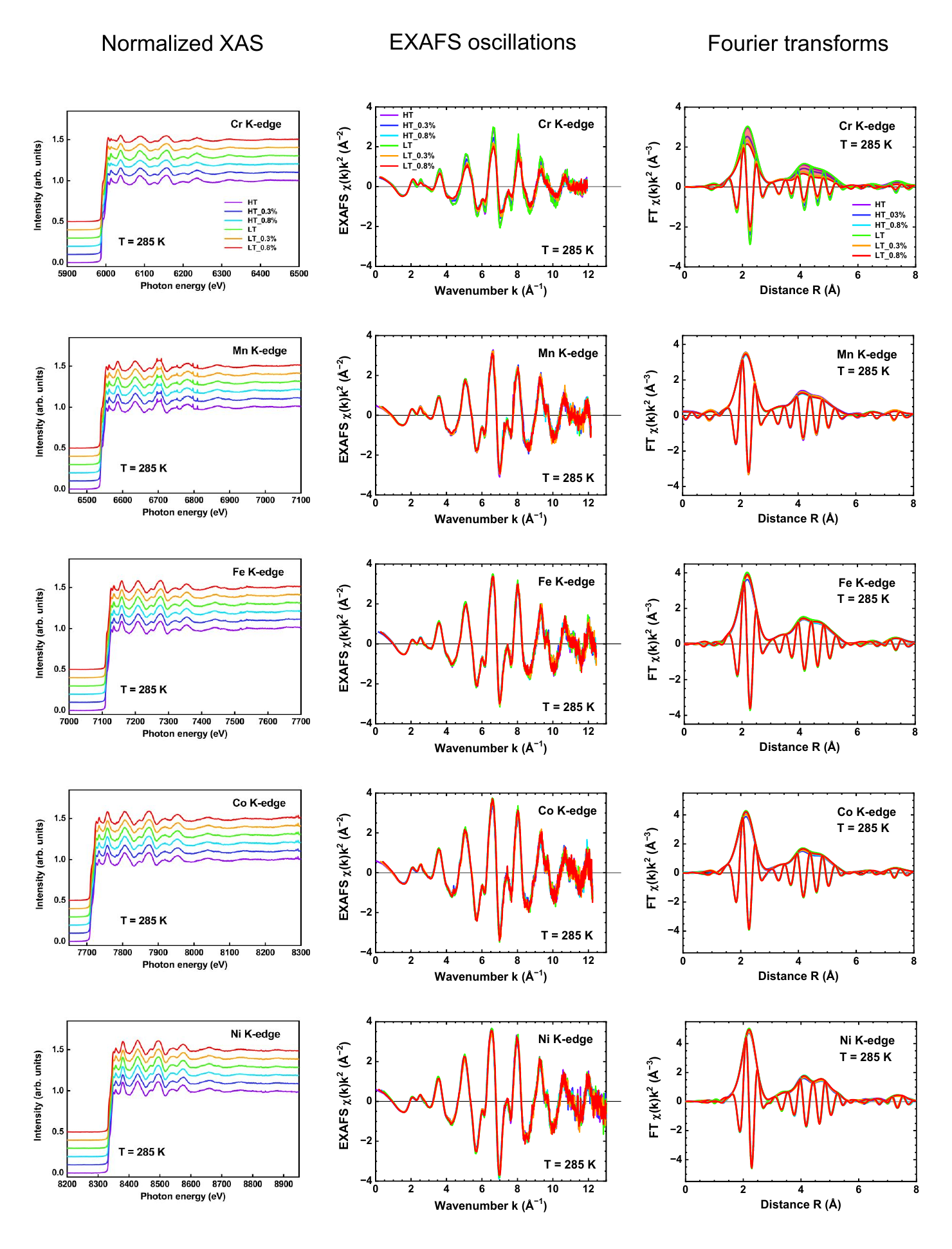}
\caption{
(left) The normalized X-ray absorption spectra collected at the K absorption edges of Cr ($E_0$ = 5.9~keV), Mn ($E_0$ = 6.5~keV), Fe ($E_0$ = 7.1~keV), Co  ($E_0$ = 7.7~keV), and Ni ($E_0$ = 8.3~keV) at 285~K using fluorescence yield and 45$^{\circ}$  incidence geometry for the (CrMnFeCoNi)$_{1-x}$C$_x$ ($x$ = 0, 0.3 and 0.8~at.\%) alloys. The spectra are normalized to unity and shifted vertically for clarity. The color scheme is the same for all plots. (middle) The experimental EXAFS spectra $\chi(k)k^2$ and (right) their Fourier transforms (FTs) for the (CrMnFeCoNi)$_{1-x}$C$_x$ alloys collected at the K-absorption edges of Cr, Mn, Fe, Co, and Ni at 285~K. The most pronounced changes upon carbon doping are seen in the $\chi(k)k^2$ EXAFS oscillations and the FTs for Cr component.}
\label{fig:SM_EXAFS285K}
\end{figure*}

\begin{figure*}[tbh]
\centering

\includegraphics[width=0.35\textwidth]{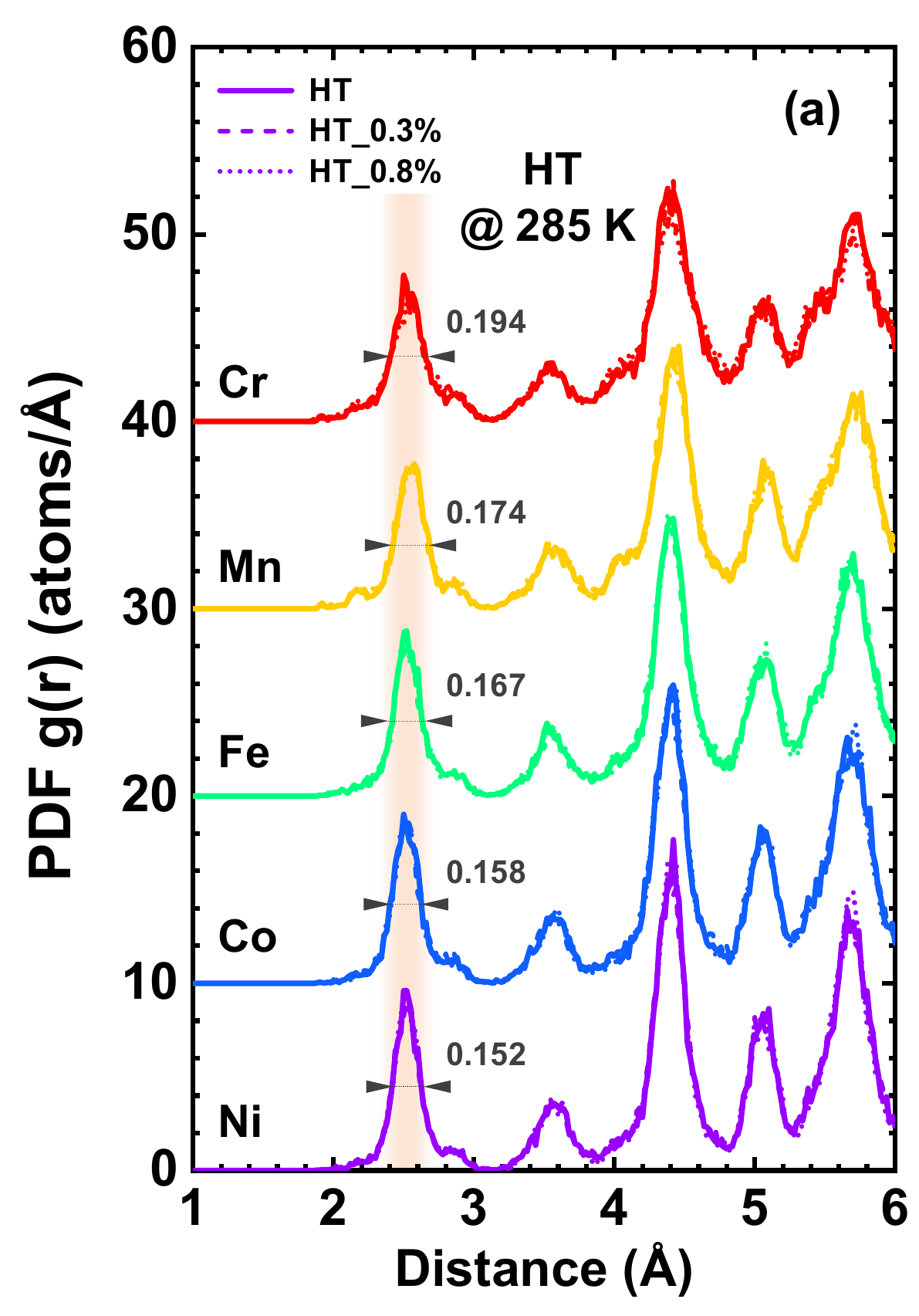}
\includegraphics[width=0.35\textwidth]{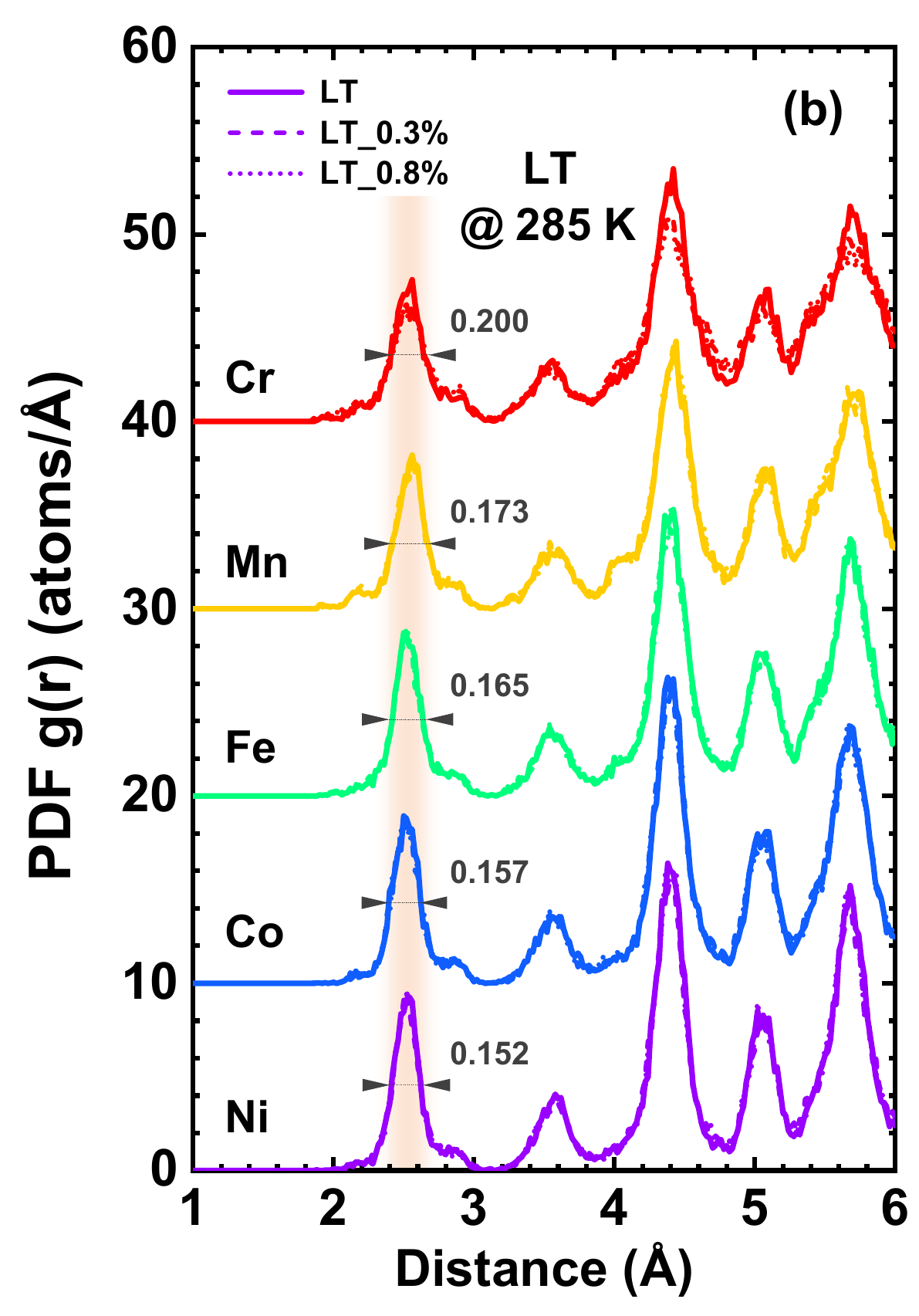}

\caption{Effect of carbon doping on the total pair distribution functions (PDFs) in the (CrMnFeCoNi)$_{1-x}$C$_x$ alloys ($x$ = 0, 0.3 and 0.8~at.\%) at 285~K in the HT (a) and LT (b) states: 
 HT and LT (solid lines), HT\textunderscore0.3\%, LT\textunderscore0.3\% (dashed lines), HT\textunderscore0.8\% and LT\textunderscore0.8\% (dotted lines). The most pronounced changes upon carbon alloying are seen in PDFs of Cr atoms for the alloy in the LT state. The peaks related to the first coordination shell around a particular type of absorbers are highlighted; the values for the averaged MSRDs taken from Table\ \protect\ref{table2} (in \AA) are indicated for a quick cross reference.}
\label{fig:PDF-carbonRT}
\end{figure*}

The total PDFs calculated from the EXAFS data collected at 285~K for the alloys under study are shown in Fig.\ \ref{fig:PDF-carbonRT}. The comparison of PDFs peak intensities reveals that all coordination shells of Cr atoms are visibly affected upon interstitial carbon alloying and demonstrate strong variations for all distant shells contrary to the PDFs of other 3d components.

\begin{figure*}[tb]
\centering
\includegraphics[width=0.42\linewidth]{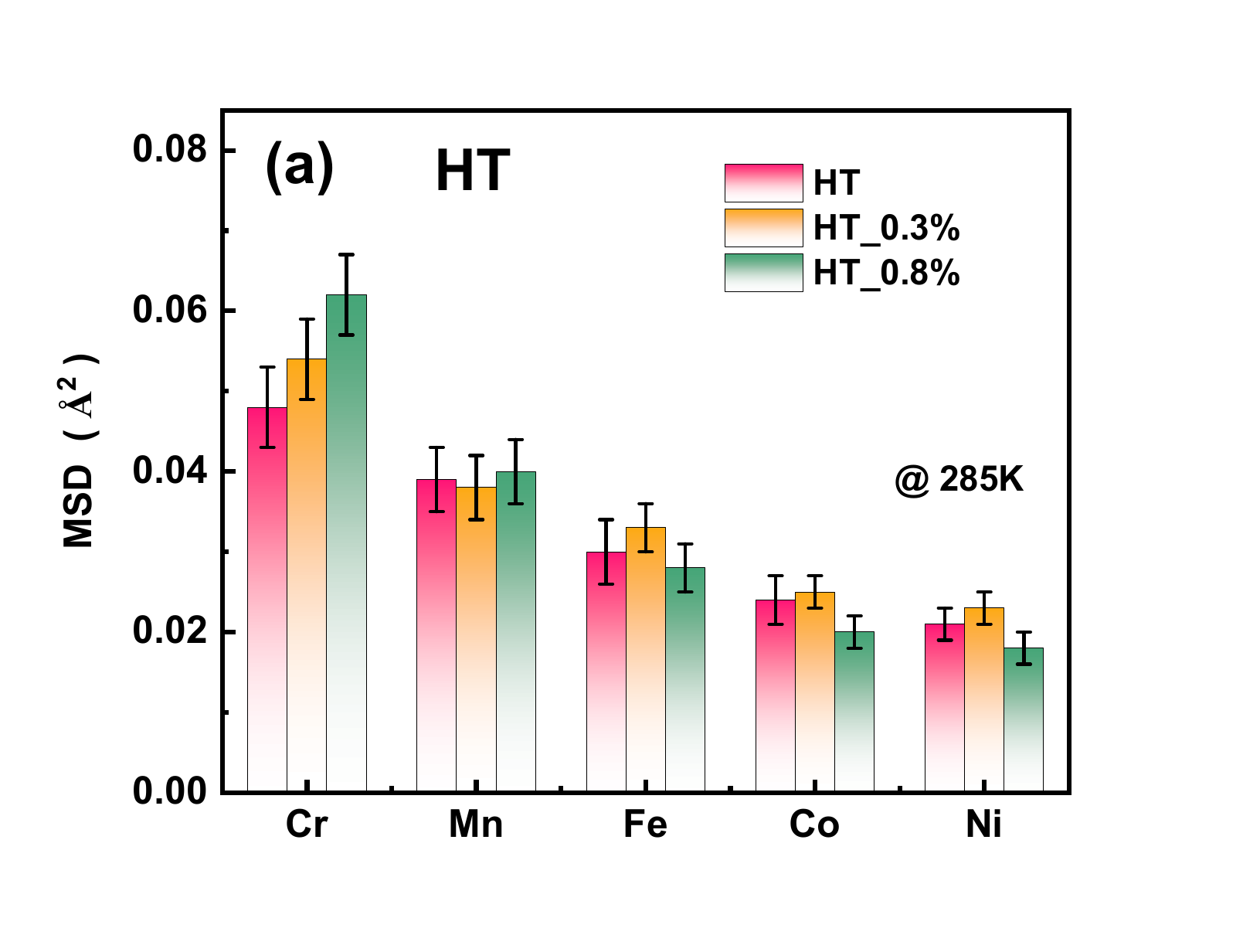}
\includegraphics[width=0.42\linewidth]{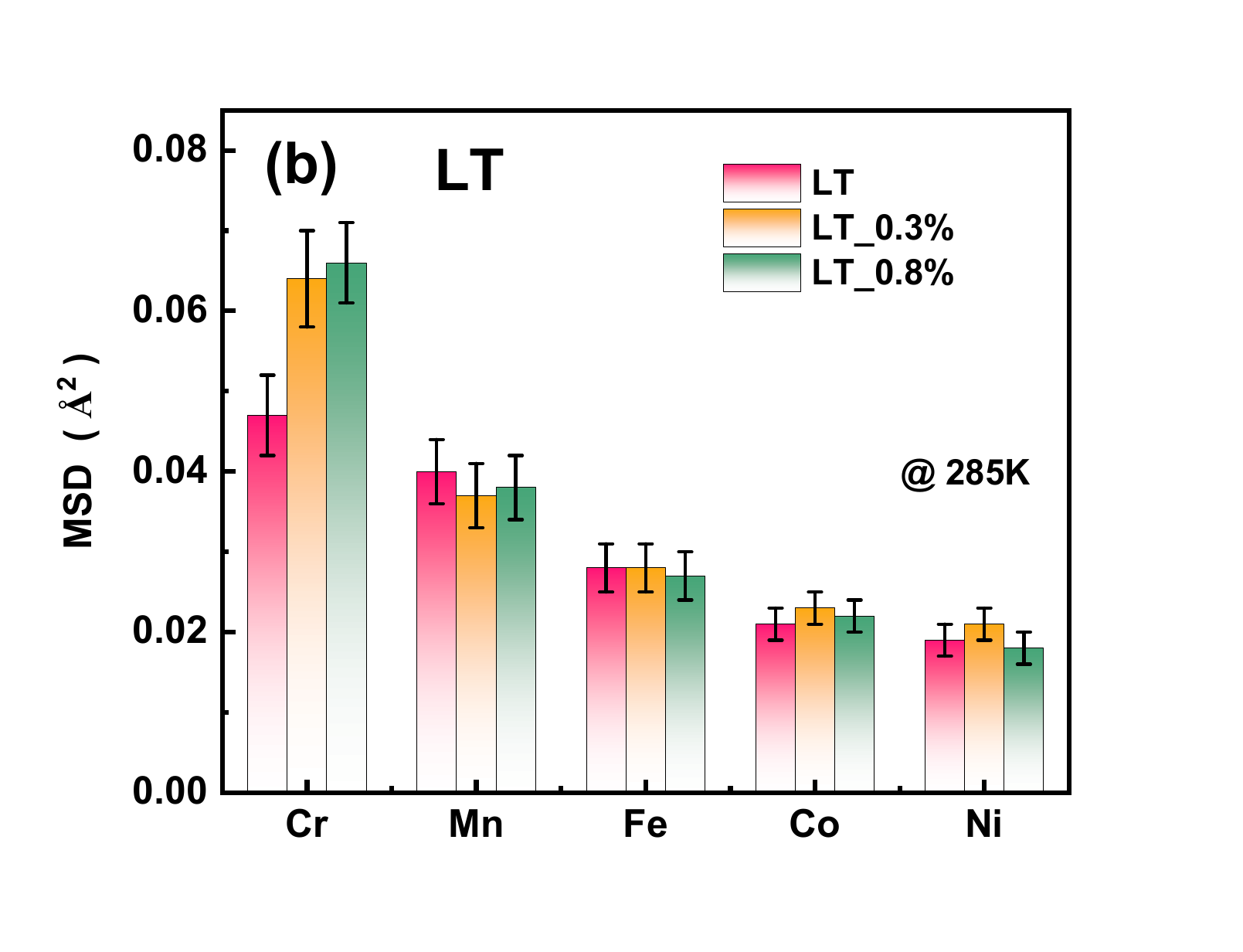}
\caption{Impact of carbon alloying on the component-dependent MSDs extracted from the experimental $\chi(k)k^2$ EXAFS spectra recorded at 285~K from the (CrMnFeCoNi)$_{1-x}$C$_x$ HEAs in the HT (a) and LT (b) states. The monotonic increase of MSDs was found solely for the Cr component.}
\label{fig:CarbonContent285K}
\end{figure*}

In Fig.\ \ref{fig:CarbonContent285K} the component-dependent MSDs extracted from the experimental $\chi(k)k^2$ EXAFS spectra recorded at 285~K from the carbon-free and carbon-doped HEAs are presented. The monotonic increase of MSDs was found solely for the Cr component, whereas other components demonstrate different non-monotonic behaviors.  

\begin{figure*}[tb]
\centering
\includegraphics[width=0.42\linewidth]{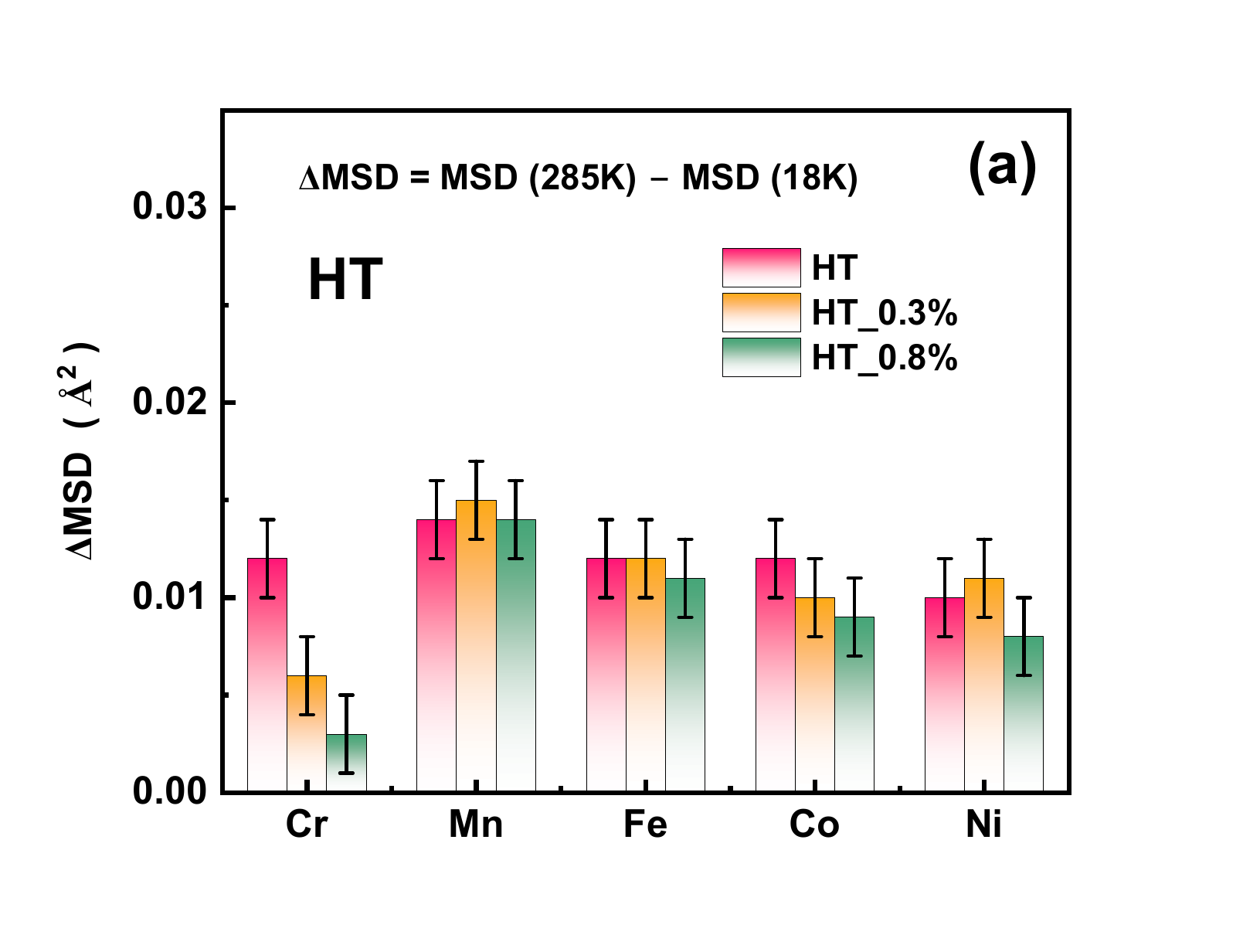}
\includegraphics[width=0.42\linewidth]{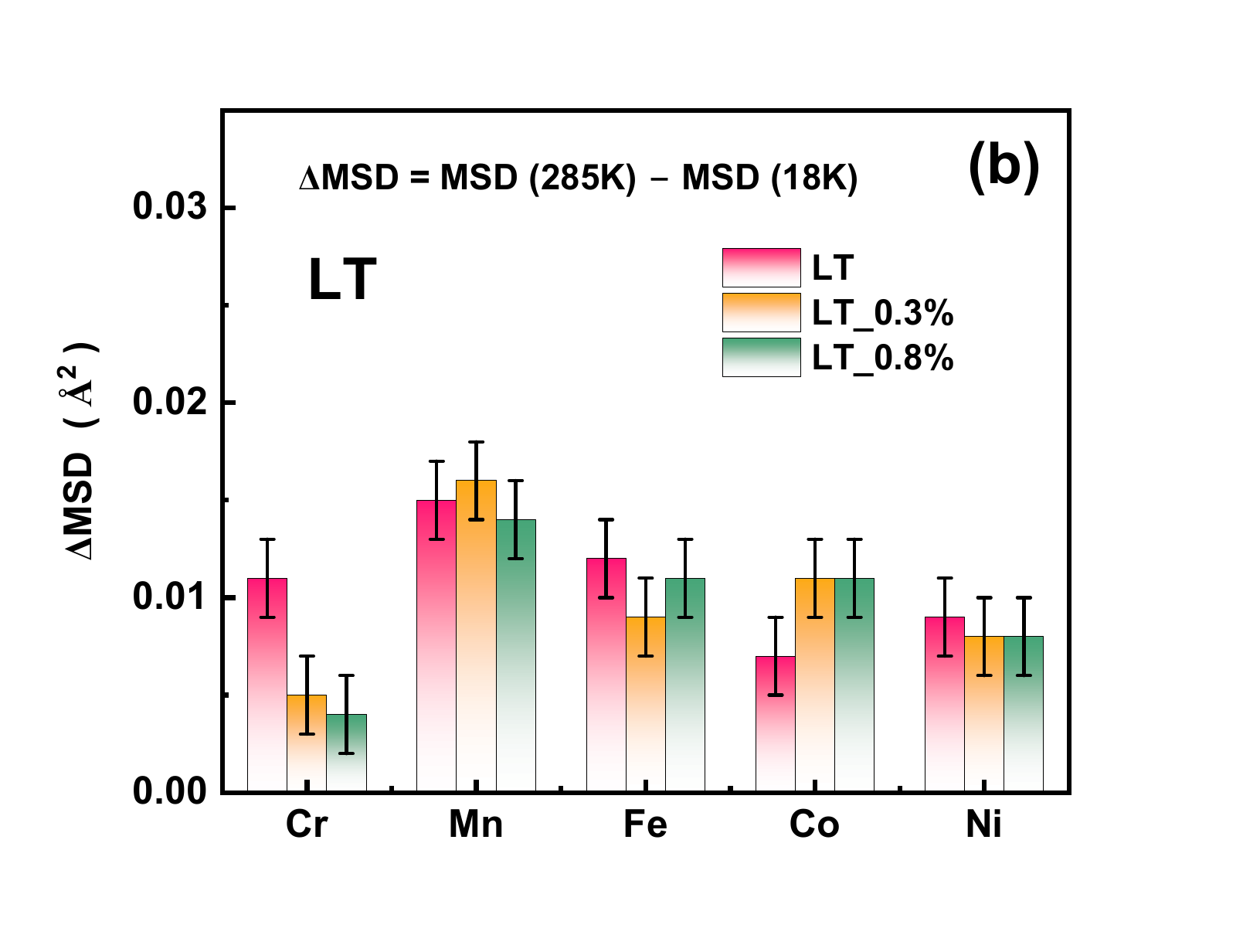}
\caption{The effect of thermal (dynamic) disorder expressed through $\Delta$MSDs for the (CrMnFeCoNi)$_{1-x}$C$_x$ HEAs in HT (a) and LT (b) states. The effect is significantly reduced specifically for the Cr constituent.}
\label{fig:deltaMSDHTLT}
\end{figure*}

\begin{figure*}[tb]
\centering

\includegraphics[width=0.32\textwidth]{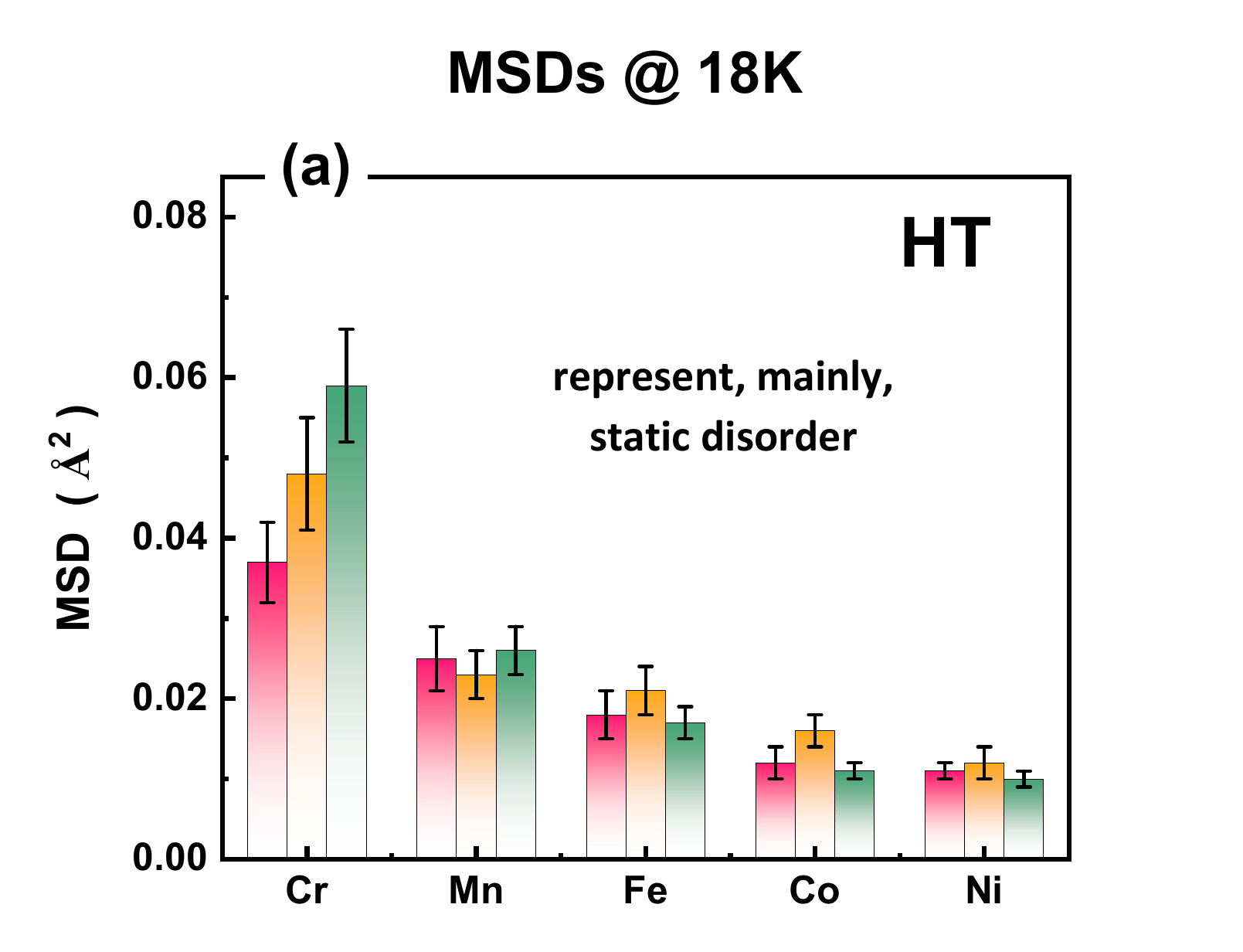}
\includegraphics[width=0.32\textwidth]{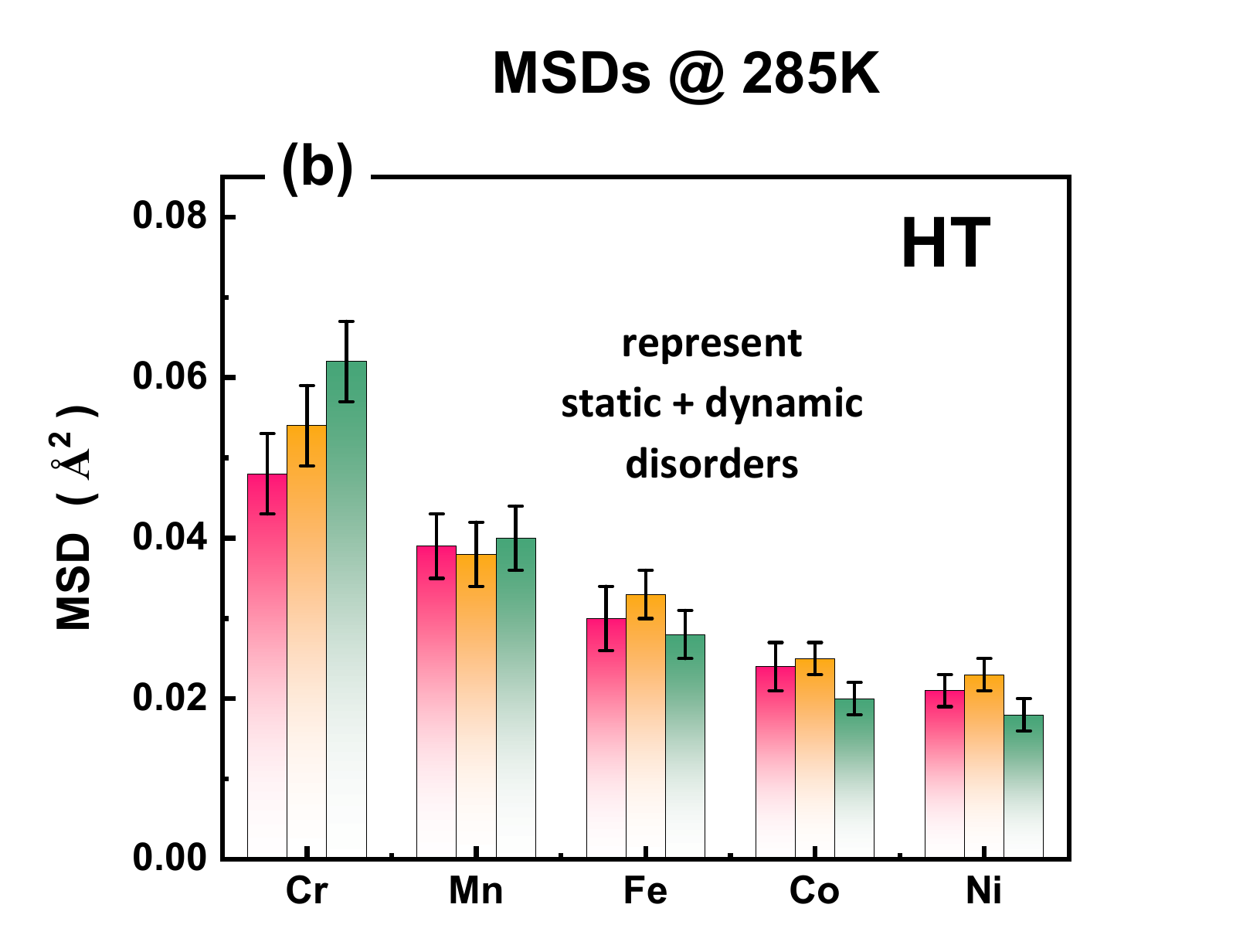}
\includegraphics[width=0.32\textwidth]{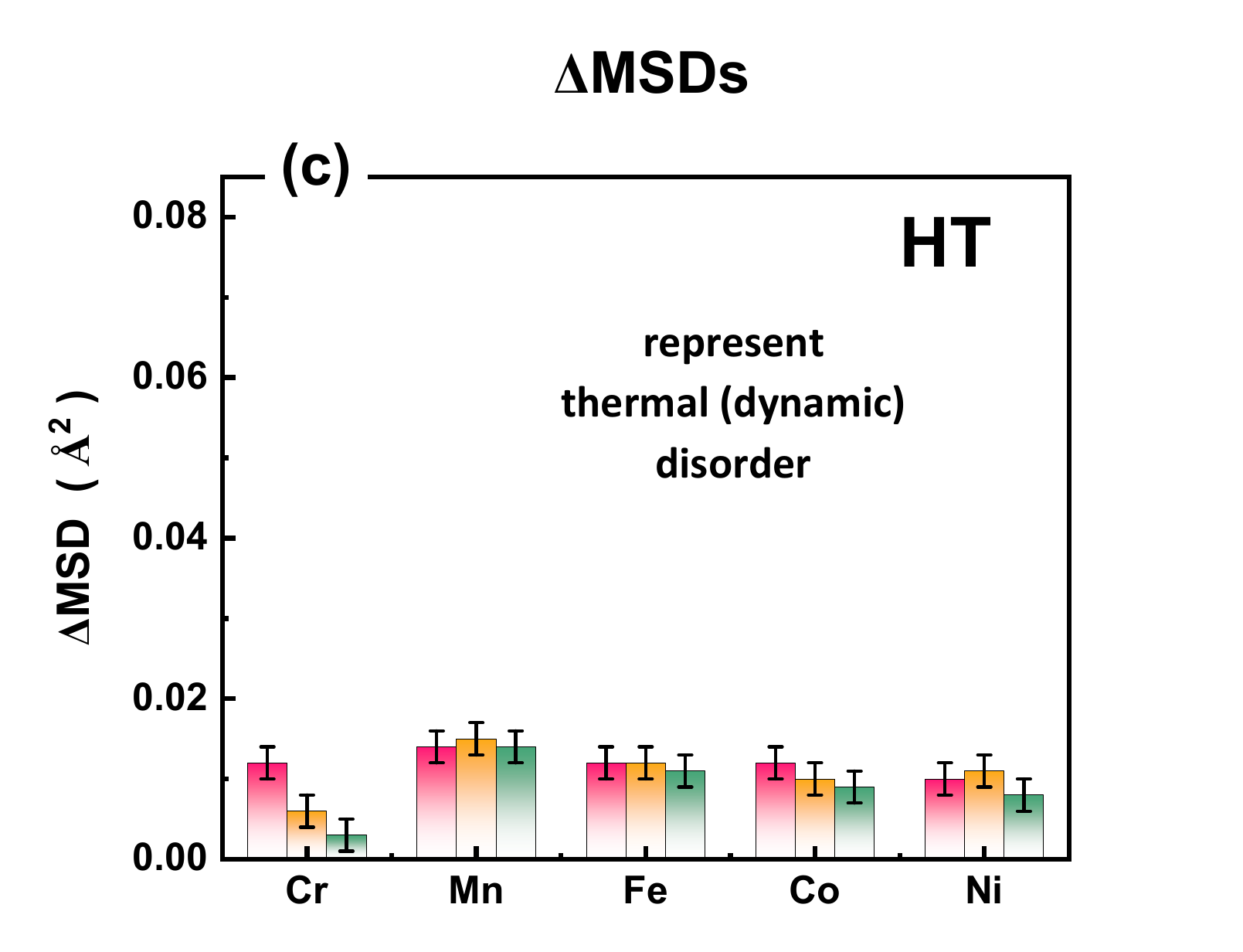}
\hfill
\includegraphics[width=0.32\textwidth]{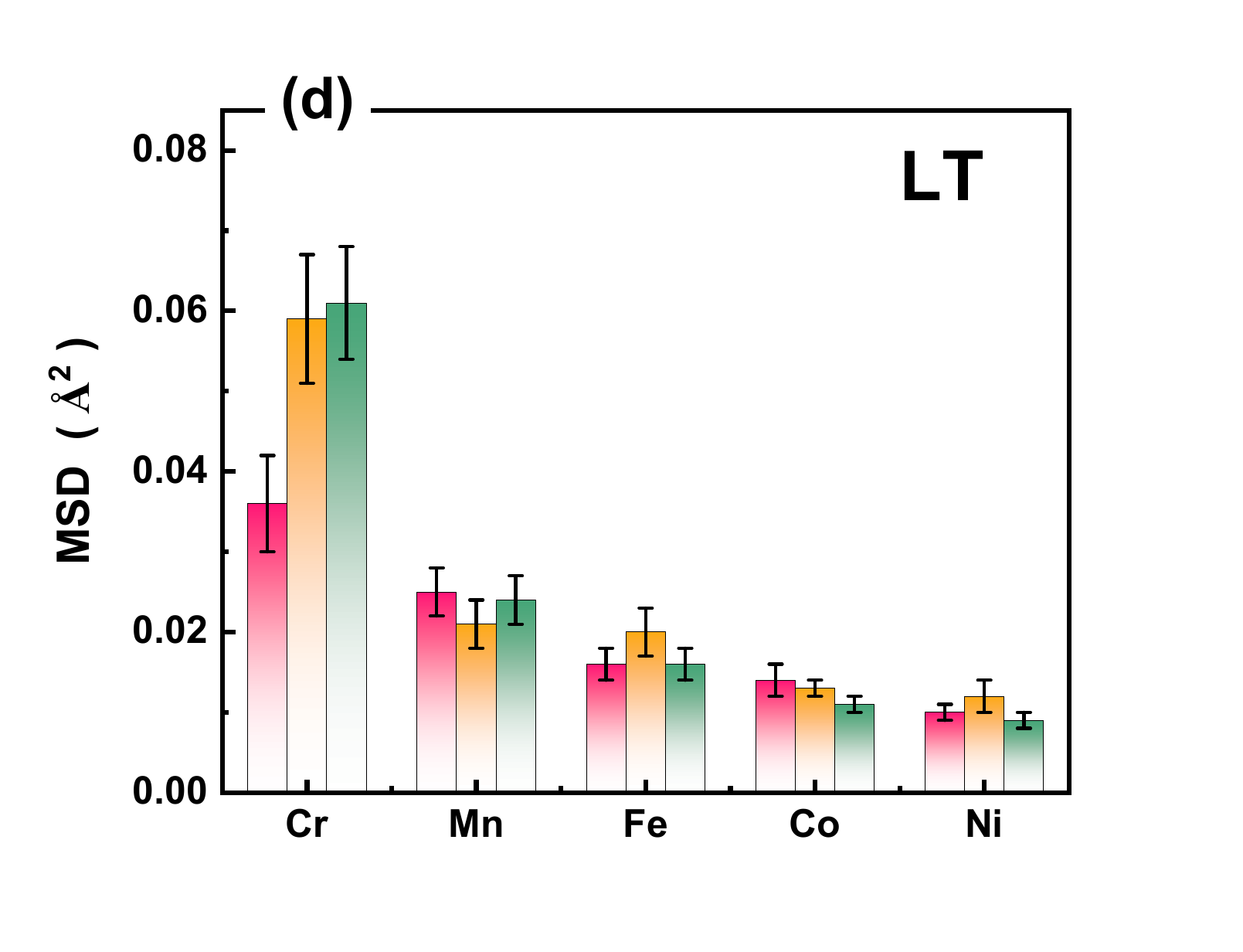}
\includegraphics[width=0.32\textwidth]{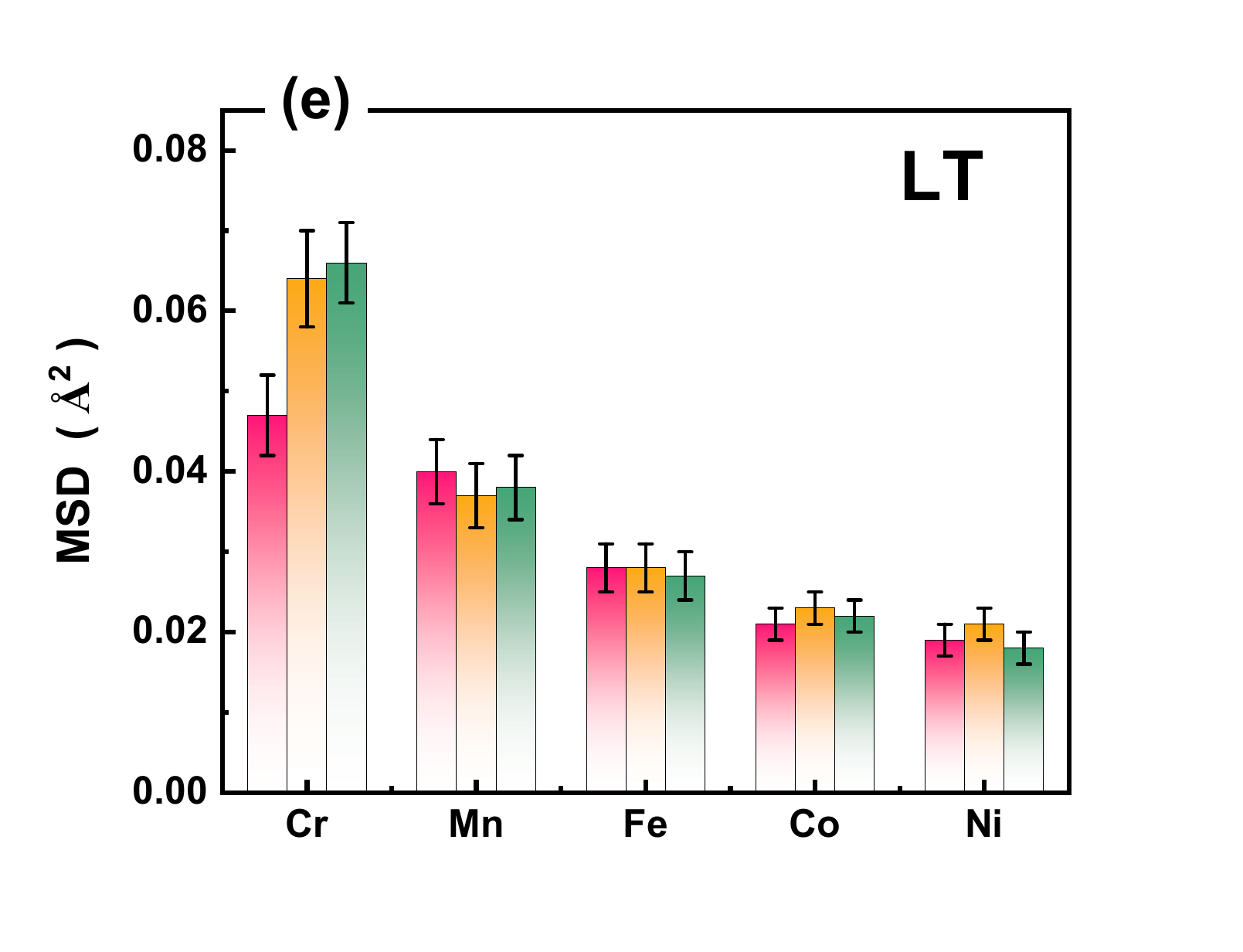}
\includegraphics[width=0.32\textwidth]{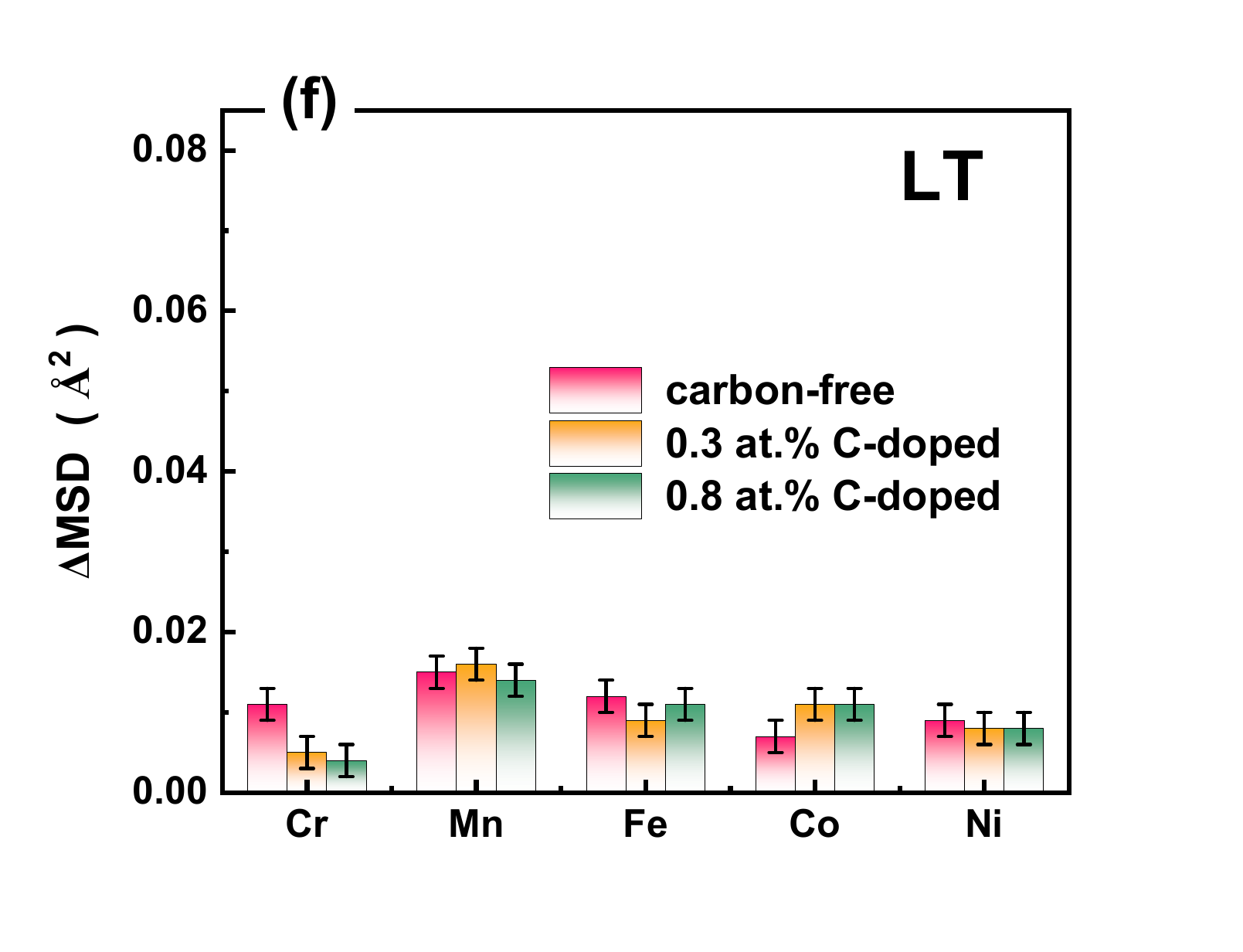}

\caption{The consolidated plot of experimentally obtained element-resolved MSDs and $\Delta$MSDs for the (CrMnFeCoNi)$_{1-x}$C$_x$ HEAs in HT (a, b, c) and LT (d, e, f) states for both temperatures (18~K and 285~K).}
\label{fig:consolidatedPlot}
\end{figure*}

The effect of thermal (dynamic) disorder expressed through $\Delta$MSDs is shown in Fig.\ \ref{fig:deltaMSDHTLT} for all alloys studied. The effect is significantly reduced specifically for the Cr constituent, whereas other components demonstrate the variations within the error bars of the method. The consolidated plot with all element-resolved MSD and $\Delta$MSD values obtained with a use of RMC-based analysis from the experimental EXAFS spectra of (CrMnFeCoNi)$_{1-x}$C$_x$ HEAs is presented in Fig.\ \ref{fig:consolidatedPlot}.

Throughout this work, the difference (residual) between the calculated and experimental EXAFS spectra is calculated involving their WTs as $\xi_{R,k} = \|w_{calc}(R,k) – w_{exp}(R,k)\|_2 / \| w_{exp}(R,k)\|_2$, where $w_{calc}$ and $w_{exp}$ are the Morlet wavelet transforms of calculated and experimental EXAFS spectra ($\chi_{calc}(k)k^{n}$) and $\chi_{exp}(k)k^{n}$, with $n=2$), respectively. The vertical lines $\|…\|_2$ denote the Euclidean norm. Typical values of $\xi_{R,k}$ for six samples at two temperatures for a single fit are shown in Table~\ref{tab:goodness}, and indicate a similar good fit quality for all samples at both temperatures.

\color{black}

\begin{table*}[h]
    \centering 
    \caption{ The residual difference $\xi_{R,k}$ between the calculated and experimental EXAFS spectra obtained through the comparison of their WTs. } \label{tab:goodness}   
    \begin{tabular}{lcccccc}
    \\
    \hline
         & \multicolumn{6}{c}{ $\xi_{R,k}$ } 
         \\
         & \multicolumn{3}{c}{HT state} & \multicolumn{3}{c}{LT state} \\
         & x = 0 at.\% & x = 0.3 at.\% & x = 0.8 at.\% & x = 0 at.\% & x = 0.3 at.\% & x = 0.8 at.\% \\
         & (HT) & (HT\textunderscore0.3\%) & (HT\textunderscore0.8\%) & (LT) & (LT\textunderscore0.3\%) & (LT\textunderscore0.8\%) \\
    \hline         
   18~K & 0.12  & 0.10  & 0.11  & 0.12  & 0.13  & 0.11 \\
  285~K & 0.082  & 0.085  & 0.092  & 0.086  & 0.090  & 0.092 \\
    \hline         
    \end{tabular}
    \label{tableKsi}
\end{table*}

\clearpage

\section{Convergence check in DFT}
\label{sec:ConvDFT}

In this appendix, we discuss the convergence of element-resolved MSDs in DFT with respect to the plane-wave cutoff energy, the k-point-grid density, the supercell size, as well as the MD simulation time.

For the static MSDs, a higher plane-wave cutoff energy (520\,eV) and a denser k-point grid (6 {\texttimes} 6 {\texttimes} 4) were investigated. We also computed the MSDs using 150-metal-atom cells with the ``ABCABC'' stacking of six close-packed \{111\} layers. The results are presented in Fig.~\ref{fig:convergence}. The differences from the results in the main text (Fig.~\ref{fig:DFT_MSDs_vs_T}) are at most 0.001\,{\AA\textsuperscript{2}}, and thus do not affect the derived conclusions.

\setcounter{figure}{0}
\renewcommand\thefigure{B\arabic{figure}} 
\begin{figure}[b]
\centering
\includegraphics{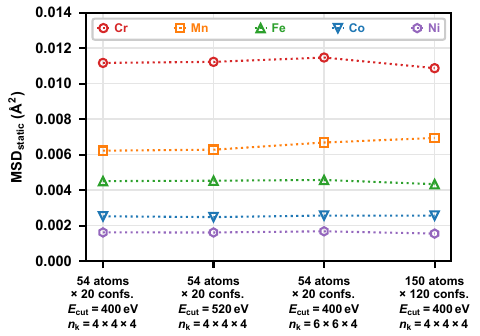}
\caption{Element-resolved static MSDs of CrMnFeCoNi without carbon obtained with DFT at 0\,K: The values from the main text (cutoff energy 400\,eV, 4 {\texttimes} 4 {\texttimes} 4 supercell size); the values obtained with a higher cutoff energy (520\,eV, the same supercell size); the values obtained with a denser k-point-grid density (cutoff energy 400\,eV, 6 {\texttimes} 6 {\texttimes} 4 supercell size); the values obtained from a larger supercell size (150 atoms) with 120 configurations (cutoff energy 400\,eV, 4 {\texttimes} 4 {\texttimes} 4 supercell size).}
\label{fig:convergence}
\end{figure}

\begin{figure}[htb]
\centering
\includegraphics{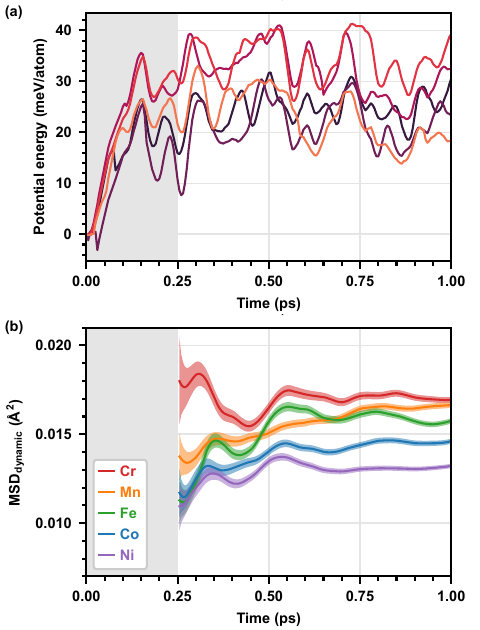}
\caption{(a) Potential energies from DFT-MD simulations of carbon-free CrMnFeCoNi at 300\,K for the five investigated configurations. The potential energy at 0\,ps is taken as the reference for each configuration. (b) Element-resolved dynamic MSDs over the five configurations and over the simulation time from 0.25\,ps onward.}
\label{fig:B2_alternative_H2}
\end{figure}

For DFT-MD at finite temperatures, the time evolution of the potential energies and the element-resolved dynamic MSDs are investigated, as presented in Fig.~\ref{fig:B2_alternative_H2}. The potential energies indicate that equilibration is achieved within the first 0.25\,ps, while the MSDs converge during the subsequent 0.5\,ps, demonstrating that the 1\,ps simulation time is sufficient to obtain converged MSD values.

\section{Impact of magnetism}
\label{sec:Magnetism}

In this appendix, we briefly discuss the magnetic moments of atoms in the DFT-MD trajectories and their relations to the atomic displacements, as presented in Fig.~\ref{fig:magmoms}. Cr and Mn atoms show both spin-up and spin-down magnetic moments, while the magnetic moments of the other elements are oriented predominantly to the spin-up direction. Cr also shows frequent spin flips during the MD simulations, whereas Mn shows much less frequent spin flips. The frequent spin flipping of Cr may also contribute to its comparatively large MSDs at finite temperatures, although no clear one-to-one correlation between the instantaneous magnetic moment and atomic displacement is observed.

\renewcommand\thefigure{C\arabic{figure}} 
\setcounter{figure}{0}
\begin{figure*}[tb]
\centering
\includegraphics[width=\linewidth]{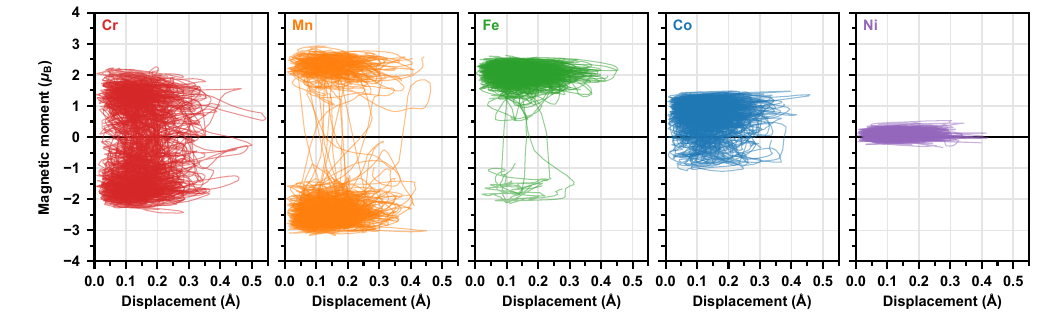}
\caption{Magnetic moments of atoms of carbon-free CrMnFeCoNi obtained from DFT-MD trajectories for the five investigated configurations as functions of their displacements from the time-averaged positions.
The local magnetic moments are obtained by projecting the plain-wave spin density onto local atomic regions and should therefore be interpreted qualitatively.}
\label{fig:magmoms}
\end{figure*}

\section{Dynamic MSDs}
\label{sec:Einstein_vs_MD}

Figure~\ref{fig:Einstein_vs_MD} presents the dynamic contribution to the element-resolved MSDs obtained from the Einstein model and MD based on DFT.
For all the metal elements, the zero-point vibrations are found to be around 0.005\,{\AA\textsuperscript{2}}, which is comparable to the static contributions (Fig.~\ref{fig:DFT_MSDs_vs_T}) and therefore non-negligible.
In contrast, already at 300\,K, the dynamic MSDs obtained from MD are substantially larger than those predicted by the Einstein model, indicating a significant contribution from collective vibrations of atoms (phonons) and/or their anharmonicity.
Assuming that the dynamic MDs obtained from MD increase approximately linearly with temperature, the contribution beyond the Einstein model is expected to be appreciable above approximately 100\,K.

\renewcommand\thefigure{D\arabic{figure}} 
\setcounter{figure}{0}
\begin{figure*}[tbh]
\centering
\includegraphics[width=\linewidth]{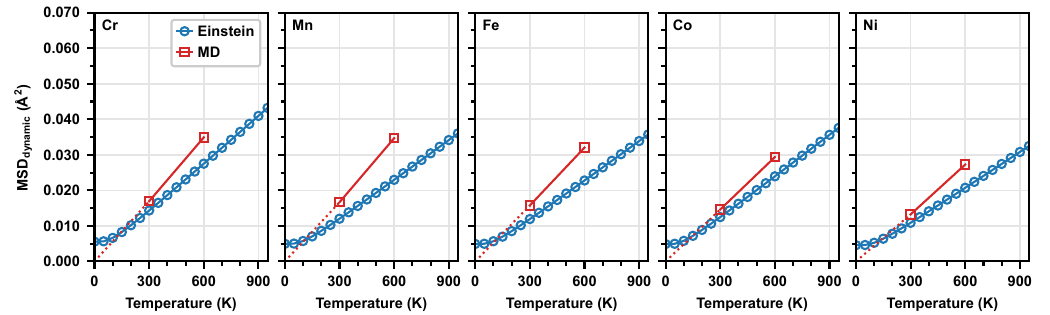}
\caption{Dynamic contribution to the element-resolved MSDs obtained from the Einstein model and MD based on DFT.}
\label{fig:Einstein_vs_MD}
\end{figure*}

\section{Impact of carbon}
\label{sec:CarbonImpact}

Figure~\ref{fig:DFT_MSDs_C_impact_all} presents the effect of carbon on the element-resolved MSDs in (CrMnFeCoNi)$_{1-x}$C$_x$ at 0\,K, 300\,K, and 600\,K obtained from DFT. The trends are essentially independent of temperature and remain the same as those observed at 0\,K (Fig.~\ref{fig:DFT_MSDs_C_impact_0.0}). Specifically, carbon primarily affects the static contribution to the MSDs, and this effect is largely confined to the 1NN metal atoms surrounding each C atom.

\renewcommand\thefigure{E\arabic{figure}}
\setcounter{figure}{0}
\begin{figure*}[tb]
\centering
\includegraphics{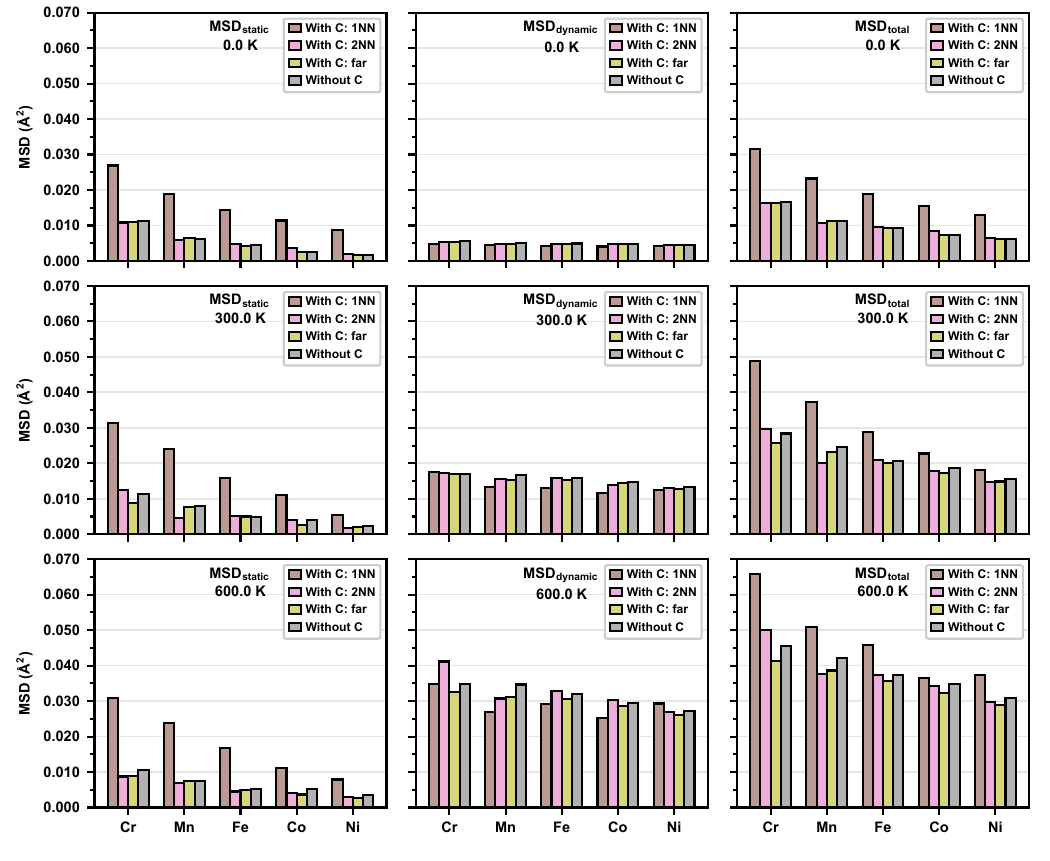}
\caption{Impact of carbon alloying on the element-resolved MSDs in (CrMnFeCoNi)$_{1-x}$C$_x$ ($x \approx 1.8$~at.\%) at 0\,K, 300\,K, and 600\,K in DFT.}
\label{fig:DFT_MSDs_C_impact_all}
\end{figure*}

\clearpage

\bibliography{lib,xafs,dft}

\end{document}